\documentclass[acmsmall]{acmart}

\AtBeginDocument{%
  }

    \newcommand{\sam}[1]{{\textcolor{black}{#1}}}
\newcommand{\yuling}[1]{{\textcolor{black}{#1}}}
\newcommand{\final}[1]{{\textcolor{black}{#1}}}

\setcopyright{acmlicensed}
\acmJournal{PACMHCI}
\acmYear{2025} \acmVolume{9} \acmNumber{2} \acmArticle{CSCW160} \acmMonth{4}\acmDOI{10.1145/3711058}

\usepackage{tabularx}
\usepackage{colortbl}
\usepackage{booktabs}  
\usepackage{makecell}
\usepackage{subcaption}
\usepackage{graphicx}
\usepackage{verbatim}

\begin{document}

\title[Technological Solutions for Community-Based Older Adult Care]{Rethinking Technological Solutions for Community-Based Older Adult Care: Insights from `Older Partners' in China}

\author{Yuling Sun}
\authornote{The first two authors contributed equally to this work.}
\email{yulingsun12@fudan.edu.cn}
\affiliation{%
  \institution{Fudan University}
  \country{China}
}

\author{Sam Addison Ankenbauer}
\authornotemark[1] 
\email{samank@umich.edu}
\affiliation{%
  \institution{University of Michigan}
  \country{USA}
}

\author{Zhifan Guo}
\affiliation{%
  \institution{Georgia Institute of Technology}
  \country{USA}}
\email{zguo381@gatech.edu}

\author{Yuchen Chen}
\email{yuchen.chen@baruch.cuny.edu}
\affiliation{%
  \institution{City University of New York}
  \country{USA}
}

\author{Xiaojuan Ma}
\affiliation{%
  \institution{Hong Kong University of Science and Technology}
  \country{Hong Kong SAR}}
\email{mxj@cse.ust.hk}

\author{Liang He}
\affiliation{%
  \institution{East China Normal University}
  \country{China}}
\email{lhe@cs.ecnu.edu.cn}

\renewcommand{\shortauthors}{Yuling Sun, et al.}

\begin{abstract}

Aging in place refers to the enabling of individuals to age comfortably and securely within their own homes and communities. Aging in place relies on robust infrastructure, prompting the development and implementation of both human-led care services and information and communication technologies to provide support. Through a long-term ethnographic study that includes semi-structured interviews with 24 stakeholders, we consider these human- and technology-driven care infrastructures for aging in place, examining their origins, deployment, interactions with older adults, and challenges. In doing so, we reconsider the value of these different forms of older adult care, highlighting the various issues associated with using, for instance, health monitoring technology or appointment scheduling systems to care for older adults aging in place. We suggest that technology should take a \textit{supportive, not substitutive} role in older adult care infrastructure. Furthermore, we note that designing for aging in place should move beyond a narrow focus on independence in one's home to instead encompass the broader community and its dynamics.

\end{abstract}



\received{January 2024}
\received[revised]{July 2024}
\received[accepted]{October 2024}

\maketitle

\section{Introduction}

The global growth of adults aged 65 and older has triggered an increasing demand for older adult care, a sector that requires critical attention \cite{DemographicCareBurden}. In the broadest sense, \textit{older adult care} encompasses a range of caring behaviors, technologies, and services aimed at meeting the physical, emotional, and social needs of older adults. \sam{Older adult care often extends across varied settings, incorporates multiple forms of infrastructure (transportation, healthcare, cultural)}, and features varied actors performing distinct roles (friends, family, neighbors, nurses, doctors, educators).

Older adult care can be institutionalized as part of the residential care industry as in assisted living facilities, nursing homes, or Continuing Care Retirement Communities (CCRC). However, these institutions have been noted for varied problems that have created a crisis of confidence in residential care \cite{powell2006social} that, coupled with a shortage of healthcare workers and the general preferences of older adults \cite{stones2016home}, led to an interest in aging in place. Aging in place \sam{refers to the enabling of individuals to age within their own homes and communities} \cite{wilesMeaningAgingPlace, vitmanschorrAgingPlaceQuality}. In response to an AARP survey, 89\% of individuals over 55 strongly or somewhat agree that they would prefer to live in their home for as long as possible \cite{bayer2000fixing}. 

However, issues with aging in place have been noted, from practical issues with older adults' homes (e.g., \cite{fausset2011challenges}) to health and wellness issues (e.g., \cite{caroux2014verification}). Technology has been presented as a solution that can enable aging in place \cite{caldeira2017senior, kim2017digital, wang2019technology}. \sam{Despite this presumed potential,} research has also noted such technologies have their limitations and distinct issues as well, from older adults being disinterested in utilizing the technologies (e.g., \cite{peek2016older}) to potential surveillance issues (e.g., \cite{berridge2020older}) to a fundamental disconnect with the needs and desires of older adults \cite{vines2015age, trothen2022intelligent}. In all, the extent to which technology \sam{should be incorporated into the structure of aging in place and into the everyday lives of older adults is contested.}

In China, 254 million people were aged 60 or over---and \sam{the country's} looming issues \sam{regarding} older adult care were recently framed by the Lancet as either a ``crisis or opportunity'' \cite{lancet2022population}. \sam{China's state government has been deeply involved in supporting its aging population with the development and implementation of policy that has technology as its center~\cite{jingxinwei, sun2023care, sun2023data}. Simultaneously, caregivers have been mobilized to support aging in place~\cite{olderpartner1, olderpartner2}.} The core philosophy of these human-centered initiatives is to mobilize social forces, especially retired individuals and volunteers, to participate in older adult care within the community as \textit{care partners}, to provide regular social interaction and emotional support, as well as to help older adults live at home more safely and comfortably \cite{thomas2020s, friedman2019aging}. \sam{Together, we refer to this assemblage of government policies, assistive technologies, care partners, and neighborhood-built environments that support the everyday living of older adults in the community as \textit{care infrastructure}. In our work, we show that the mobilization of care infrastructure in China is a sociopolitical move \cite{butler2006evaluating} supported by human-driven care programs that support the practical implementation of aging in place \cite{laohuobanpolicy}, while funding, policy support, and social resources have mainly been distributed to the private tech sector.}

In all, \sam{this care infrastructure} raises a series of critical questions: \textbf{1) How and why do existing technology interventions designed for aging in place work (or do \textit{not} work) to support aging in place?} \textbf{2) \sam{How and why do existing human interventions designed for aging in place work (or do \textit{not} work) to support aging in place?}} and \textbf{3) \sam{How can technology-driven and human-driven approaches to aging in place inform one another?}} Taken together, these questions can shed light on how to appropriately design, deploy, and support interventions for aging in place in the larger care infrastructure \cite{AIPdigitalcaretech1, kim2017digital, Lazar2018}).

In this work, we answer the above research questions by conducting a long-term field study in Shanghai, China. China has one of the fastest-growing aging populations in the world \cite{ChinaAging}, and Shanghai has one of the highest percentages of older adults per overall population in China \cite{shanghaiaging}. According to a report by the Shanghai Municipal Health Commission \cite{Shanghaireport}, the proportion of the population aged 60 and above in Shanghai is close to 40\%, entering into a ``severely aging society''---as such, the pressures on care infrastructure are extreme. In Shanghai’s case, over 97\% of \sam{older adults} age in place \cite{shanghai9073} and, as such, the Shanghai government is in the process of tackling the attendant issues of how to care for this vast population. Simultaneously, Shanghai is actively undergoing a citywide digital transformation \cite{shanghaidigitaltrans}. Given this confluence of aging, technology, \sam{and infrastructural development}, the Shanghai government has promoted the use of digital technologies to reform and empower home-based older adult care \cite{shanghai145}.

Encouraged by the Shanghai government, a majority of residential communities have installed and implemented digital technologies with the aim of building aging-friendly environments and supporting older adults to live independently within their own homes and communities \cite{shanghaitech4older}. Meanwhile, the Shanghai government and relevant social services departments have deployed the ``Older Partner program'' that encourages retired older adults who are ``younger'' (aged 75 and below) to be paired with ``older'' adults (aged 75 and above) who live alone. 
The stated goal of the program is to support all older adults living alone in the community through this mutual assistance network. The implementation of this program has been advertised throughout the neighborhoods of Shanghai \cite{olderpartner1, olderpartner2}. In this way, Shanghai provides us with an ideal context to understand aging in place, large-scale aging in place programs, caregiving in the context of older adult care partners, and the technologies that support caregiving.

We employed an ethnographic study that took advantage of multiple qualitative methods, including: semi-structured interviews with a diverse range of stakeholders; participant observations of 6 communities in Shanghai that have employed both technological interventions and the Old Partner program in supporting the care of older adults aging in place; and participant observation at 3 companies that primarily focus on developing elderly care technologies. Our research reveals that while these technologies are promising in some areas, such as improving management efficiency, they also encounter various challenges in effectively supporting aging in place, \sam{as when technological solutions used to monitor care recipients break down or when these solutions seek to supplant portions of extent care provider duties.} Additionally, our research also highlights the significance of older care partners who play a crucial and irreplaceable role in this context, emerging as pivotal figures in the overall care infrastructures of Shanghai.

Our work contributes to HCI and CSCW in three ways. First, and in keeping with calls to research human-computer interaction in ``diverse international sectors'' to ``better understand how diverse communities adopt and interact with [...] computing technologies and how they might, in turn, be leveraged to drive new interactions'' \cite[p.3066]{kumar2017hci}, we provide an empirical look at a unique problem/solution space---Shanghai \sam{with} its notable aging in place population\sam{, evolving infrastructure,} and exemplar caregiving program that does not just aid older adults but is informed and enacted by older adults. In doing so, we provide a particular \sam{context} through which to view aging in place.

Second, and in keeping with CSCW’s interest in incorporating coordinative practices into computational artifacts and vice versa \cite{ciolficscw, wallace2017technologies}, we draw on science and technology studies (STS) perspectives on infrastructure---particularly its focus on labor, relationships, and practices---to evaluate existing care infrastructure in our context, highlighting its successes, ongoing maintenance, and points of breakdown \cite{mattern2018maintenance, edwards2007understanding, jackson2014rethinking}. \sam{We bifurcate care infrastructure into human- and technology-driven assemblages. We do so not because they are separate in practice but, instead, to trace their respective histories, successes, contingencies, relations, and tensions that inform the greater project of aging in place.} In doing so, we critique a \final{vision of care} that supplants existing relationships, behaviors, and artifacts with technologies---such as when a system is designed to substitute an interaction between members of a community. We suggest that technology should be \sam{\textit{supportive, not substitutive}} \sam{within the larger care infrastructure. We then offer design considerations for ``supportive'' aging in place technologies, as informed by} our participant observation. 

Third, we consider the interactions between older adults, older adult care partners, and technology, as well as the nature of ``independence'' and ``dependence'' to consider the value of technology and community within the aging-in-place context. We move away from framing aging in place as a pursuit of "independence" (understood as transitioning from residential care to aging at home) and instead reframe it as a shift toward \textit{aging in community}. Accordingly, we advocate for future research to explore and design around the values and dynamics of aging in community. \sam{This change can be seen as a move away from health and wellness technologies that monitor individuals within their given homes towards technologies that inform community interaction.}

\section{Related Work}

\subsection{The Turn Towards Aging in Place and Its Challenges}

As a sociocultural, moral, political, and economic issue, older adult care has not one approach but many, each with distinct challenges. In the 20th century, older adult care was formalized within institutions---assisted living facilities, nursing homes, or Continuing Care Retirement Communities (CCRC) \cite{caldeira2017senior, zimmermanAssistedLivingNeeds2001}. Such facilitates \sam{were} part of a larger \sam{orientation} that framed aging as a medical issue, one that required institutional solutions \cite{powell2006social,  brown2007historical}. \sam{This industry faced} a crisis of confidence at the turn of the millennium (see: \cite{brown2007historical, mezuk2015suicide}). In the Chinese context, \sam{the residential care industry has developed in response to the} growing older adult population and \sam{the increasing distance between aging parents and their mobile children who leave for social and economic reasons \cite{zhang2006family}.}

Aging in place has been forwarded as an alternative aging orientation. As per the National Institute on Aging, ``many people want the same things as they get older: to stay in their own homes, to maintain independence for as long as possible, and to turn to family and friends for help when needed'' \cite{aging}. Research has shown that the predominance of older adults prefers to age in place \cite{stones2016home}, with local ties, social capital, neighborhood satisfaction, and more influencing the decision to remain at home \cite{clark2023place}. This turn towards aging in place has been recommended by international organizations (e.g., the World Health Organization \cite{world2017}) and governments alike. 

As aging in place is a fairly new orientation towards aging in our contemporary moment, there is a present need to develop models for aging in place \cite{semke2003older}. As institutional aging dominated, there is presently a lack of infrastructure designed to meet the needs of a population who desire to age in place \cite{gardner2003meeting}. Infrastructure needs to change to support the dynamic needs of older adults aging in place---local government policy, transportation, community design, and technology each need to be designed and developed to support an aging in place agenda \cite{lehning2012city}. Often, the things found within residential care---for instance, at-hand care and a consolidation of space---are not found in aging in place contexts and vice versa (for instance, despite the presence of a social community, older adults in residential care face a higher percentage of depression \cite{anstey2007prevalence}.) 

It should be noted that there is not an exact right answer or solution to these problems, but an individual and dynamic interplay of contexts and situations, informed by location, government funding, age, disability, informal care networks, and more. As such, the personal or family decision to age in place can be difficult. In China, this decision has been noted for its practical and ethical complexities \cite{zhang2006family}. Overall though, aging in place is considered a cheaper alternative for individuals and families, a solution to a crisis of care workers, and a more meaningful personal and social practice \cite{clark2023place, wilesMeaningAgingPlace}. \sam{Still}, issues of aging in place persist: practical issues with the homes of aging in place older adults \cite{fausset2011challenges}; a lack of structural and policy support to allow older adults to remain in place \cite{lau2007health}; an overreliance on informal social networks and family, which may not exist or may not form an effective basis for community caregiving \cite{bulmer2015social}. 


\subsection{Technology \sam{as Mediating Aging in Place}}

The development of technologies has been offered as a way of combating the above issues facing older adults aging in place. In CSCW and its neighboring fields, technology has been framed as a potential tool in ``enabling'' aging in place \cite{caldeira2017senior, kim2017digital, wang2019technology}. For instance, increasingly advanced, accessible, and affordable assistive technologies, sensing technologies, and data-driven healthcare technologies have been presented as artifacts with which to support and facilitate aging in place for older adults \cite{sun2015reliving, sun2017method, zhou2018data, ni2019human,xiao2024chinese}. Researchers and practitioners in various industries, government \sam{agencies}, and academic communities have speculated on the significant potential of these technologies in enabling care reform and addressing \sam{the} existing challenges of aging in place. Take, for instance, the following:

\begin{quote}
    [Advancing aging in place] will require substantial innovation in the incorporation of advanced technologies [...] Recent and ongoing innovations in digital health technologies, in particular, have great potential to have a transformative impact on diagnosing, preventing, monitoring, and treating a wide range of conditions. \cite[p.26]{kim2017digital}
    \end{quote}

Here, technologies support the agenda of aging in place---in particular, digital health technologies support the diagnosis and monitoring of health conditions \cite{li2023dynamic}. In these ways, aging in place continues the tradition of the medicalization of aging. In HCI and CSCW, the predominance of research on aging in place \sam{investigates} the design and implementation of digital health technologies, turning the home into a ``living \sam{laboratory}'' of health and wellness devices \cite{ holbo2013safe, consel2015unifying}. For instance, research has been interested in the development of health monitoring technologies for older adults (e.g., \cite{shen2023privacy, guo2022caremap, wang2022lightweight}.) In these and other works, aging in place is made possible through technological innovation. Technologies can ``increase independence in daily living to allow ageing in place (not having to move to an institution)'' \cite[p.4]{maciuszek2005help}. Older adults with dementia run the ``risk of relying on [their] own abilities'' if not supported adequately with technology \cite{holbo2013safe}. When older adults aging in place are monitored with the help of technology, ``deviations are a warning sign of degradation'' and early warnings can support continued aging in place \cite[p.43]{caroux2014verification}.  

A large body of literature has noted the issues and limitations of these technologies. Research has observed that older adults sometimes have limited acceptance of such technologies, suggesting that they are not relevant to their lives---or otherwise ageist in their assumptions and design \cite{felt2016handbook, peek2016older}. Technologies that allow for communication and monitoring have been investigated for their potential as surveillance tools, allowing younger family members and other caregivers to have an unequal distribution of power over older adults \cite{berridge2020older, piper2016technological, vitak2021designing}. Some have noted that these technologies demand certain behaviors out of older adults, tools in a larger biopolitical apparatus \cite{petrakaki2018between}. Others still have suggested that such technologies actually do not speak to the needs and desires of older adults \cite{trothen2022intelligent}. 

In keeping with previous CSCW and HCI research, we consider the introduction of technologies into domestic settings to have important (and sometimes unintended) consequences (e.g., \cite{ oogjes2018designing, soubutts2021aging}). \sam{As such, it is important to understand already extant practices and interactions within a given context} before designing technology on a large scale. We investigate both current human- and technology-driven care infrastructures for aging in place to consider this point further. If the turn towards aging in place is a series of practical challenges and technology is central to this push, how successfully does technology attend to these challenges? Can and should it supplant other methods of attending to aging in place, including care relationships?

\subsection{\sam{Infrastructure and an Orientation Toward Aging in Community}}
The lens of infrastructure endows us with relational thinking and orientation towards the social in approaching care and aging in place, which matters to CSCW as a field dedicated to ``sustained social activities'' and interactions \cite{ciolficscw}. STS scholars Star and Ruhleder note it is less ``\textit{what} is an infrastructure'' but rather, ``\textit{when} is an infrastructure.'' \final{This means that infrastructure is enacted and recognized by different people in different contexts, who take different things as infrastructure with respect to their everyday experiences \cite{star1994steps}. Categorizing and discussing what is or can be considered care infrastructure, then, requires one to consider infrastructure as both shaping social relations and as ``fundamentally social assemblies'' in and of themselves \cite{larkin2013politics, anand2018public}. In this paper, we turn our focus to the social relations and networks that sustain care infrastructures and are continuously (re)constructed through infrastructural practices \cite{jackson2014rethinking, ribes2010sociotechnical}.} \final{We thus foreground the understanding of infrastructure as an infinite regress of relations, instead of a ``thing'' \cite{star1994steps, bateson2000steps}.} In our case, this line of inquiry orients us to socialities in practice and unveils the work that develops, organizes, implements, and repairs technology \cite{jackson2014rethinking} and helps sustain the everyday functioning of care infrastructures. Such an orientation provides a social understanding of the implementation of technology in our health and caregiving context---for further work in this area, see the work of Gui and Chen \cite{gui2019making} and Kaziunas et al. \cite{kaziunas2019precarious}. 

\sam{Our focus on the value of the social is informed by infrastructure studies but also extends to our interests in aging in place.} As noted above, aging in place is often considered a lateral move: independent individuals in a given home get to remain independent and within that home. This is often the stated appeal of aging in place research (e.g., \cite{vacher2015evaluation}): in this way, ``Rather than experience a loss of independence, we remain masters of our own domain'' \cite[p.13]{thomas2009moving}. While this is significant, it should also be noted that one of the positive aspects of aging in place is aging in a place one knows and this extends beyond a given home \cite{clark2023place}. Aging in place is also aging in a given community, alongside the people and the places one knows. This aspect of aging in place is sometimes underemphasized in technology research that investigates primarily technologies of the individual domestic sphere. 

The importance of community for older adults has been noted by other research (e.g., \cite{trothen2022intelligent}) as in, for instance, a study exploring the value of community services in addressing the needs of older adults aging in place \cite{greig2019transforming}. Aging in community has been investigated for its sustainability, positive focus on interdependence, and promotion of community participation and engagement \cite{thomas2009moving}. It should also be noted that if the adoption of technology is a social process \cite{peek2016older}, one's community at large can influence which technologies get used and which do not. Informed by this work, we take this opportunity to explore aging in community and aging alongside other community members. In exploring our community setting, in which older care recipients, older partners, and neighborhood committee members each inform aging in place, we highlight aging in community as a significant orientation and an under-explored area of aging in place research within CSCW.

\sam{Additionally, a relational and infrastructural lens helps foreground the paradoxical nature of care infrastructure: there exists a tension between local, intimate, and situated use \cite{suchman1987plans, chen2023maintainers} on the one hand and a need for standardization and scalability on the other \cite{star1989institutional, star1994steps}. To examine the tension between the situated nature of care and the desire for care to scale, we consider two assemblages of care infrastructure: one predicated on technology-driven care and one predicated on human-driven care. These assemblages begin with different considerations of what care should look like but, in practice, are themselves intertwined. Through them, we can consider the relations that inform them, the relations that they inform, and the tensions between situated and standardized care. Below, we unpack our context and these assemblages further.}

\section{Research Context: Care Infrastructure for Older Adults Aging in Place in Shanghai}

Our investigation is rooted in the context of communities and districts in Shanghai, China, where both human- and technology-driven care infrastructures have been established to support older adults aging in place. 
As one of the cities with the highest degree of aging \cite{shanghaiaging} and the benchmark city for digital transformation in China \cite{shanghaidigital}, Shanghai has established a comprehensive and complex sociotechnical older adult care infrastructure to address the significant pressures facing home-based older adult care, which provide\sam{s} us with an ideal context to identify and research various technologies and care providers in their natural environments. Below, we unpack further our specific context, providing legibility \sam{and situating} information for those who may not know about \sam{this} Chinese context. We break up care infrastructure into two sections: technology- and human-driven care infrastructures. In practice, these are not easily separated. We do so here to trace the relations and practices of different approaches to care, an approach that is taken throughout the paper.

\subsection{\yuling{Technology-Driven Care Infrastructures}}
\begin{figure}[ht]
 
\centering
\includegraphics[scale=0.40]{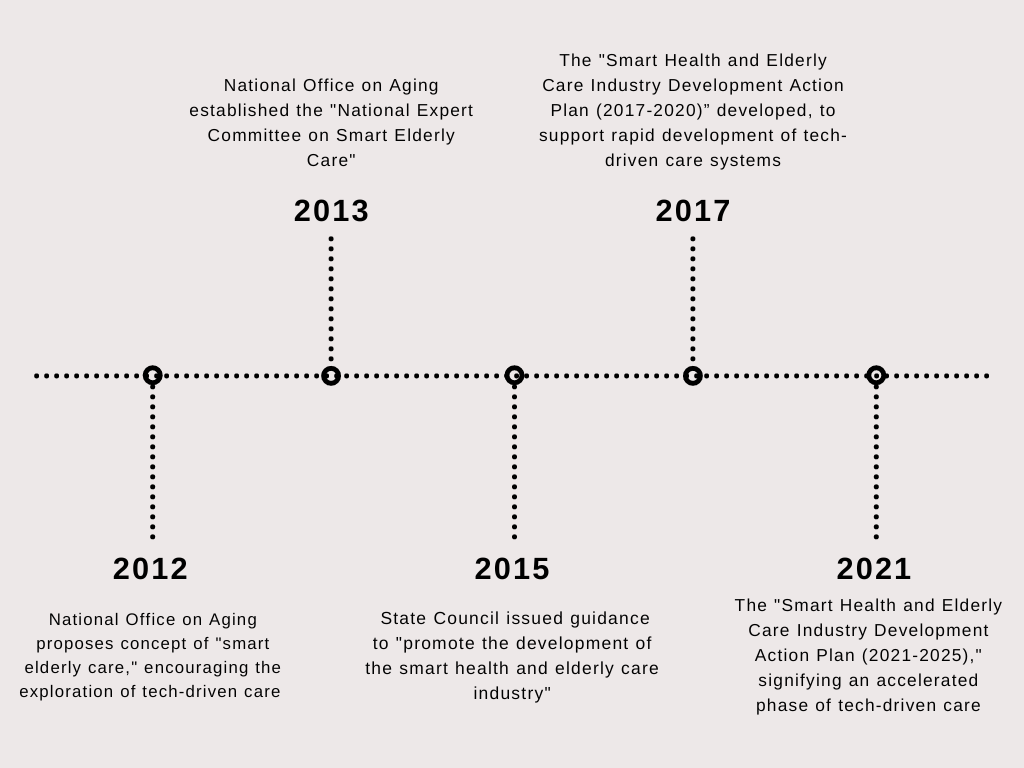}
\caption{\sam{A timeline of a national push for technology-driven care infrastructure~\cite{xiaohui}}}
\Description{A timeline with five years (2012, 2013, 2015, 2017, 2021), each signifying growth and change in China's technology-driven care infrastructure for older adults, from the National Office on Aging proposing the concept of "smart elderly care" to the "Smart Health and Elderly Care Industry Development Action Plan", an acceleration of tech-driven care in China.}
\label{Fig:timeline}
\end{figure}

With the trend of citywide digital transformation \cite{shanghaidigitaltrans}, technology-related departments of the Shanghai government, such as Shanghai Municipal Commission of Economy and Informatization and Shanghai Municipal Science and Technology Commission, have issued policies and work guidelines \cite{jingxinwei, weijianwei}, encouraging and \final{supporting} the use of various Information and Communication Technologies (ICTs) and Artificial Intelligence (AI) technologies to facilitate home-based older adult care and supporting older adults to live at home safely and independently. One policy, for instance, issued in 2022, encouraged science and technology committees, economic committees, health committees, civil bureaus, and relevant corporate and institutional interests to aid in the implementation of these technologies \cite{jingxinwei}. \sam{This follows a decade-long national push to develop technology-driven care infrastructure---see Figure \ref{Fig:timeline}.}

\final{The propositions, guidances, and plans espoused by various state and municipal departments found in Figure 1 enact a certain kind of vision for care. This vision informs ``technology-driven care infrastructure,'' the assemblage of technologies, systems, actors, and environments that collectively shape caregiving practices and in which technology plays a central and integrative role.} Based on \sam{national and city} government encouragement, various types of technologies for aging in place have increasingly been developed and deployed. An illuminating example of a tech-driven infrastructural assemblage for aging in place can be found in the ``Home-based Virtual Nursing Home'' initiative \cite{virtualbed1}, \sam{a plan to supplant the nursing and care services from care facilities into the homes of older adults aging in place.} In this initiative, various monitoring and data-driven technologies and devices are used to sense and assess the health and safety conditions of older adults at home, to ultimately identify whether the older adult is safe and healthy \cite{4tech1, 4tech2}. \sam{The basic technologies used in this instance include life status sensing technologies, smoke detectors, gas detectors, door magnetic sensors, infrared body sensors, and more \cite{4tech1}. Based on this general foundation of technology-driven care infrastructure, various additional sensing technologies could be added to an older adult's home---wearable technology for vital sign monitoring, systems for emergency calling---and when anomalies or dangers are detected, these systems can automatically issue alerts and dispatch relevant care providers to deliver assistance, according to the presumed needs of older adults.}
\raggedbottom

\subsection{\yuling{Human-Driven Care Infrastructures}}
Meanwhile, the civil affairs service departments of the Shanghai government, such as the Shanghai Civil Affairs Bureau, have rolled out relevant policies (e.g. \cite{human2, human1}) to encourage the establishment of human-driven care infrastructures for older adults aging at home. \final{We call this vision for care ``human-driven care infrastructure,'' taken to mean the assemblage of systems, actors, and environments that collectively shape caregiving practices, with individual actors playing a central and guiding role.} In these policy briefs, various social forces and individuals within the community, especially retired individuals and other volunteers, are mobilized to participate in daily services, helping older adults live at home safely and comfortably. In essence, these policy briefs present \final{an alternative vision for what care, in this context, should look like,} with similar stated aims as technology-driven care but with different ways of achieving their converging goals. 



One representative initiative is the ``Older Partner program'' \cite{olderpartner1, olderpartner2, olderpartner3}, a nationwide and vigorously promoted community-based mutual aid initiative dedicated to older adult care. In this initiative, retired older adults who are ``younger'' (often aged 75 and below) are paired with ``older'' older adults (aged 75 and above) who live alone to engage in a community-based assistance network that provides various kinds of care services---regular home visits, policy advocacy, health education, companionship, and emotional support \cite{olderpartner3}. Often, each \sam{``younger''} older adult is paired with 5-10 \sam{``older''} older adults. \sam{Local districts organize older partner initiatives} through neighborhood committees and non-profit social welfare organizations with organizers providing limited economic or material subsidies to the younger older adult---such as a small monetary stipend or \sam{gifts of} essential \sam{items} during the holidays. The overall purpose of this initiative is to cover all older older adults living safely at home. The Older Partner program and similar initiatives have been developed over the past decades in China \cite{olderpartner4}, with the Older Partner program itself being launched in 2012, the same year the National Office on Aging proposed \sam{research into future} ``smart elderly care.'' In this way, human-driven care infrastructures predated technology-driven care infrastructures and play a significant role in China's home- and community-based older adult care service system. \final{In the case of these distinct visions of care, we trace how they are enacted in practice. We then critique the assemblages that result and make research and design recommendations based on these critiques.}

\section{Methods}
\sam{Our study aims to gain an in-depth empirical understanding of human- and technology-driven care infrastructures in practice.} We are particularly focused on understanding: 1) how current technological and human infrastructures work in real-world situations, 2) the respective roles and challenges inherent in these systems, and 3) any collaborative efforts or interactions between them. We explore our research question\sam{s} through a long-term two-year ethnographic research \sam{project}. 
The primary data collection was conducted by the first author, a native Mandarin speaker, and the subsequent analysis involved collaboration among three co-authors. In the following sections, \sam{we} unpack our research methods, including participant observation and semi-structured interviews, \sam{and discuss our} participants and data analysis \sam{process}.




\subsection{Data Collection: Participant Observations}
We began with long-term participant observations in six local communities within three districts of Shanghai, China, from June 2021 to December 2023\footnote{This research was put on hold for nearly one year due to pandemic lockdowns that precluded participant observation, especially in the case of vulnerable older adult populations. Due to Shanghai's size and population density, there is no single downtown district in Shanghai---instead, there are 16 districts within Shanghai proper and each district contains multiple subdistricts and communities. There are 4,463 listed communities within Shanghai.}. As is common with participant observation, the purpose of this ethnographic approach is to gain a thorough understanding of a setting through the eyes of an observer within the space itself---in particular, we observed how the current \sam{care infrastructure works to support home- and community-based care for older adults in practice.}

\subsubsection{Recruitment and Criteria}
The first author was affiliated with the X\footnote{The name of the department has been anonymized as per our IRB policy} Department's older adult care project as a working member of the institution. This department has issued a letter of approval for the research conducted within this work. Aside from the year-long lockdown during the COVID-19 pandemic, the majority of older adult service centers and neighborhood committees in Shanghai were open to the public. After obtaining IRB approval, the first author, carrying X Department's introductory approval letter, visited older adult service centers and neighborhood committees in different districts of Shanghai. The criteria for potential participation was: 1) the served neighborhood had a relatively high degree of aging populations, 2) the center or committee had already implemented at least one kind of older adult care technology, and 3) the center or committee had already implemented the Older Partner program to support daily older adult care services in the community. In this criteria, we did not specify specific technologies or how long the Older Partner program had been implemented so that the observed communities could have various technology-driven older adult care systems at play and various levels of engagement in human-driven older adult care systems.


After identifying potential communities for participation, the first author actively reached out to the staff in the communities' older adult service centers, disclosing our intention of doing research about older adult care in the community, requirements for participation (including long-term participant observations and interviews with stakeholders), IRB approval and ethical considerations, and our commitment to data privacy and safety, and obtained their permission for the study. In all, we observed six communities in three different districts of Shanghai. Each of these communities had employed a series of technology interventions for older adult care. Each of these communities also had built robust community mutual aid networks primarily composed of older partners. In this way, each space was itself ideal as a research site for our particular line of inquiry.

\subsubsection{Participant Observation}

\yuling{After obtaining this permission, the first author started participant observation. 
To obtain a thorough understanding of \sam{how existing infrastructure works to support} care for older adults aging at home, she volunteered alongside old partners and staff members in the six communities listed to provide services for older adults at home. }During this work, she observed and documented the practical caregiving processes and behaviors of relevant stakeholders, and the technologies, tools, and materials they employed. She also actively interacted with relevant stakeholders to understand their experiences, feelings, encountered issues, and solutions in existing older adult \sam{caregiving}. Meanwhile, employing the behavioral observation method \cite{bakeman2012behavioral}, she specifically observed and recorded the daily work processes and activities of older partners.
In addition, she also followed various care cases and tracked \sam{meaningful events that occurred during her research through memoing}. Handwritten field notes, photographs, and videos were used to record data.

Through nearly two years of participant observations, accounting for over 100 days of volunteer work, 60 participant observation \sam{activities concerning} older adult care events in the six communities, 80 pages of notes, as well as a large amount of image and video data, we acquired a thorough \sam{understanding of what existing care} infrastructures looked like in these communities, and how they worked (and didn't work) \sam{towards} supporting real-world care for older adults aging in place. These observations were recorded as they occurred and were \sam{later} analyzed, leading to the overall structure of this paper, as discussed below.

\subsection{Data Collection: Semi-structured Interviews}

Another method of the ethnographic process, the semi-structured interview, allows the researcher direct access to the intentions, opinions, and beliefs of their participants, and one can triangulate this data with results from participant observation. To deeply investigate 1) how older partners, neighborhood committee members, and project managers understand the current technologies that support aging in place and caregiving, 2) older adults' attitudes, perceptions, and expectations of future technology interventions in supporting aging at home, and 3) the caregiving practices, experiences, and encountered challenges of older partners and other related stakeholders, we conducted in-depth semi-structured interviews with 24 stakeholders. 

\subsubsection{Recruitment and Participants}

Our participants were primarily recruited during participant observation \cite{taylor2013ethnography}. While conducting the observations, the first author actively reached out to stakeholders who were involved in the care process, primarily the \sam{``younger''} older adult care partners and neighborhood committee members. Meanwhile, during the months of participant observations, technology suppliers frequently visited the communities to deploy and assess the usage of their technologies. The first author, therefore, got to know \sam{and interview} several project managers of technology suppliers, seeking to understand their existing products and the ongoing deployment and assessment of these technologies.

\final{This work focuses on the perspectives of caregivers---care partners, neighborhood committee members, and those who make technology-driven older adult care systems.} In all, we recruited 24 stakeholders, including 15 older partners, 6 workers from neighborhood committees, and 3 project managers of technology-driven older adult care systems. \sam{This diversity of stakeholder roles speaks to the assemblage of actors that develop, inform, and implement ``care.'' To review these stakeholder roles: 1) The older partners are predominately retired older adults who are ``younger'' (aged 75 and below) and paired with ``older'' (aged 75 and above) adults who live alone.} \sam{Older partners engage in a community-based assistance network that provides various kinds of care services---regular home visits, policy advocacy, health education, companionship, and emotional support. 2) The staff in neighborhood committees are government employees who act as mediators between policy and action. They are responsible for enacting the Older Partner program and bringing care work into the local community. 3) Project managers work for smart home and care technology companies. They work in tandem with national policy to support the burgeoning tech-driven care infrastructure and can be found developing and deploying their technologies locally in Shanghai. Participants' demographic information is shown in Table 1, including location in Shanghai, age, gender, and years of experience in their given role.}
    
\begin{table}[h]
\centering
\begin{tabular}{
>{\columncolor[HTML]{FFFFFF}}l 
>{\columncolor[HTML]{FFFFFF}}l 
>{\columncolor[HTML]{FFFFFF}}l 
>{\columncolor[HTML]{FFFFFF}}l 
>{\columncolor[HTML]{FFFFFF}}c 
p{0.2in}cclll}
\rowcolor{black} 
\cline{1-10}
\cline{1-5} \cline{7-10}
\cline{1-5} \cline{7-10}
\cline{1-5} \cline{7-10}
\cline{1-5} \cline{7-10}
\cline{1-5} \cline{7-10}
\cline{1-5} \cline{7-10}
\cline{1-5} \cline{7-10}
\cline{1-5} \cline{7-10}
\cline{1-5} \cline{7-10}
\cellcolor[HTML]{D0CECE}\textbf{ID} & 
\cellcolor[HTML]{D0CECE}\textbf{Age} & 
\cellcolor[HTML]{D0CECE}\textbf{Gender} & 
\cellcolor[HTML]{D0CECE}\textbf{Location} & 
\cellcolor[HTML]{D0CECE}\textbf{Experience} &  & 
\cellcolor[HTML]{D0CECE}\textbf{ID} & 
\cellcolor[HTML]{D0CECE}\textbf{Age} & 
\cellcolor[HTML]{D0CECE}\textbf{Gender} & 
\cellcolor[HTML]{D0CECE}\textbf{Location} \\ 
\cline{1-5} \cline{7-10} 
\cline{1-5} \cline{7-10} 
\cline{1-5} \cline{7-10}
\cline{1-5} \cline{7-10}
\cline{1-5} \cline{7-10}
\cline{1-5} \cline{7-10}
\cline{1-5} \cline{7-10}
\cline{1-5} \cline{7-10}
\cline{1-5} \cline{7-10}
\cline{1-5} \cline{7-10}

\multicolumn{5}{l}{\cellcolor[HTML]{FFFFFF}\textbf{Older Partner}}                                                                                                                                             &  & 
\multicolumn{4}{l}{\cellcolor[HTML]{FFFFFF}\textbf{Staff in Neighborhood Committee}}  

\\ \cline{1-5} \cline{7-10} 
\cline{1-5} \cline{7-10} 
\cline{1-5} \cline{7-10}
\cline{1-5} \cline{7-10}
\cline{1-5} \cline{7-10}
\cline{1-5} \cline{7-10}
\cline{1-5} \cline{7-10}
\cline{1-5} \cline{7-10}
\cline{1-5} \cline{7-10}
\cline{1-5} \cline{7-10}
Mei                                  & 69                                   & Female                                  & Cyang                                     & 14                                          &  & \cellcolor[HTML]{FFFFFF}Wei          & \cellcolor[HTML]{FFFFFF}36           & \cellcolor[HTML]{FFFFFF}Female          & \cellcolor[HTML]{FFFFFF}Cyang             \\
Hui                                  & 69                                   & Female                                  & Cyang                                     & 12                                          &  & \cellcolor[HTML]{FFFFFF}Yu          & \cellcolor[HTML]{FFFFFF}57           & \cellcolor[HTML]{FFFFFF}Female          & \cellcolor[HTML]{FFFFFF}Jiangwa           \\
Xiao                                  & 65                                   & Female                                  & Beixin                                    & 7                                           &  & \cellcolor[HTML]{FFFFFF}Jing          & \cellcolor[HTML]{FFFFFF}43           & \cellcolor[HTML]{FFFFFF}Female          & \cellcolor[HTML]{FFFFFF}Xihua             \\
Fang                                  & 69                                   & Female                                  & Beixin                                    & 12                                          &  & \cellcolor[HTML]{FFFFFF}Guo          & \cellcolor[HTML]{FFFFFF}38           & \cellcolor[HTML]{FFFFFF}Male            & \cellcolor[HTML]{FFFFFF}Beixin            \\
Yang                                  & 64                                   & Female                                  & Beixin                                    & 6                                           &  & \cellcolor[HTML]{FFFFFF}Jun          & \cellcolor[HTML]{FFFFFF}41           & \cellcolor[HTML]{FFFFFF}Male            & \cellcolor[HTML]{FFFFFF}Shuyang           \\
Li                                  & 66                                   & Female                                  & Beixin                                    & 10                                          &  & \cellcolor[HTML]{FFFFFF}Tian          & \cellcolor[HTML]{FFFFFF}36           & \cellcolor[HTML]{FFFFFF}Male            & \cellcolor[HTML]{FFFFFF}Kojiang           \\ \cline{7-10} \cline{7-10} \cline{7-10} \cline{7-10} \cline{7-10} \cline{7-10} \cline{7-10} \cline{7-10} \cline{7-10} \cline{7-10} 
Qiu                                  & 68                                   & Female                                  & Shuyang                                   & 15                                          &  & \cellcolor[HTML]{FFFFFF}            & \cellcolor[HTML]{FFFFFF}             & \cellcolor[HTML]{FFFFFF}                & \cellcolor[HTML]{FFFFFF}                  \\ \cline{7-10} \cline{7-10} \cline{7-10} \cline{7-10} \cline{7-10} \cline{7-10} \cline{7-10} \cline{7-10} \cline{7-10} \cline{7-10} 
Lan                                  & 65                                   & Female                                  & Shuyang                                   & 7                                           &  & \multicolumn{4}{l}{\cellcolor[HTML]
{FFFFFF}\textbf{Project Manager}}                                                                 \\ \cline{7-10} \cline{7-10} \cline{7-10} \cline{7-10} \cline{7-10} \cline{7-10} \cline{7-10} \cline{7-10} \cline{7-10} \cline{7-10}
Jing                                  & 43                                   & Female                                  & Shuyang                                   & 2                                           &  & \cellcolor[HTML]{FFFFFF}Lei          & \cellcolor[HTML]{FFFFFF}             & \cellcolor[HTML]{FFFFFF}Male            & \cellcolor[HTML]{FFFFFF}Putuo             \\
Ya                                 & 64                                   & Female                                  & Kojiang                                   & 7                                           &  & \cellcolor[HTML]{FFFFFF}Zhi          & \cellcolor[HTML]{FFFFFF}             & \cellcolor[HTML]{FFFFFF}Male            & \cellcolor[HTML]{FFFFFF}Yangpu            \\
Ming                                 & 75                                   & Female                                  & Kojiang                                   & 11                                          &  & \cellcolor[HTML]{FFFFFF}Kai          & \cellcolor[HTML]{FFFFFF}             & \cellcolor[HTML]{FFFFFF}Male            & \cellcolor[HTML]{FFFFFF}Xuhui             \\
Shu                                 & 67                                   & Female                                  & Xihua                                     & 9                                           &  & \cellcolor[HTML]{FFFFFF}            & \cellcolor[HTML]{FFFFFF}             & \cellcolor[HTML]{FFFFFF}                & \cellcolor[HTML]{FFFFFF}                  \\
Xiang                                 & 66                                   & Female                                  & Xihua                                     & 8                                           &  & \cellcolor[HTML]{FFFFFF}            & \cellcolor[HTML]{FFFFFF}             & \cellcolor[HTML]{FFFFFF}                & \cellcolor[HTML]{FFFFFF}                  \\
Yun                                 & 63                                   & Female                                  & Xihua                                     & 5                                           &  & \cellcolor[HTML]{FFFFFF}            & \cellcolor[HTML]{FFFFFF}             & \cellcolor[HTML]{FFFFFF}                & \cellcolor[HTML]{FFFFFF}                  \\
Chun                                 & 71                                   & Female                                  & Jiangwa                                   & 12                                          &  &                                     &                                      &                                         &                                           \\ \cline{1-10} 
\end{tabular}
\caption{Summary of study participants}
\Description{Table 1 displays information about the participants who were in this study. Included are their participant identification number and information regarding age, gender, location, and years of experience.}
\end{table}
\raggedbottom

All 15 of the care partners were female. All except Jing were over 60 years old, with the average age of our older partner participants being 65.6. Our older partner participants were from 6 communities in Shanghai---2 from Cyang, 4 from Beixin, 3 from Shuyang, 2 from Kojiang, 3 from Xihua, and 1 from Jiangwa. Between them, our older partner participants had 137 years of volunteer experience in the older partner program, with an average of 9.1 years. Notably, 12 older partner participants had been working as older partners for over 7 years. Of the 6 staff on the neighborhood committees, 3 were male and 3 were female, with their age ranging from 36 to 57. They came from six different neighborhood committees. All 3 project managers were male. All 3 project managers were male---and each came from districts outside of the 6 participant observation communities\sam{, but were still situated within Shanghai}.

\subsubsection{Semi-structured Interview}
In semi-structured interviews with older partners, we asked questions about 1) their personal history of being an older partner, 2) their daily caregiving procedure, practices, experience, encountered challenges, and reactions towards the challenging situation, if any, 3) technologies, tools and materials they employed during the caregiving processes, 4) their usage and challenges of existing technological interventions for supporting older adults aging in place, and 5) their attitudes, perceptions, and expectations related to their role as older partners and the vision of technology-mediated aging in place. \sam{In semi-structured interviews with} the neighborhood committee members, we included questions similar to those with older partners, but from the perspective of a government worker. \sam{In semi-structured interviews with} project managers of technology-driven older adult care systems, we primarily included questions about 1) what kinds of technologies they had designed and deployed, a general description of these technologies and their respective utilities, 2) their general working process, 3) users' adoption, usage, and feedback related to these technologies, and 4) their perceptions of future technology-mediated aging in place. All interviews were conducted in Chinese and in person by the first author; each lasted between 40-60 minutes. \sam{In regards to all three groups and with each participant’s permission, each interview was audio recorded and later transcribed in Chinese and, then, English, verbatim for analysis}. In addition, we also obtained other forms of data from participants with their express permission, including screenshots of their mobile app pages and related photos or videos they shot, to further triangulate participants' responses. 


\subsection{Data Analysis}
\sam{To analyze our data, w}e conducted \sam{thematic analysis}, an inductive process for analysis \cite{strauss1990basics, riger2016thematic}. During the open coding phase, three researchers started independently analyzing the data while it was being collected while collaborating and developing an initial set of codes. With this initial set of codes, the researchers conducted preliminary coding exercises and discussed their results. This process was then iterated upon by generating an initial codebook from the collected data, classifying different codes into similar themes, regularly discussing these themes with each other to ensure reliability, and checking and elaborating these codes and themes as new data were obtained. The data collection process stopped once all the core variables reached saturation. 

By the end of the open coding phase, we generated an initial set of codes, capturing: existing technology interventions (``tools,'' ``technologies,'' ``working process,'' ``promised functions,'' ``encountered challenges in practice,'' ``reasons for challenges,'' etc.); human care providers' daily work (``motivations,'' ``daily experiences,'' ``emotional experiences,'' etc.); the collaboration and interactions between technology-driven older adult care systems and human-driven older adult care systems (``tools used,'' ``materials used,'' ``encountered challenges,'' ``reasons for challenges,'' ``resolutions,'' etc.) and more. Based on our preliminary codebook, we then iterated further on our codes and themes by examining the contextualized meanings of emergent themes by reviewing relevant Shanghai-based policy literature and related theoretical literature. \sam{This review helped inform our critical orientation and focus on the practices of ``younger'' older partners in supporting human- and technology-driven care infrastructures.}

While the codebook and coding of interviews were being done, participant observation data was being collected and analyzed. Fieldnotes were written frequently, presenting lived experiences and observations from the field as they occurred. For instance, the duties of daily volunteer work, the location of volunteer work, the older adults and care providers encountered, and the situations therein. These fieldnotes were coded through our interview codebook. In keeping with Emerson, Fretz, and Shaw, the first and second authors conducted memo writing to generate themes, interpret the day's observations and volunteer work, and explore emergent concepts \cite[p.103]{emerson1995processing}. These memos can be used to consider new ideas, mull over data, define relations, and gain a sense of reflexivity with the data itself \cite{charmaz2000grounded}. Through this process, all data and concepts gradually converged on three interrelated themes: care infrastructure based on technology, care infrastructure based on human labor, and older adults aging in place/community. These themes informed the way in which our findings and discussion have been presented. Below, we present the details of our findings, using representative quotes and illustrative cases. These were translated from their original Mandarin into English by the authors. 

\subsection{Ethical Considerations}
We obtained research ethical approval from the Ethics Committee of the authors' institution to conduct all the procedures of our study involving human subjects. 
All researchers of this study received formal ethics training, and were granted official certificates from ethics committee in their institutions. 
Throughout the research process, we took careful steps to protect user rights and privacy. 
Before starting, we informed all participants our intentions and background information, our promise that all collected data would be used only in this study, and got their permission. 
All collected data for this study was anonymized and all personally-identifiable information was stricken from recorded data. In this way, there are no links between the data collected and the identity of each informant.

\section{Findings I: Technology-Driven Care Infrastructure}
In this section, we present the overall structure of the existing \yuling{technology-driven} infrastructure employed across our research sites, as well as care providers' experience in using \sam{these technologies. This includes an exploration of the technologies' intended functions and capabilities, and an assessment of their real-world effectiveness in supporting older adult care for aging in place. This assessment, in particular, revolves around the firsthand experiences and perspectives of frontline care providers, encompassing both older partners and neighborhood committee members.}

\subsection{Overview of Existing Technology-Driven \sam{Care} Infrastructure}
\begin{figure}[ht]
 
\centering
\includegraphics[scale=0.45]{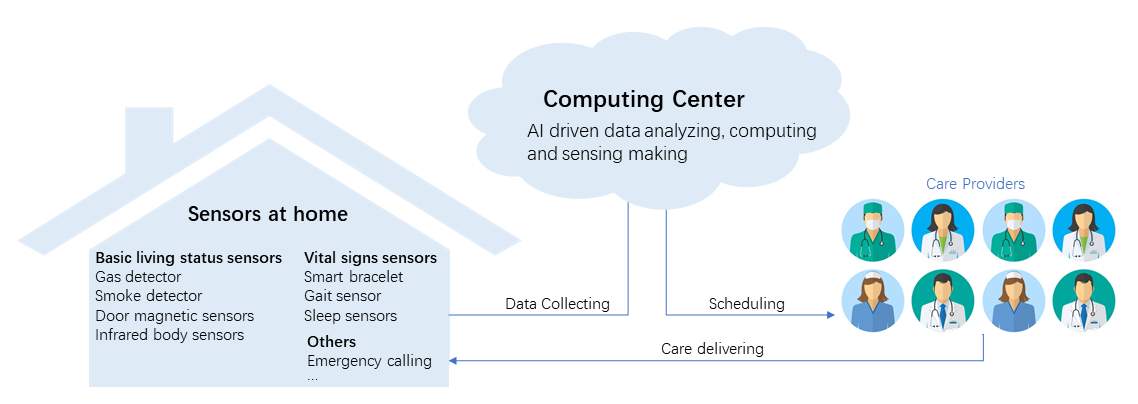}
\caption{The structure of technology-driven infrastructure for aging at home}
\label{Fig:techinfra}
\end{figure}

The existing technology-driven infrastructure for aging in place in Shanghai primarily contained three layers, shown in Figure \ref{Fig:techinfra}. To be specific: 1) ``Sensors at home,'' including but not limited to basic living status sensing technologies (e.g., smoke detectors, gas detectors, door magnetic sensors, and infrared body sensors), vital signs sensing tools (e.g., smart bracelet for physiological sign sensing, sleep sensors, gait sensor), as well as others like emergency calling \sam{systems}---these were used to monitor older adults' living conditions at home and identify whether \yuling{they} were living safely; 2) ``Computing centers,'' including AI or algorithmic processes used to collect, analyze, and make sense of the data coming from the older adult's home, \sam{both coming from and going to disparate informational sources}; and 3) ``Guiding and scheduling centers,'' leading to care actions taken in practice that were informed by the results of the processes above\sam{, as, for example, when anomalies or dangers are detected or calculated, the systems will automatically issue alerts and dispatch relevant care providers to deliver assistance.}


\sam{For this technology-driven care infrastructure, the primary goal is to support older adults aging in place by making sure they are active in a safe home environment. According to Lei, a project manager on sensing technology projects, this is often understood through basic measurements,} 

\begin{quote}
    {\sam{The four types of sensors [I work on] are gas sensors, door sensors, human sensors, and smoke detectors. These sensors are relatively mature in the industry and can accurately determine if an elderly person is still `alive,' such as if they haven't gone out for several days, haven't used gas for several days, or if the infrared sensor has no data feedback. Therefore, the primary goal of installing these sensors was to ensure that the government mandate of `don't let elderly people die at home without being noticed' is met.}} 
\end{quote}

\sam{Here, we can see a certain kind of sensing technology, one that produces data that presents itself as a baseline for activity within the home. For project manager Lei, whose technology has been installed in around 3,000 care recipients' homes, the technologies connect policy desires, corporate enterprises like his own, family members who receive sensing information if anything out of the ordinary is detected, and care recipients themselves.}

\subsection{Experiences \sam{With} Technology-Driven Care Infrastructure}
In our interviews, we focused on two types of frontline care providers: older partners and neighborhood committee members.\footnote{\sam{See methods section 3.3.1 for further explanation of these groups and their given roles within the community}} Within our corpus of transcripts, there was limited information about the benefits of existing technology perceived by frontline care providers. Some neighborhood committee members mentioned that digital management systems improved their daily work efficiency. For instance, as per neighborhood committee member Wei,

\begin{quote}
    We used to conduct cognitive impairment screenings manually, and it took 10 minutes for each elderly person. If we were to screen 2000 older adults, it would require 20,000 minutes, which is unmanageable. Additionally, privacy is a concern in this matter, as multiple people cannot undergo screening in the same space at the same time, as my conversation with you might affect the people nearby. At such times, \textbf{using existing screening data tools significantly improves efficiency and greatly reduces labor costs.} 
    \end{quote}

Except for this benefit, our participants\sam{, especially older partners,} primarily reported challenges and limitations \sam{in their interactions with} these \yuling{technologies that} support aging in place. \sam{As older partners were front-line volunteers supporting aging in place for care recipients, they often encountered technology-driven care infrastructure from an outside perspective, as neither the users of, say, monitoring systems nor neighborhood committee members who monitor the data produced by its users. As such, older partners provide a unique perspective on these technologies, as mediators of their implementation and continued use, as bespoke repair people fixing these technologies upon breakdown, as gossip partners registering perceptions of technologies during their in-person visits to care recipients' homes, and as the next generation of potential users. Below, we reflect on their interactions with and perceptions of technology-driven care infrastructure.}

\subsubsection{Factors That Impact Adoption}


The incorporation of technologies into the older adult care process often involved service fees---fees incurred by older adults. For instance, in neighborhood committees where participants Wei, Jing, Guo, and Jun were located, the technologies implemented for aging in place were initially provided by the government as a form of social welfare. This meant the older adults who received these aging in place technologies did not have to incur any expenses. After one year, this responsibility shifted to the users themselves---thus, older adults had to begin to cover the cost of their continued use. According to these committee members and numerous older partners, most care recipients they encountered refused to pay for these technologies when the burden was shifted to them, even after the price was discounted after initial outrage. Even during the first free year---during which both installation and use were free---many older adults refused them due to concerns related to wasting electricity (Li) or  internet bandwidth (Yun).

Additionally, some social beliefs among older adults influenced their acceptance of technology-driven care infrastructures. For instance, older partners reported their care recipients rejected the technology as a way to lessen their belief that they were aging (to ``defy aging'' as per Shu) or to rebuke their presumed status in a group that needed support (an ``unwillingness to be treated as a special group'' as per neighborhood committee member Yu). In particular, this feeling that the technological interventions were a mark of dependency and set them apart from ``regular'' people was felt by many care recipients---the interventions labeled them as ``needing special care'' or ``unable to take care of themselves'' (Yang). For instance, and in reference to the ``Home-based Virtual Nursing Home'' initiative \sam{in which nursing home care services are implemented in the homes of older adults aging in place}, some care recipients ``consider[ed] care beds like `hospital beds' for [sick] patients'' (Yang). Thus, even as neighborhood committee members presented these interventions as beneficial, care recipients refused to adopt them, \sam{often complaining about these technologies to older partners}. 

\subsubsection{Complexities That Impact Value}

\begin{figure}[ht]
 
\centering
\includegraphics[scale=0.22]{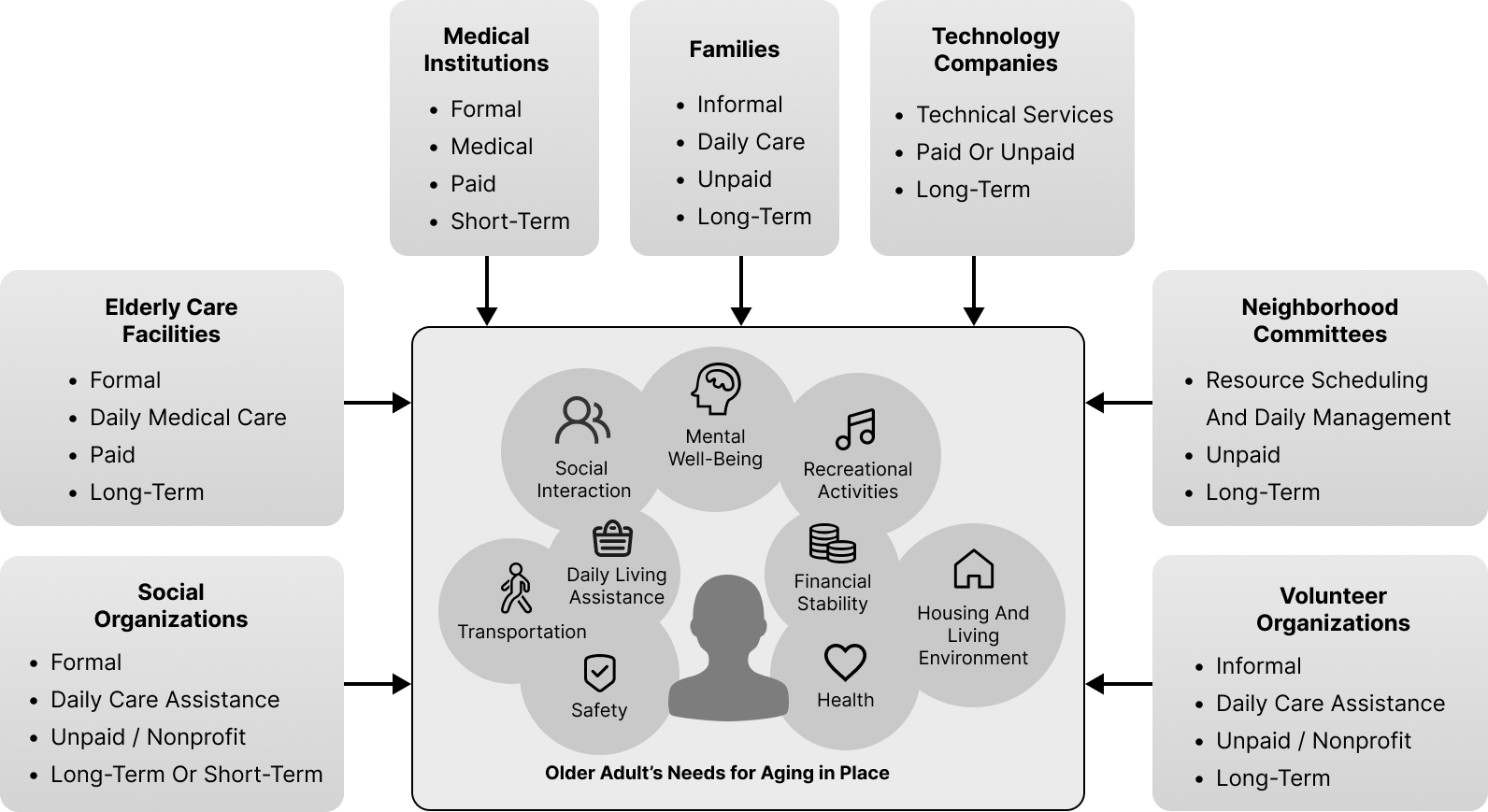}
\caption{The general structure of care service systems for aging in place}
\label{fig:caresystem}
\end{figure}

Beyond \sam{monetary and ageist beliefs influencing the adoption of these technologies,} \sam{c}ontextual factors within the lived environment and overall care complexities \sam{impacted the value of these technologies when they were used}. These challenges manifested \sam{due to} the complexity of older adults’ needs within their home environment, the complexity of care service systems, and the complexity of matching care needs to care services, as shown in Figure \ref{fig:caresystem}.

Specifically, as noted by the circles surrounding the older adult in the center \yuling{in Figure \ref{fig:caresystem}}, the needs of older adults aging in place were highly diverse, spanning from health and wellness to social, mental, and emotional needs. Most of these \yuling{were impossible to be collected, sensed, and computed by the existing developed and deployed }technology-driven care infrastructure which, as noted above, was predominately designed to support \yuling{health and activities of daily living (ADLs) monitoring} in the form of sensing technologies, \yuling{shown in Figure \ref{Fig:techinfra}}. 

\yuling{Further, the practical caregiving process involved a complex collaboration among multiple care providers} such as \sam{government institutions and policy,} families, medical institutions, technology companies, senior care facilities, neighborhood committees, social organizations, and volunteer organizations. Each of these care providers had \yuling{their own ways of approaching ``care'', in terms of }modalities, actors, \sam{desires,} funding, and specialties. For instance, in terms of caregiving, families, care facilities, social organizations, and volunteer organizations offered different forms of daily care assistance, whereas medical institutions offered professional medical care, technology companies offered various technical services, and neighborhood committees were primarily responsible for scheduling and daily management. These services were also differentiated by their service fees and formality, with some being free and others being paid, some care providers being formal and others being informal. The details can be found in Figure \ref{fig:caresystem}. \yuling{Such complex collaboration posed significant challenges to the data-driven guiding and scheduling within the technology-driven care infrastructure, making them struggle to accurately and effectively determine which care providers should be scheduled. }



\subsubsection{Usability and Data Limitations That Impact Value}

Technology-driven care infrastructure was often ineffective \yuling{and unusable}, giving, for instance, inaccurate descriptions or outcomes, or difficult to use due to design and utility. For instance, Xiao noted that the smart-sensing smoke detector ``malfunctions: sometimes, it’s too sensitive, sometimes, it goes off just because someone is cooking.'' These experiences reduced the utility of such sensors and created additional problems---\sam{older partners now needed to understand the systems well enough to fix them, turning older partners into bespoke repair and maintenance workers.}

Further, current technology-driven care infrastructure had not formed a complete technology and application ecosystem, with incomplete data-driven loops in real-world scenarios that also caused limited usability of technology-driven infrastructure. Specifically, data was scattered among various stakeholders without effective circulation and application. Each stakeholder only recorded data related to their own specific motivations. For example, \sam{a health-monitoring system recorded heart rate, a bed monitor noted movement and length of time in bed, medical institutions maintained their medical service records dashboard, and community workers recorded various demographic and management data, such as electricity usage.} \yuling{T}hese individual pockets of data were scattered across different data systems and were not aggregated, linked, or shared among different stakeholders to harness the expected and stated digital efficiency of these initiatives. This lack of data circulation even resulted in redundant data collection efforts, adding extra pressure to the workforce. As neighborhood committee member Yu said, 

\begin{quote}
The basic demographic statistics of older adults we manage are publicly available, such as the number of older adults in each age group and the male-to-female ratio; this is all public information. The problem currently lies in the lack of data sharing between different platforms. Our platform (the platform of the civil affairs system) doesn't share data with the public security system, the health commission's system, or data from third-party enterprises.
\end{quote}

\begin{figure}[h]
  \centering
 \begin{subfigure}[b]{0.494\textwidth}
    \includegraphics[width=\textwidth, height=4cm]{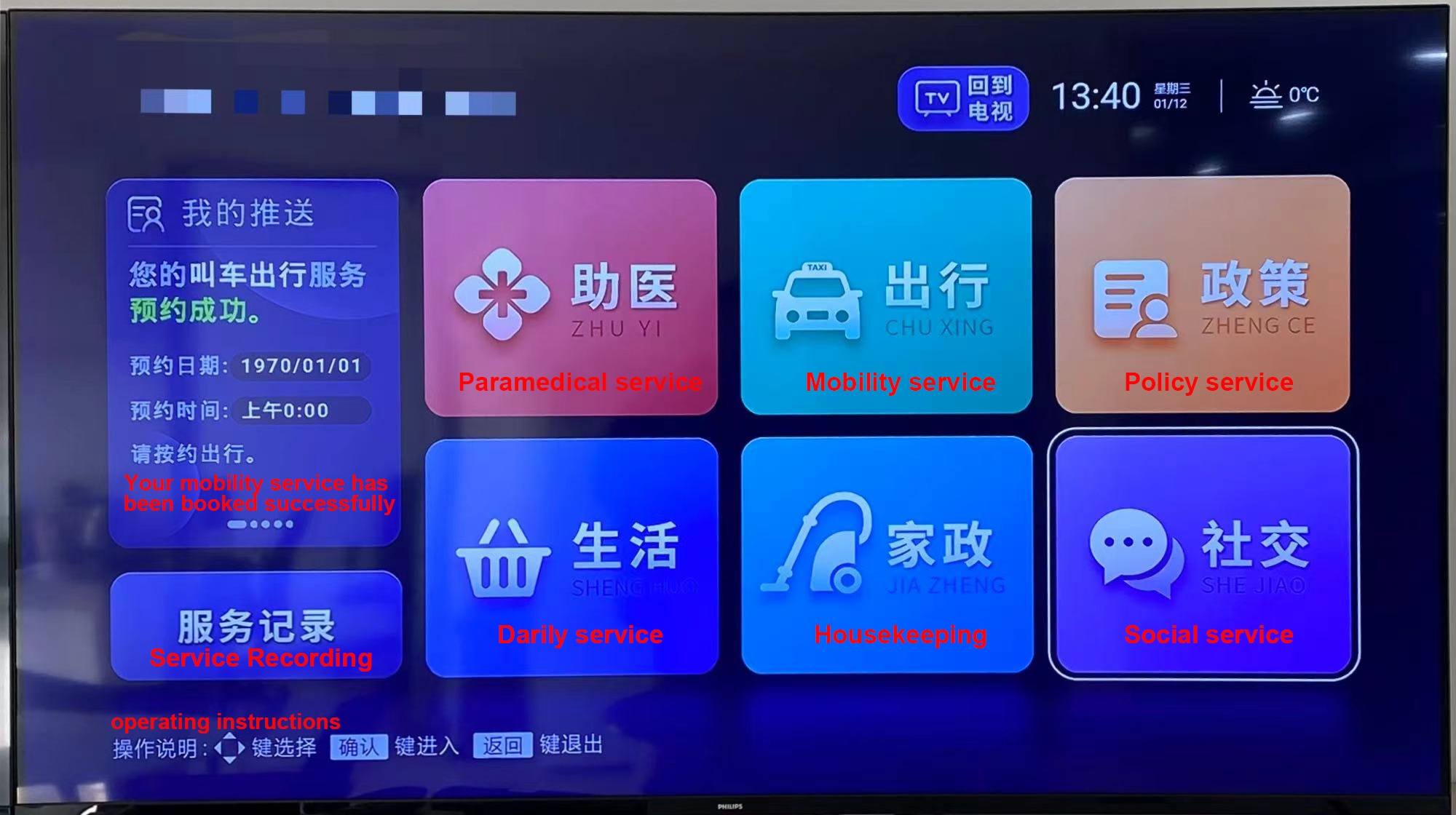}
    \caption{}
    \label{fig:data1}
  \end{subfigure}
  \hfill
  \begin{subfigure}[b]{0.494\textwidth}
    \includegraphics[width=\textwidth, height=4cm]{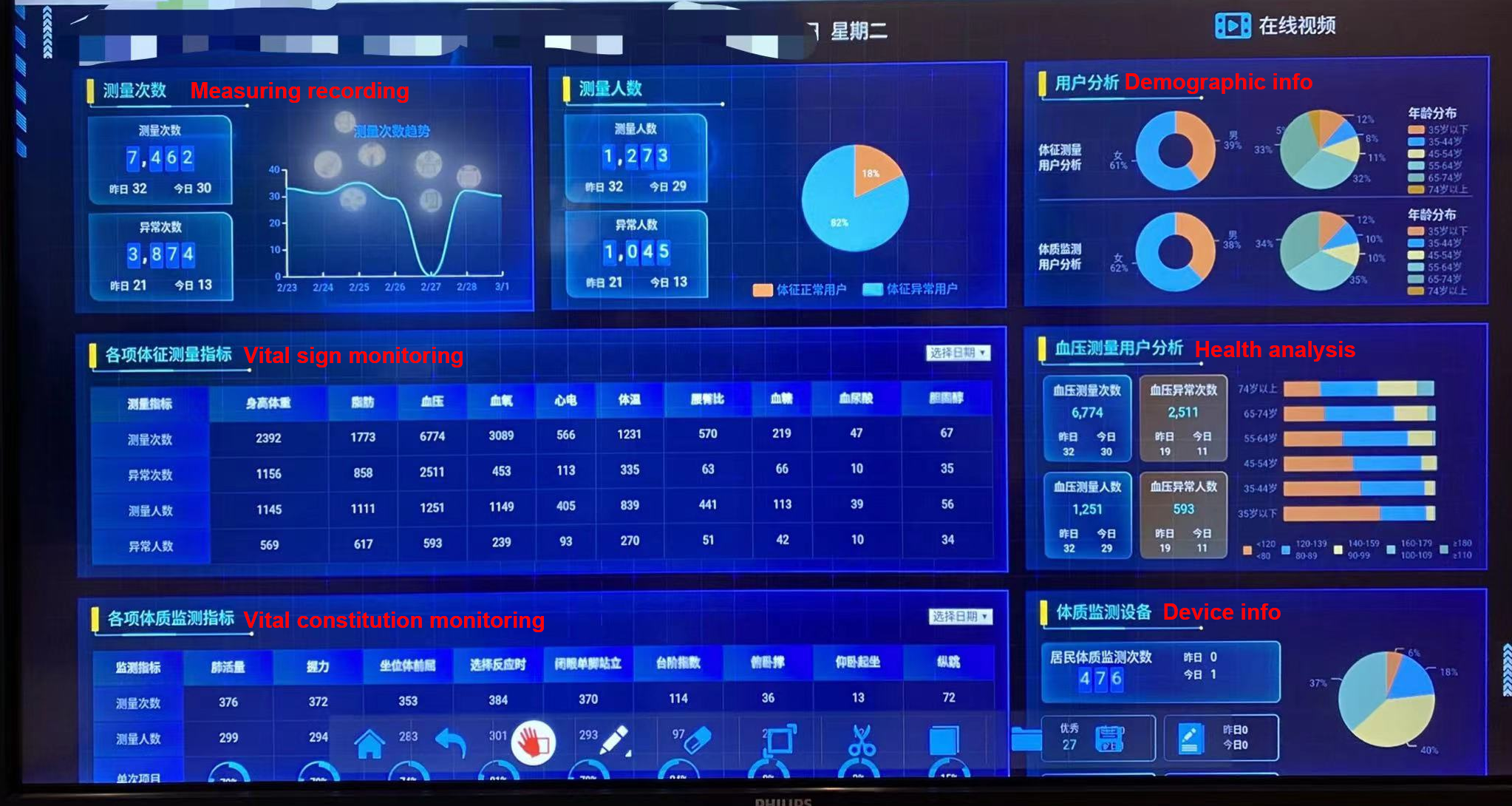}
    \caption{}
    \label{fig:data2}
  \end{subfigure}

   \begin{subfigure}[b]{0.494\textwidth}
    \includegraphics[width=\textwidth, height=4cm]{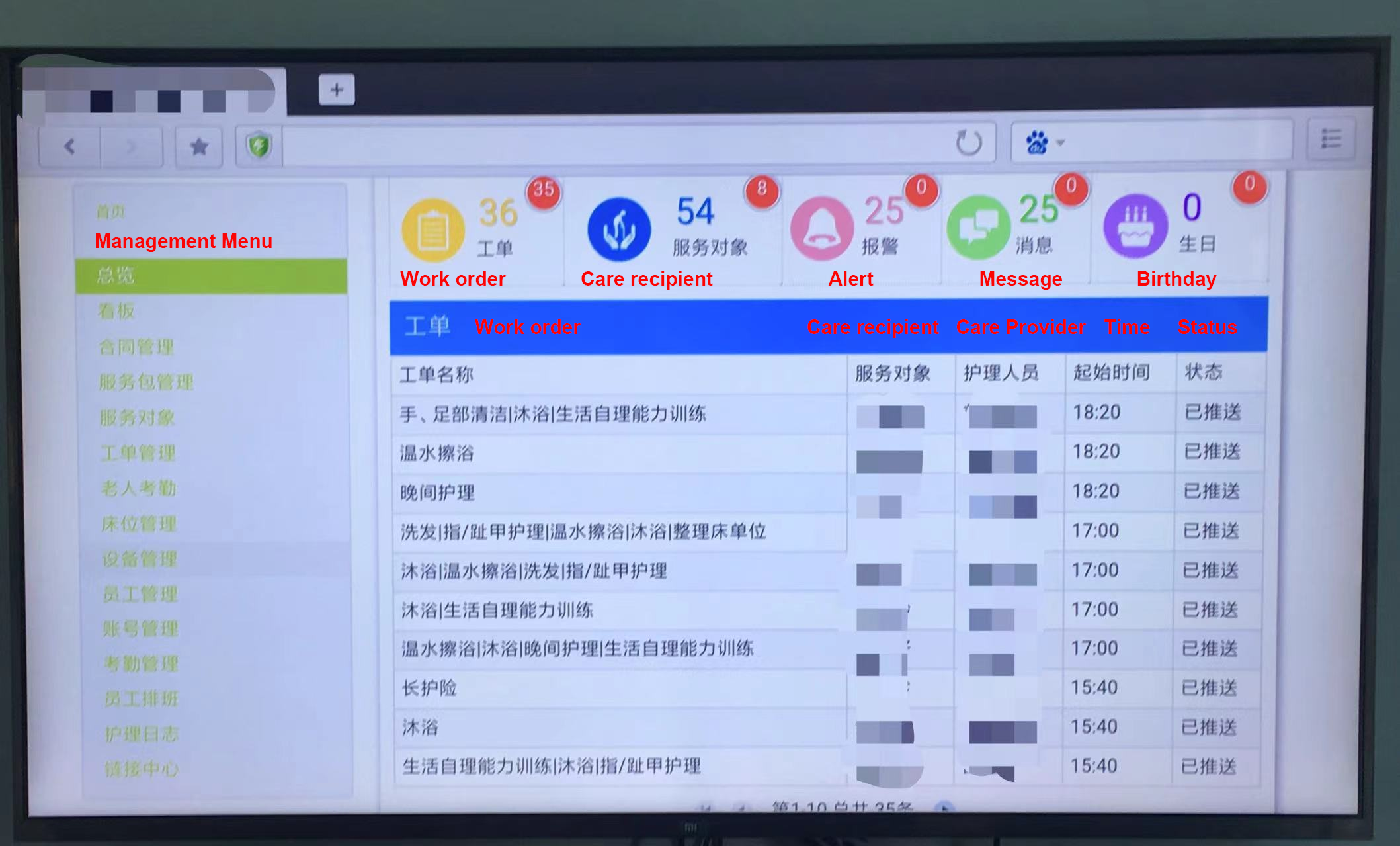}
    \caption{}
    \label{fig:data3}
  \end{subfigure}
  \hfill
  \begin{subfigure}[b]{0.494\textwidth}
    \includegraphics[width=\textwidth, height=4cm]{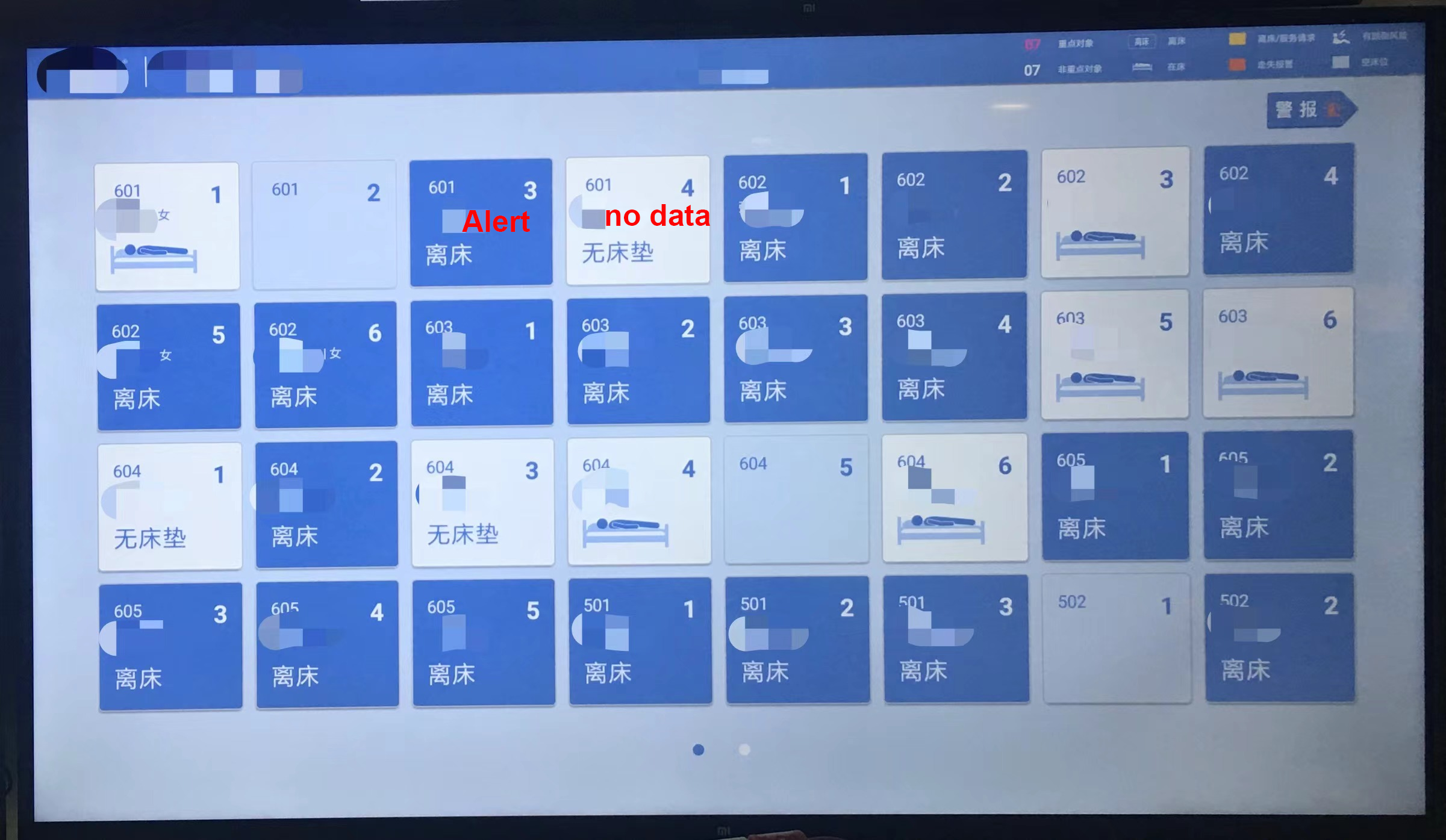}
    \caption{}
    \label{fig:data4}
  \end{subfigure}
  \caption{\yuling{Data dashboards used by different care providers}}
  \Description{}
  \label{fig:variousData}
\end{figure}

\yuling{Among different technology suppliers, there was also limited data sharing or cooperation. Based on our observation study, multiple technology and care suppliers were involved in the assemblage of technology-driven care infrastructure, but each technology supplier was striving to build its own ecosystem, and, as such, each deployed their own sensing devices to collect data from older adults. Figure \ref{fig:variousData} shows system dashboards of different technology suppliers we observed. Figure \ref{fig:variousData}-a was the dashboard of the service requesting system, which was installed on older adults' televisions. After older adults made a service request through it, the system would schedule the corresponding service providers. Figure \ref{fig:variousData}-c was the corresponding management dashboard, visible for employees at the company that provided the requesting dashboard but not visible to older adults. Figure \ref{fig:variousData}-b and Figure \ref{fig:variousData}-d were data dashboards located in the neighborhood committee and care center, respectively, displaying the vital signs and sleeping status of monitored older adults. This data was collected through the sensing technologies installed at older adults' homes. However, there was also no data sharing or cooperation among these systems, resulting in a large amount of complex and even redundant data collection but limited actual efficiency. What's worse,} the collected data was often not fed back into the system of data scheduling system suppliers to support their scheduling service providers. Thus, a device collecting medical data could not inform a scheduling service about recent changes in health---effectively, such data was for the company and not for the older adults, healthcare providers, or other stakeholders. 

Sometimes, inconsistencies emerge between data in different systems. For instance, when an older adult aging in place uses a scheduling service to request the need for medical service, the service should connect to medical resources, like a hospital. However, as \sam{a third-party company mediated the service}, some requests could not be scheduled---in the case of a hospital visit, a scheduling service did not have permission to book a hospital visit, which would need to be taken up elsewhere. Project manager Lei, who worked for a company that supplied digital scheduling systems for aging in place in one district of Shanghai, explained to us the challenges of using data systems to schedule medical resources:

\begin{quote}
    \textbf{The idea is excellent, but its implementation is quite challenging.} There are various issues when it comes to actual execution. For instance, features like one-click calling and one-click appointments in hospitals may sound simple, and even building the system seems quite straightforward. However, in practice, it's quite challenging. For example, if an older person wants to book an appointment at a top-tier hospital, our company might not be able to achieve that without the local authorities opening up the resources of these hospitals for our platform to connect with.
\end{quote}

\sam{In this quote, we can see the limitations of technology. As noted by project manager Lei, systems that help with tasks such as scheduling are only as good as their access and interconnection with other services and systems. When access and interconnection are limited, the value of these systems is limited as well. In these instances, the systems rarely live up to their ideation stage. This point was echoed by project manager Zhi, who said advanced sensing technologies ``sound good in theory. I think management personnel can use it... but it is not practical for frontline workers.''}

\sam{However, w}hen such inconsistencies emerge, frontline workers like older partners serve as data workers to sift through the collections of data, attempting to make sense of what is right and what is wrong. Our participants reported that this data work was burdensome. For instance, neighborhood committee member Yu used an example of delivering milk to older adults aged 90 and above to illustrate this burden:

\begin{quote}
    We provide the welfare of delivering milk to older adults aged 90 and above, and this falls under the responsibility of the civil affairs department, but the data [of age, home address, and recent deliveries] comes from the public security system. However, the data in the public security system is not updated in real-time, and it does not accurately record which older adults reside in which community because the population is constantly changing. \textbf{As a result, the data is never accurate, and we often have to change it manually, which is truly exhausting.}
\end{quote}

As neighborhood committee member Yu notes, in practice, repairing these breakdowns, interoperating between stakeholders, and sharing data were all challenges in making the technology-driven care infrastructure work. An additional consideration was noted by older partners, neighborhood committee members, and project managers: concerns about user privacy and security. In some cases, these \sam{disparate actors thought} data should \textit{not} be shared across platforms, even in an ecosystem dedicated to the accumulation and parsing of data. However, these privacy concerns were often provisional, with rules on data privacy as diverse and dynamic as the data itself. As a result, current data-driven technologies generally do not fully realize their promised function and, as a result, can negatively affect stakeholders and care work itself.



In sum, we illustrated a series of challenges existing within the present technology-driven care infrastructure. Among these challenges, some could be addressed by improving technical performance, design, and interoperation. However, some others, such as fulfilling care recipient's needs, were difficult to resolve through current technologies. \sam{At present}, human infrastructure addresses care recipients' needs but in different practices---we now turn our attention to human infrastructure within our research sites to demonstrate how human-driven care infrastructure operates.

\section{Findings II: Human-Driven Care Infrastructure}
The human-driven care infrastructure for aging in place identified within our research involved multiple stakeholders that collaborated to support older adults aging in place (see Figure \ref{fig:caresystem}). Among them, we focus on the older partner who played a crucial role in human-driven care infrastructure. In this section, we first present a brief description of human-driven care infrastructure for aging in place (Section \ref{Section:overviewofHumanIn}). Following that, we investigate the role of older partners, elaborating on their motivation and experience of being an older partner (Section \ref{Section:beingAnOlderPartner}), and how they worked effectively to address the challenges encountered by technology-driven infrastructure (including work practices, tools, and materials) (Section \ref{Section:WorkofOlderPartners}).

\subsection{Overview of Human-Driven \sam{Care} Infrastructure}
\label{Section:overviewofHumanIn}
\begin{figure}[ht]
\centering
\includegraphics[scale=0.25]{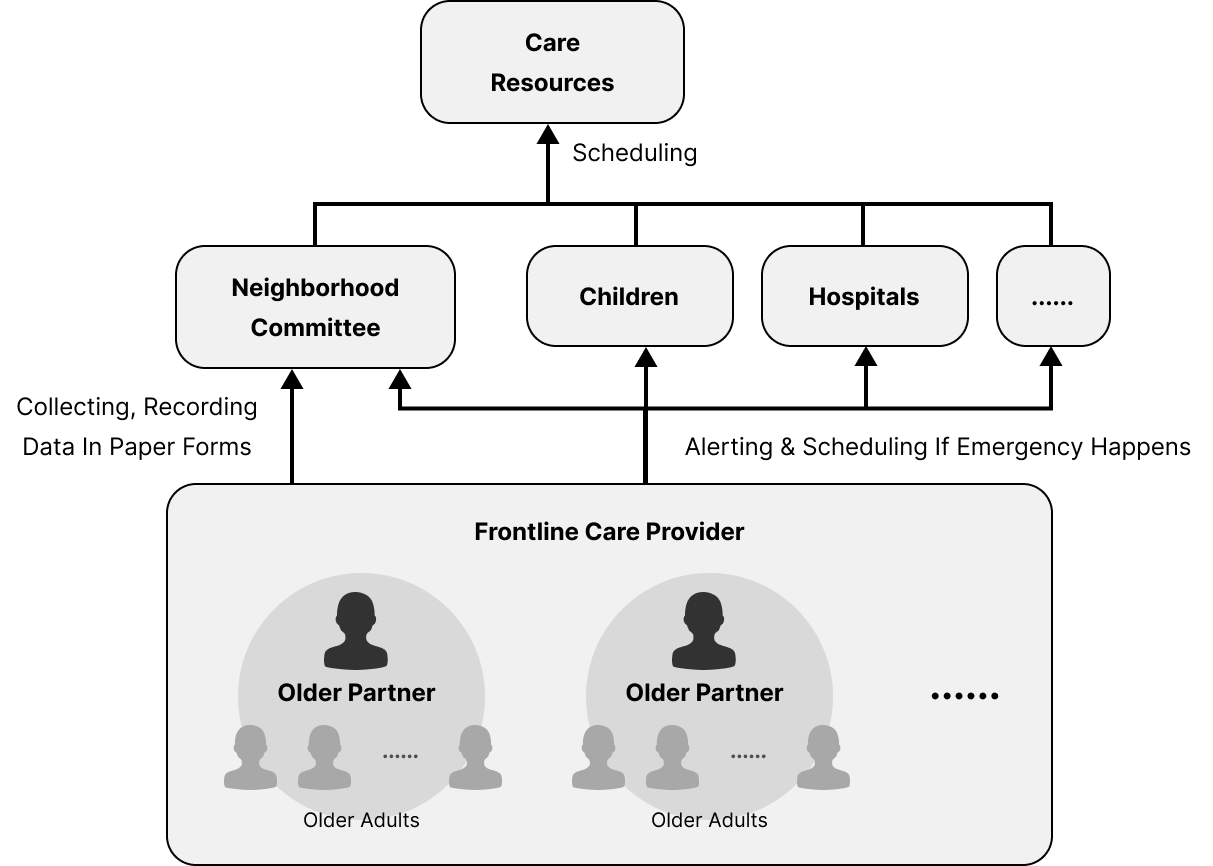}
\caption{The structure of human-driven care infrastructure for aging in place}
\label{fig:humansystem}
\end{figure}

The structure of human-driven care infrastructure is illustrated in Figure \ref{fig:humansystem}. The foundation of this infrastructure was the frontline care providers, who were primarily composed of retired ``older partners'', as described in 3.1.2. They formed a community mutual aid network with a one-to-many pairing approach (typically, one older partner was paired with 5-10 care recipients), covering many ``older'' older adults (aged 80 and above) aging in place within the community. They provided various types of care services regularly to care recipients to support their safe and comfortable aging at home. If there was an unexpected situation, the older partners would liaise with other care providers, such as the care receiver's children, hospitals, or neighborhood committees, based on the care recipients' specific situations. Serving as a type of human sensor, these older partners often played the role expected of technology-driven care infrastructure (as shown in Figure \ref{Fig:techinfra})\sam{: they played the role of data collectors of health and wellness and schedulers of doctor's appointments, and when the technology broke down, they acted as mediators who helped make the technology work in practice and informed its use in the homes of ``older'' older partners.} 

Based on our participants' descriptions, this mutual aid network was not fixed but rather scalable. Additional informal stakeholders could be connected in some way, forming small, adaptable mutual aid networks tailored to specific care recipients. For instance, neighbors of care recipients might proactively assist the caregivers by providing relevant information. Older partner Chun mentioned that the neighbor of one older adult she cared for often helped her check on this older adult's situation when Chun couldn't reach him: ``Sometimes when I knock on his door, and no one answers, it could be that he's gone out to buy groceries or for a walk. So, I then asked his neighbor next door to keep an eye out, and if he returned, the neighbor would inform me.'' As such, the structure of Figure \ref{fig:humansystem} is dynamic, with stakeholders coming in and out of the care infrastructure \sam{as needed.}

\subsection{Being an Older Partner: Motivation, Experience, and Perception}
\label{Section:beingAnOlderPartner}
We first elaborate on the motivations of our participants to become an older partner, as well as the perceptions of these individuals and other stakeholders, particularly neighborhood committee members, regarding their role in human-driven care infrastructure for aging in place.

\subsubsection{\sam{Motivations: Contributing to Society}}
During our research, we found that older partners were a highly proactive, enthusiastic, and dedicated group. By their own admission, they eagerly responded to the government's (via the neighborhood committees') calls to participate in the Older Partner program and, by all accounts, cared deeply for each care recipient paired with them. Most of the older partners encountered joined the Older Partner program immediately upon retirement---of the 15 older partners we interviewed, 12 had been engaged in the Older Partner program for over 8 years. Instead of seeking monetary or other material compensation, all older partners we interviewed expressed their motivation for being an older partner as an emotional one: they had a sense of dedication and a need to fulfill a social responsibility.

In interviews, our participants considered this sense of dedication as something inherent and unique to their generation of Chinese citizens born in the 1940s, '50s, and '60s. By their own admission, they had experienced many cultural events and historical processes, such as the Cultural Revolution and the Movement of Chinese Educated Urban Youth to the Countryside in the 1960s and 70s, as well as the early retirement and reemployment due to the policy of reform and opening in the 1990s---decades of political and social upheavals that led them to be referred to as ``the lost generation'' \cite{Rosen2014}. Meanwhile, their lives paralleled the continuous development of China. By the time they retired, China's economy had experienced rapid development, and they generally had sufficient pension funds to support their daily life. This journey from hardship to monetary and social comfort made them grateful to society, and they had a strong desire to give back---especially to the generation before them. Additionally, during their upbringing, China consistently stressed the core values of `unity, friendship, and dedication to society' in its national ethos \cite{contribution}, which profoundly influenced the values of this generation. In this way, their role as older partners was said to be a form of dedication and social responsibility. For instance, Lan said,


\begin{quote}
    [My generation was] influenced by the deep sense of transformation our parents' generation experienced and their selfless spirit of dedication. Our generation had a quality of virtue and hard work ingrained in us from birth. \textbf{We take the older partner [program] as a spirit of dedication, regardless of money or reward.}
\end{quote}

Here, Lan regards their volunteer labor as a ``spirit of dedication'' that is not influenced by money. Some neighborhood committees provide limited material subsidies (e.g., money or essential living supplies), but most of our participants said they \sam{do not keep these for themselves, instead giving} these supplies to the care recipients they cared for. For instance, Mei said, ``I used the subsidies from the neighborhood committee to buy things for the older adults [I care for], like groceries or daily necessities. It doesn't matter to us. We're happy to do it.''



\subsubsection{\sam{Experiences: Not just a ``Job''}}
Although most of the older partners were ``younger'' older adults (aged 70 and below), they faced the common challenges of older adults, such as a desire to find meaning post-retirement or a concern about becoming socially isolated. For them, volunteer work served as another form of ``employment'' and a continuation of their established self-worth and social value. Many of our participants had been engaged in this role since their retirement---for them, this ``job'' had become a part of their lives, providing them "satisfaction", "a sense of accomplishment", "happiness", "self-fulfillment", "a purpose in life", and a connection to the greater society. As described by Yun,
\begin{quote}
    I find great joy in doing this work. When I have free time after meals, I go for a walk in the park in the morning and then buy some groceries. After that, I return to check on the older adults. It makes me incredibly happy and keeps me engaged. (...) I don't feel disconnected from society.
\end{quote}

Affirmation and recognition from care recipients and their families were also important sources of happiness for older partners, giving them a sense of value. As Fang mentioned: ``We help the elderly solve various problems, and they, as well as their children, are very grateful to us. A simple `thank you' from those older adults makes us extremely happy.''

At the same time, older partners we interviewed expressed that this volunteer work was more than just a ``job'' \sam{that traditionally traded effort for} monetary reward. Rather, they viewed the care recipients they cared for as neighbors living in the same community, integral members of the neighborhoods \sam{and their lives}. In this way, the older partner viewed themselves as providing companionship to neighbors instead of a service between a care provider and care recipient within a working relationship. As Xiang described, 

\begin{quote}
    My [care recipients] are not strangers; they are people I already knew well before. For me, I just visit and chat with them from time to time. They always inform me if they have something going on. When they're about to go out, they'll say, `You don't need to come today; I won't be at home.' Sometimes, even their children call me, showing concern, or bringing me some food.
\end{quote}

\sam{For Xiang, her care recipients were individuals she already knew from being in the community, people she already felt a connection to. In the eight years since beginning her work with the older partner program, she has become an important part of not just the lives of her care recipients but also their children, who may live at a distance that precludes the kind of in-person care that older partners can provide.}

\subsubsection{\sam{Perceptions: Significant and Irreplaceable}}
For neighborhood committee members, older partners played a crucial and irreplaceable role in their daily older adult care services and management, realizing a human-driven care infrastructure \sam{laid out by the policies noted in 3.1.2}. As they understood the care recipients the most, older partners identified and responded to the needs or abnormalities of older adults first and often scheduled care resources to support them. As noted by neighborhood committee member Tian, ``every older partner is very familiar with the care recipients paired with them. They know the detailed situation of each older adult, her/his family relationships, children's situations, etc. They will know (...) who has been hospitalized, who has been discharged, what is missing in whose home, who has difficulty moving up and down the stairs, and who may need assistive devices. They don't need to use pen and paper to record; it's all in their minds.'' In this way, the older partners collect and sort through information about their care recipients, acting as integral components of the care infrastructure.

Meanwhile, our participants also noted that the presence of older partners provided important emotional support for older adults, which was something that technology could not replace. As neighborhood committee member Wei expressed, "Older adults still need to communicate with humans. We are in one community. We are all neighbors. We can see each other by just walking around. Using technology and such is too modern...". The neighborhood committee members we interviewed also generally expressed that these older partners were important links in reshaping and improving neighborhood relationships, as when Guo said older partners ``brought this building and these families together like one family, wherein everyone shares what's good to eat and helps each other.''

Further, care recipients trust older partners, and this trust is crucial for the acceptance of new policies and technological interventions. Staff in our study expressed that trusted older partners played a significant role in conveying and explaining policies, assisting in communication, and more. If, as noted in 4.2.1, care recipients are suspicious of \sam{technology within the care assemblage, due to it being perceived as} surveillance or otherwise emblematic artifacts of personal frailty, older partners can be mediators between technology and these presumptions. Per neighborhood committee member Jing,

\begin{quote}
     Sometimes, \textbf{when communicating with older adults about certain issues, the role we play may not be as significant as the role older partners play}. They are most familiar with the older adults, they know best how to communicate with them, and sometimes older adults trust them more. So, many matters in the community are assisted by these older partners.
\end{quote}

\subsection{The “Work” of Older Partners: Practices, Encountered Challenges, and Reactions}
\label{Section:WorkofOlderPartners}
We now turn to the work of our older partners and elaborate on how our participants worked daily to address challenges encountered by newly implemented technology-driven infrastructure, highlighting specifically their work practices, encountered challenges, and reactions to these challenges.

\subsubsection{“Work” Duties: Broad Roles, Expanding Practices}
In the original government document of the Older Partner program (e.g., \cite{olderpartner3, olderpartner2, laohuobanpolicy}), the role of the "older partner" had clear task boundaries: "providing home mutual assistance services, including disability prevention, health education, emotional and social support, and basic life assistance services, to older adults, focusing on promoting a `Healthy Lifestyle', ensuring their safety, preventing or reducing the occurrence of risks, as well as enhancing the quality of life and social interaction of older adults.” \cite{laohuobanpolicy} From the perspective of government workers, such as neighborhood committee staff in our study, the work of the old partners is ``quite simple''. As neighborhood committee member Guo mentioned, "There are no high demands, just knocking on older adults' doors every day, checking if they are well, and informing us if there are any issues. It's quite simple. We don’t want a situation where an emergency happens at home, and we don't even know."

However, in practice, due to the emotional connection between older partners and care recipients, defining clear task boundaries became more challenging. What older partners actually did often went beyond the initial work description. They would proactively inquire about the needs of \sam{the} older adults they cared for. In addition to confirming their safety, they sometimes needed to assist them with various daily tasks, such as "grocery shopping," "picking up medication" (Ming, Shu, Qiu), and "appliance repair” (Ming), among others. As Ming said, "At the beginning, I just knocked on their doors. Once we got closer, I began assisting them with various daily tasks, for instance, helping them with medication; or if their gas was not working at home, I would help them call someone for repairs.” Our participants reported various types of care ``work'', for example, when no family members were available to support a care recipient temporarily, Li prepared meals for him/her; when older adults at home had difficulty going out to pick up medications, Shu took their medical insurance card to assist them in purchasing medication; when older adults were not familiar with smartphones, Xiang assisted them in operating these devices; when the issues facing older adults were beyond what older partners could handle, most of them compelled other stakeholders, like the neighborhood committee or their family, to help. 

For our participants, this "work" was not required by the Older Partner program, but because of the long-term relationship and emotional connection they had with their care recipients, they voluntarily offered such assistance. Qiu, who assisted a care recipient in picking up medication every week, mentioned:

\begin{quote}
    [Picking up medicine] is not mandatory, there's no such requirement. But as I build a connection with him over time, if I don't visit him, I start missing him, and he also misses me. So I check on him daily, exchange greetings, and observe his mental and physical state. Knowing that [my care recipients] are doing well brings peace to my heart.
\end{quote}

The long-term caregiving process also fostered strong trust relationships between older partners and older adults, as well as their family members. Many older adults also turned to older partners for assistance with various household matters. For example, Li said, “Some older adults had a particularly strong trust in me. For instance, when they stayed at their children's homes for some time, they'd leave their keys with me and ask me to come over from time to time to air out the place and water the plants.” Some older adults' children also heavily relied on older partners and entrusted them with the care of their parents. Mei mentioned such kind of case to us: "there was an older adult who, due to illness, was bedridden and unable to move during that time. Her daughter entrusted me to measure their blood pressure every day and inform her via WeChat. I informed her every day---today's blood pressure, whether it was normal or not."

Additionally, many of them had become valuable assistants for the neighborhood committee, helping conduct older adult-related management tasks such as gathering older adults' demographic information and disseminating relevant policies. As mentioned by Hui: ``We inform older adults about the newest elderly care policies and welfare, such as when they can get free check-ups or rent assistive devices at half price. Sometimes, we assist with paperwork, gather signatures, and implement notifications, etc.'' Explaining and implementing policies to older adults are also part of the older partners' daily "work". As Yang noted, "Some older adults may not understand the policy even if you provide them with the concrete text. They need someone to explain it face-to-face, step by step, and write the notes down on paper for them.” \sam{In this way, older partners help keep their care recipients informed, taking their time to make sure each partner understands policy changes in the modality and language that best supports them.}

\subsubsection{Encountered Challenges}

The interactive attributes of human-to-human interaction not only fostered positive emotional connections, as mentioned above but also brought about various challenges that our participants described as the main challenges during their daily work. Particularly, older adults aging in place had diverse and personalized health characteristics, personal preferences for communication and connection, family relationships, and more. These factors pose certain challenges to the daily work of older partners. For instance, some older adults might not appreciate being "disturbed" every day (Chun); some were "uncooperative" (Shu) to older partners' daily work and found them "bothersome" (Ming); some might be "difficult to handle" (Lan); some might have specific requirements, e.g., "insisting on our visit at a specific time every day" (Chun). \sam{In response to these challenges,} older partners had created some implicit ways of task execution. That is, after understanding the behavioral patterns of the older adults they cared for, they incorporated these patterns into their daily lives, forming their own daily routines to complete older partner tasks while minimizing their ``disturbance'' to older adults.

\begin{quote}
    [My care recipients] generally go to buy groceries around 6 a.m. and sit in the park in the afternoon around 2 or 3 p.m. So, I come out during these two time periods every day. As I walked along, I would see them. Seeing them reassures me that they are safe. If I haven't seen one in the morning and afternoon, I check in the evening to see if his/her light is on; if yes, it means everything is fine. If there is no light in the evening, I quickly visit in person.
\end{quote}

Here, Shu uses context clues and information gathered from years of service to check on her care recipients without visiting them. Similar to buying groceries in the morning, daily information drawn from life was frequently used by older partners to fulfill their roles. Some participants had formed a tacit understanding with older adults they cared for. For instance, Chun reached an agreement with their care recipients to ``hang a towel on the[ir] balcony to signal'' they were safe and doing well. In these and other instances, care was dynamic and unfolding, with interactions and information being exchanged between older partners and care recipients, as lines were drawn and the definition and responsibilities of ``care'' were assembled.

In addition to diverse personalities and preferences, some older adults had disability or health-related issues that made it more difficult to drop in, such as older adults who were hard of hearing or older adults with limited mobility. As Qiu said,

\begin{quote}
    Many [of my care recipients] with advanced age have poor hearing. They don't open the door when we knock. If this continues for several days, I must communicate with their children or the neighborhood committee. Sometimes, when their children rush back to check, they find the loved one is actually at home, but due to poor hearing, they don't hear us knocking.
\end{quote}

For these care recipients, everyday activities may also pose potential safety risks, because "if they fell while opening the door for us, what should we do?" (Yang). Moreover, some older adults had complex family relationships, leading to situations where "children are unwilling to take care of older adults when problems arise" (Ming), presenting certain challenges to the work of older partners. When they are the only care stakeholder available or willing to help, they sometimes add additional labor to their already busy schedule. Such instances can lead to ``unclear situations'' (Yun) or even ``good intentions that cause trouble'' (Xiao). Yun elaborated on how good intentions and an unclear or elevated role in the care infrastructure can cause trouble:

\begin{quote}
    One older partner helped a [care recipient] buy medicine during the pandemic, but the [care recipient] claimed that this older partner used his social security card to buy her \textit{own} medicine, making the situation difficult to explain clearly. In reality, it was because the price of the medicine had increased, but the older adult insisted that their old partner had bought her own medicine for herself with his information.
\end{quote}

Chun reported another similar case, as one older partner was accused of stealing items from a care recipient by their family. To avoid potential risks like this, some older partners gradually developed behavioral guidelines: ``Do not enter the home, do not touch money, chat briefly at the door'' (Ming).

Moreover, the task of, as stated in the Older Partner program, ``confirming the safety of the older adult every day'' also brought about certain mental pressure for older partners. Many older partners expressed that they were anxious when they couldn't get in contact with their care recipients or when they were late on their checking-in. Chun, who has been a partner for 12 years, mentioned that over these years, five of the older adults she cared for had passed away, creating an invisible psychological pressure:

\begin{quote}
    Sometimes, when an ambulance comes to our community, I ask my husband to help me look. If it's not parked below the apartments where my older adults live, then I am relieved. There is an invisible, self-imposed pressure and fear. When I see the older adult, it's comforting, and I feel that the task for this day is completed. If I don't see them, I keep worrying about them.
\end{quote}

In summary, our participants reported a series of challenges encountered during their daily work. Yet, despite acknowledging these challenges, most of our participants did not perceive them as hardships. Their responses leaned more towards expressions like "it's nothing" (Mei), "we've gotten used to it" (Xiao), and "considering the age of these older adults, it's understandable" (Chun). To address these challenges, older partners have utilized certain technologies, as detailed in our final section below.

\subsubsection{\sam{Technology Use}}

The primary technological tool employed by our participants was WeChat\sam{, the Chinese social media and messaging platform that has over one billion monthly active users}. Some older partners, like Fang, Xiang, and Yun, used WeChat to communicate with older adults, ensuring their daily safety. As Xiang said, ``Nowadays, some older adults use smartphones. We exchange greetings on WeChat every day. They sometimes cooperate with our work, actively sending me messages, saying, `Today is fine, rest assured'.'' Some older partners even created WeChat groups for older adults they cared for to execute daily care tasks. Fang, who had worked for 12 years, shared how she used WeChat group to assist in her older partner duties:

\begin{quote}
    I currently have 14 older adults. I formed a group of older adults who can use smartphones and their children. We send messages every morning. If there are replies, there's no need to visit; if there are no replies, I'll go visit. It's like this every day, starting at 7 a.m. and ending at 10 a.m.
\end{quote}

However, the use of WeChat was not suitable for all older adults, as they ``don't know how to use WeChat'' (Fang). \sam{Despite the use of social media platforms like WeChat, older partners mentioned that} even if they greeted older adults on WeChat, they still visited them in person because ``personally visiting provided more reassurance'' (Yun) \sam{and the act of physically visiting provided more social value for both them and their care recipients.}

\section{Discussion}

In keeping with our first contribution and research question, our findings provided a look at aging in place and care infrastructure from a unique context to ``better understand how diverse communities adopt and interact with [...] computing technologies and how they might, in turn, be leveraged to drive new interactions'' \cite[p.3066]{kumar2017hci}. We provided an empirical look at a unique problem/solution space---Shanghai, China, its notable \yuling{aging in place} population, and its caregiving programs that do not just aid older adults but are often informed and enacted by older adults. \final{In doing so, we trace two distinct visions of care that are enacted through particular assemblages: human- and technology-driven care infrastructures.}

\sam{In this context, multiple government agencies attempt to support aging in place but approach it from different perspectives. Technology-related departments, such as the Shanghai Municipal Science and Technology Mission, issue policies and guidelines to facilitate a technology-driven care infrastructure \cite{jingxinwei}, that is, using predominately sensors at home to collect and analyze older adults' home health and safety data that can then guide materials to and schedule resources for older adults in need.
Meanwhile, civil affairs departments, such as the Shanghai Civil Affairs Bureau, issue policies and guidelines to facilitate a human-driven care infrastructure \cite{olderpartner3}, wherein 
``younger'' older adults work not just to collect and analyze home health and safety data, guide materials to, and schedule resources for older adults in need, but also as a host of other practices that include minor activities like explaining a new policy or repairing the household to major activities like mediating care stakeholders and acquiring food and medications from stores. }\sam{Building on the lens of infrastructure studies that foreground sociotechnical contexts and relations, below, we advocate for turning to aging in \textit{community} and rethinking technology's role in future CSCW research on aging and care infrastructure at large. }

\subsection{\sam{Rethinking Technological Solutions for Aging in Place}}


\sam{Previous research has suggested that technology can play a key role in the lives of older adults, with technology seen as the ``enabling'' factor that allows for aging in place to occur \cite{kim2017digital}. For instance, previous research has shown how digital technologies can record activities of daily living (ADLs), a process of monitoring routines that can chart one's status quo or downfall: ``[Older adults'] activities do not need to be recognized but should rather be verified. Deviations are a warning sign of degradation'' \cite[p.43]{caroux2014verification}. As long as one does not degrade, according to the parameters set by a verification system, one can age in place as ``aging in place is desirable.'' The enabling possibilities of technology are shared in our empirical case by the national- and city-wide push to develop technology-driven care infrastructures, such as in the "Home-based Virtual Nursing Home" initiative, that allows ``older adults to enjoy `institution-like' professional care services at home'' through the implementation of digital technologies like ``vital sign monitoring, alarms, bed sensors'' \cite{virtualbed1}. Such technology-driven pushes were then developed further by project managers like Lei, who discussed these mandates as informing the trajectory of his health-sensing technologies.}

\sam{We do not intend to minimize the importance of health and safety technologies for aging in place. Aging in place is itself a deliberate, ongoing practice that needs careful consideration. An older adult's relationship to their environment is ``not [in] a static state'' but exists instead in ``dynamic tension that can vary day to day, month to month, year to year'' \cite[p.122]{golant2015aging}. The book \textit{Aging in the Right Place} by Golant, for instance, discusses the virtues of many living arrangements and suggests that aging in place is not always the right move, depending on one's contextual self. Sometimes, one should not age in place, and similarly, sometimes, health and safety technologies provide the assurance and support that is necessary for aging in place.} 

\sam{However, what we do want to question is the centrality of aging in place technologies, as currently designed. These technologies tend to focus on one aspect of the lives of older adults aging in place---namely, quantified measures of health, as taken in the form of heart monitoring readings, movement in bed and around the home, and metrics of a salubrious home environment, free from smoke and carbon monoxide. The centrality of these technologies also ignores the practical limitations of the technologies themselves and the alternative enabling factors already in practice. In this section, we focus on those limitations of technology, before unpacking the alternative enabling factors already in practice.}

As noted in our findings, we have uncovered that in this context, the role played by current technology-driven systems in practical caregiving scenarios to support older adults aging in place is remarkably limited. Data is often collected for data's sake and is often siloed into the data repositories of respective companies. Despite some excitement about the potential for technology to improve management efficiency, the majority of our participants highlighted the limitations and challenges of existing technology interventions in supporting aging in place. Breakdowns occurred between technology systems, care recipients considered some technologies to be surveillance artifacts that marked them as frail, and stakeholders were required to repair broken infrastructure and change information manually.

The inherent contextual complexity of aging in place poses a series of technological challenges for increasingly standardized and institutionalized data-driven care technologies \cite{sun2023data,sun2023care, thach2023key} in aging in place scenarios. \sam{Often, these technologies were not necessarily built to support the needs of older adults---they were instead built to support a kind of health monitoring. The complexity} of older adults' aging-in-place needs \cite{sun2023data} (see Figure \ref{fig:caresystem}) is not considered\sam{, and non-quantifiable needs are difficult for }digital technologies to accurately and effectively perceive. 

Additionally, the technology-driven care infrastructure often mediates different care providers, as noted in 4.2.2. These different stakeholders, including medical care providers, daily care providers, government service providers, and more, can be differentiated by their formality (formal and informal), service content (medical care, daily care, social support, emotional care), funding source, cost, frequency, and more (see Figure \ref{fig:caresystem}). \sam{The diversity of stakeholders adds further complexity and makes it difficult for data-driven systems to quantify and consolidate pertinent information for} each stakeholder---recall the hospitals that do not allow booking on third-party services or systems that do not share information. These examples align with previous CSCW research on the fragmentation of resources and institutional and inter-organizational complexity \cite{pine2012fragmentation, toombs2018sociotechnical, star1994steps}. 

These issues led to limited practical utility and effectiveness\sam{, even to those project managers who developed these technologies (recall Zhi who said advanced sensing technologies ``sound good \textit{in theory''}}). It should be noted that the technologies mentioned in our research were developed and implemented locally, with project managers and technologies coming from Shanghai. In previous work, local artifacts have been noted for their ``crucial'' role in promoting access, informed design, and increase in use \cite{manda2018inclusive}. While that is certainly true in some contexts and a step in the right direction, here, this has not been the case. 

\sam{These findings warrant careful attention and reflection from the CSCW community about current technologies for aging in place and how such technologies interact with the larger care infrastructure. Such attention is aligned with Bell et al., who recommended a critical stance and delicate touch when designing domestic technologies \cite{bell2005making}. For Bell et al., designers should ``actively reflect on, rather than passively propagate, the existing politics and culture of home life and to develop new alternatives for design'' \cite[p.150]{bell2005making}. In reflecting on the limitations of current aging in place technologies, our study aligns with prior claims in CSCW and HCI that argue that the predominant conception of older adults as merely needing technology to compensate for the decline in their biological capabilities is inaccurate \cite{sun2014being, lindley2008designing, durick2013dispelling, vines2015age, felt2016handbook}. This can be seen in research that finds older adult participants respond negatively towards self-management health systems \cite{d2019ageing}. In this research, a perceived lack of value, ``frustration towards a society that demands more and more [tech] savviness,'' and more are noted as reasons for non-adoption and non-use \cite[p.290]{d2019ageing}. In response, we turn to the human-driven care infrastructure, an alternative enabling factor, to unpack new directions for aging-in-place technologies.}

\subsection{\sam{Human-Driven Care Infrastructure as Enabling Aging in Place}}

\sam{In keeping with our second research question, we have investigated the roles care partners play in a human-driven care infrastructure. This human-driven care infrastructure \final{is an assemblage of systems, actors, and environments that collectively shape caregiving practices, with human actors playing a central and guiding role}---in our case, older partners. It is also important to note that the Older Partner program and its loose, informal counterparts in everyday life \cite{cheung2019changing} predate technology-driven care infrastructures. Human-driven care infrastructure, then, is aligned with previous work on people as infrastructure \cite{simone2004people, simone2021ritornello, cheon2023powerful}. This corpus ``render(s) people’s actions as technical, an infrastructure generating possibilities of acting in concert beyond the explicit intention or planning of any individual or group'' \cite[p.1341]{simone2021ritornello}. This work suggests the ``when'' of infrastructure in Star and Ruhleder, as infrastructure is enacted by and through relations that have temporal and spatial considerations \cite{star1994steps}.}

\sam{The Older Partner program and human-driven care infrastructure, in general, are compelling because of what they are informed by and how they are perceived.} Prior literature \cite{harvey2009ageing, sun2014being} has argued that generations are shaped by their specific historical, social, and cultural context. Our paper demonstrates how particular social and cultural processes and historical moments in China, such as the Cultural Revolution and the Movement of Chinese Educated Urban Youth to the Countryside, have influenced our participants' perceptions of aging and responsibilities towards those who are aging\sam{---recall our findings} in 5.2.1 that suggest care work is a social responsibility. Traditional Chinese notions of aging at home \cite{zhang2006family}, filial care \cite{lancet2022population}, and government sponsorship of older adult care \cite{olderpartner2} each significantly influenced attitudes and informed what kind of caregiving was preferred. From the perspective of older partners, their care-providing practices are also significantly influenced by Chinese sociohistorical background. All participants joined as older partners after retirement, worked for years as care partners, and devoted their retirement time to caring for and accompanying care recipients, without expecting any tangible returns (returns that were, whether expected or not, minimal). \sam{Still, our participants expressed happiness, a sense of self-worth, and a strong belief that their foundational years  \cite{contribution} growing up had shaped their current practices as older partners and the desire to support the previous generation in aging in place.} 

\sam{Additionally, the Older Partner program is unique in that its practices are enacted by older adults for other older adults. Put another way, the articulation of care circulates within the older adult populations of Shanghai. These caregiving practices are emergent and contextually situated, not to be understood as predefined or predetermined \cite{suchman1987plans}. Recall Xiang, who collaborated with other stakeholders like family members in her caregiving practices when necessary, or Chun, who navigated testy care recipients with tacit understandings of behaviors, like the draping of a towel to signify continued health. In this way, the Older Partner program wove together sociohistorical circumstances, peer support, and community and contextual knowledge into an assemblage of care that was meaningful to both care partners and care recipients, and was understood to be significant by local officials and family members alike. The Older Partner program, then, supports the everyday functioning of care infrastructure and facilitates aging in place.}

\subsection{\sam{The Role of Technology for Aging in Place: Supportive, not Substitutive}}

\sam{From this perspective, the promotion of technological solutions for aging in place presupposes their own usefulness, ignoring infrastructural formations on the ground to substitute their value with metrics of health and safety. By taking a small subsection of Figure \ref{fig:caresystem}'s ``older adult's needs for aging in place,'' the technologies supplant the situated action of the Older Partner program's participants with the metrics of Figure \ref{fig:variousData}. In previous infrastructure studies work, the assemblage of infrastructure is delicate, and one small alteration---like an increase of water flow to one space \cite{anand2018public}---can have knock-on effects that resonate throughout the infrastructure. What does the introduction of technology change about our given infrastructural arrangements? It can support (or deny) aging in place, as assessments of health and wellness can be tied to medical and social services \cite{Services2022}. It can create a sense of safety (or surveillance) for older adults aging in place and their families \cite{kim2017digital, vines2015age, doyle2012yourwellness}. Such technology can support (or prohibit) the desires of older adults, changing not only the lives of care recipients but also the lives of caregivers, who find meaning in daily caregiving.} 

\sam{Thus, we recommend for aging in place technologies to take a \textit{supportive, not substitutive} role in the larger assemblage of care infrastructure. By this, we recommend that such technologies support the functioning arrangements that are already in place or, otherwise, support older adults' needs. In thinking with Figure \ref{fig:caresystem}, safety and health are but two needs of care recipients---others include mental well-being, recreational activities, and social interaction. By tying aging in place to health and safety, a deficit model of aging emerges that embeds assumptions of frailty and declining health to the development of technologies.} This line of research concludes that if older adults are to accept these technologies, ``more efforts need to be made to ensure that the benefits [of technology] are considered as irrefutably useful'' \cite[p.294]{d2019ageing}. We propose a shift in the design of aging in place technology and suggest that health and safety technologies, while significant, should not be the sole or even the predominant focus of technology for aging in place. Specifically, technologies should support older adults' fundamental human needs, such as autonomy, dignity, and value \cite{czech2023independence}. Such a suggestion also aligns with research that questions the ultimate design logics of assistive technologies like monitoring software and AI chatbots, questioning if older adults would be better supported by technologies that speak to ``spiritual needs'' including the need to find self-worth and the need to find belonging in community \cite{trothen2022intelligent}. In this way, we suggest the role of technology should be to better support older adults in fulfilling these human needs \cite{durick2013dispelling}, rather than, for instance, substituting human check-ins with monitoring technologies.

From the perspective of older partners, the shift from substitutive to supportive is also crucial. Our older partners generally fall under the conventional concept of ``old age'' (i.e., aged over 60 or 65) \cite{agingdefination}. In our case as well as others (e.g., \cite{li2023any, nurain2021hugging}), older adults are active participants in communities, and they play a vital role in supporting aging in place. In our case, the older partners mediate aging in place in local communities for both care recipients and neighborhood committees; their role and value surpass current aging in place technologies. And yet, current technologies seek, on some level, to substitute their work. If such human-driven care infrastructure for aging in place were to be fully substituted by the incomplete and siloed data of sensing technologies that tend to distill an older adult's worth into quantifiable health and safety statistics, both the ``younger'' and ``older'' older adults would lose out on valuable interactions within their community and feelings of self-worth and accomplishment. In considering the above \cite{trothen2022intelligent} and our findings section 5.2.1, the human-driven care infrastructure supports a feeling of belonging within a community. Such a feeling is supported by neighborhood committee members---recall Jing's declaration that older partners play a significant role in the dynamics of the community itself.

\sam{Of course, in practice, human- and technology-driven care infrastructures are not separate but intertwined. Consider, for instance, how older partners utilized WeChat and other social media services to support their check-ins, incorporating technology into their daily practices. We imperfectly bifurcate care infrastructure as a way to trace the relations between different approaches to care. Often in our empirical case, when human-led and technology-led care meet in practice, older partners have been tasked with supporting the newly implemented technology and not the other way around. These older partners often mediate the responsibilities of technology, complicating their own responsibilities and taking up their time. Recall how participants repaired monitoring technologies upon breakdown and aided in appointment-making when the limits of systems had been reached. Thus, older partners were embedded within the technology-driven care infrastructure, making it work as expected.}

\sam{In sum, we present two considerations for supportive, not substitutive, technologies. First, technology should be informed by and support on-the-ground arrangements already in place, as well as the needs of older adults, beyond health and safety. In our case, technologies} could support the work of older partners, helping them address the practical difficulties they encounter, reducing aspects of their workload they find troublesome, and allowing them more time and energy to deliver human-centered services. This proposition is focused on actual practices---for instance, for a care recipient who is unwilling to be ``disturbed every day'' by an older partner, proactive sensing technologies could be provided; for situations \sam{where care recipients have limited mobility, movement monitoring could offer solutions}---with the permission of care recipients briefed on the value of the technology and its possible privacy/security risks \cite{murthy2021individually}. \sam{Second, researchers and designers should consider the limits of technology as necessary tools to support aging in place. In our research, we can see how monitoring technologies imperfectly attempt to make quantitative and scalable that which is already being done by older partners. }What is the technological equivalent to agreeing that a towel on a balcony is a sign of well-being? Could such understandings be translated to a digital environment, could they be scaled \sam{up} \cite{lu2023shifting}, and would that \sam{then} be preferable to the cultivated relationships that older partners \sam{and} care recipients currently share? 

\subsection{\sam{A Direction for Technology-Driven Care Infrastructure: Supporting Aging in Community}}

From the perspective of quality of life and community development, we also believe that technology should better support neighborly interactions, rather than substituting them in the name of more precise data and services. The Older Partner program we studied illustrates how, in a rapidly digitized and commercialized city like Shanghai, neighborhood relationships are maintained and contribute significantly to the development of community efficacy \cite{kropczynski2021towards, carroll2005collective}. Our study contributes to the existing literature on how to enhance community efficacy in the digital age \cite{mosconi2017facebook, fedosov2021dozen}, providing a unique case of older adults supporting community efficacy, specifically for other older adults. Aligning with current literature, we advocate for technology that better supports these neighborhood interactions. Put another way: in opposition to the global aspirations of contemporary technologies, we orient our design implications toward the strengthening and facilitation of local communication and interaction (see: \cite{vivacqua2018personal}). 

Moreover, the presence of older partners also serves as a crucial and useful solution \sam{for} alleviating social issues arising from the surge in the aging population \cite{chamie2007population} and enhancing the self and one's social worth of individuals after retirement. Most of our participants started this older partner work immediately upon retirement, viewing the program as an integral part of their daily lives rather than strictly a ``job'' that must be done. The ``older'' older partners they care for become part of their social relationships, like their ``own parents.'' Through this work, they derive significant self-fulfillment, self-worth, and satisfaction. We consider this care infrastructure to be highly beneficial for the sustainability and healthy development of an aging society. 

This speaks to a larger issue with aging in place technologies---these technologies often support aging in place through support of ``independence.'' Aging in place is often equated with independence, something potentially fleeting: ``Aging in place research seeks to augment self-care in older adults to enable them to retain their independence for longer, and delay the need to live with a caregiver or move into an assisted living institution'' \cite[p.1606]{caldeira2017senior}. Here, dependence (in \sam{the form of a} move into an institution) is ``delayed'' by dependence (on the daily use of technological artifacts). In this way, aging in place is often conflated with being independent in one's home with a modified home environment \sam{that (in some cases) simulates the institutional environment}. However, and in keeping with human-computer interaction literature on interdependence \cite{bennett2018interdependence}, aging in place is not solitary---a more appropriate term might be \textit{aging in community}. As such, technology-driven care infrastructures for aging in place that only consider \sam{each} individual \sam{as a fragmented data point} instead of \sam{one part of a} larger community miss a significant portion of why older adults want to age in place in the first place. 

\sam{The question, then, is how can CSCW and its related fields support aging in community? Beyond the critical need for empirical work done within the community, with vested community interests' participation and older adults at the center, two recent polls in the U.S. might illuminate paths forward. In one poll, more than 4,000 adults over 50 were asked about their familiarity with programs for older adults in their area \cite{national2024most}. These programs, like Area Agencies on Aging, offer services for aging in place and are critical mediators of community care infrastructure. Less than 10\% of those polled stated they were ``familiar and have used'' these programs. In a second poll, a national sample of adults aged 50 to 80 were asked a series of questions related to aging in place. In this poll, 40\% of respondents did not know of organizations or spaces in their community ``that are welcoming for people their age to socialize, exercise, or participate in activities'' \cite{robinson2022older}.} 

\sam{In both cases, the issue is not necessarily a lack of meaningful services, programs, organizations, or spaces. There exists an informational gap between older adults and already existing aging in community resources. As a field grounded in coordination, collaboration, and the dissemination of information, CSCW is well-positioned to bridge this gap by supporting older adults’ connections to existing community services and spaces. As questioned by Trothen \cite[p.9]{trothen2022intelligent} in relation to aging and technology, ``Is part of what we are searching for a `something more' that may not be satisfied by technology?'' If this something more may not be satisfied by technology, technology may at least play a supportive role in finding it.}

 \section{Conclusion and Future Directions}

In this paper, we reported findings from a long-term ethnographic study about \sam{human- and technology-driven care infrastructures in Shanghai, China. In our findings, we highlight how technologies for aging in place are limited to designs for health and safety that are good ``in theory'' but are complicated in practice by siloed data, conflicting data collection, and resistance by care receivers. We also highlight how older adults themselves are caregivers, and their practices deeply inform care and aging in place in practice, from repairing technology to going to the store for those a generation older.}

\sam{In considering the above, our discussion makes the case that technology is not the only (or most significant) ``enabling'' factor to support older adults aging in place. We suggest technology should support, rather than substitute, current caregiving practices, and following this, we present considerations for future designers and researchers to reflect on, including the limitations of technology and a larger consideration of the varied needs of older adults who are aging in place. Finally, we point out that aging in place is not an independent but interdependent act predicated on community living, and, as such, technology should be informed by the value of aging in a community, not alone, in one's home.}


Our study \sam{can be extended by future research}. Firstly, this study, as all studies, was conducted in a specific social, cultural, and technological context---in our case, Shanghai, China. \sam{These factors need to be considered and triangulated among other methodologically-similar research into care infrastructure to see if our contributions are generalizable to other social, cultural, and technological contexts. Additionally, }our current research does not provide a comprehensive picture of the sociotechnical infrastructure \sam{but, instead, presents a segmented picture to present a narrative interrogation of human and technology caregiving practices. Future work could examine further delimited pieces of the sociotechnical assemblage, including, for instance, family members and their interactions with technology-driven care infrastructure. In addition, our study primarily employed qualitative methods.} In the future, we aim to undertake quantitative analyses, such as examining log data from current technological infrastructures, to assess care infrastructure from multiple dimensions.


\begin{acks}
We thank Shanghai Science and Technology Innovation Action Plan project (No. 21511104500) for funding this research.  
\end{acks}

\bibliographystyle{ACM-Reference-Format}
\bibliography{sample-base}


\begin{thebibliography}{134}


\ifx \showCODEN    \undefined \def \showCODEN     #1{\unskip}     \fi
\ifx \showISBNx    \undefined \def \showISBNx     #1{\unskip}     \fi
\ifx \showISBNxiii \undefined \def \showISBNxiii  #1{\unskip}     \fi
\ifx \showISSN     \undefined \def \showISSN      #1{\unskip}     \fi
\ifx \showLCCN     \undefined \def \showLCCN      #1{\unskip}     \fi
\ifx \shownote     \undefined \def \shownote      #1{#1}          \fi
\ifx \showarticletitle \undefined \def \showarticletitle #1{#1}   \fi
\ifx \showURL      \undefined \def \showURL       {\relax}        \fi
\providecommand\bibfield[2]{#2}
\providecommand\bibinfo[2]{#2}
\providecommand\natexlab[1]{#1}
\providecommand\showeprint[2][]{arXiv:#2}

\bibitem[Anand(2018)]%
        {anand2018public}
\bibfield{author}{\bibinfo{person}{Nikhil Anand}.} \bibinfo{year}{2018}\natexlab{}.
\newblock \showarticletitle{A public matter: Water, hydraulics, biopolitics}.
\newblock \bibinfo{journal}{\emph{The promise of infrastructure}} (\bibinfo{year}{2018}), \bibinfo{pages}{155--72}.
\newblock


\bibitem[Anstey et~al\mbox{.}(2007)]%
        {anstey2007prevalence}
\bibfield{author}{\bibinfo{person}{Kaarin~J Anstey}, \bibinfo{person}{Chwee von Sanden}, \bibinfo{person}{Kerry Sargent-Cox}, {and} \bibinfo{person}{Mary~A Luszcz}.} \bibinfo{year}{2007}\natexlab{}.
\newblock \showarticletitle{Prevalence and risk factors for depression in a longitudinal, population-based study including individuals in the community and residential care}.
\newblock \bibinfo{journal}{\emph{The American journal of geriatric psychiatry}} \bibinfo{volume}{15}, \bibinfo{number}{6} (\bibinfo{year}{2007}), \bibinfo{pages}{497--505}.
\newblock


\bibitem[Bakeman and Quera(2012)]%
        {bakeman2012behavioral}
\bibfield{author}{\bibinfo{person}{Roger Bakeman} {and} \bibinfo{person}{Vicen Quera}.} \bibinfo{year}{2012}\natexlab{}.
\newblock \showarticletitle{Behavioral observation}.
\newblock  (\bibinfo{year}{2012}).
\newblock


\bibitem[Bateson(2000)]%
        {bateson2000steps}
\bibfield{author}{\bibinfo{person}{Gregory Bateson}.} \bibinfo{year}{2000}\natexlab{}.
\newblock \bibinfo{booktitle}{\emph{Steps to an ecology of mind: Collected essays in anthropology, psychiatry, evolution, and epistemology}}.
\newblock \bibinfo{publisher}{University of Chicago press}.
\newblock


\bibitem[Bayer and Harper(2000)]%
        {bayer2000fixing}
\bibfield{author}{\bibinfo{person}{Ada-Helen Bayer} {and} \bibinfo{person}{Leon Harper}.} \bibinfo{year}{2000}\natexlab{}.
\newblock \bibinfo{booktitle}{\emph{Fixing to stay: A national survey of housing and home modification issues}}.
\newblock \bibinfo{publisher}{AARP}.
\newblock


\bibitem[Bell et~al\mbox{.}(2005)]%
        {bell2005making}
\bibfield{author}{\bibinfo{person}{Genevieve Bell}, \bibinfo{person}{Mark Blythe}, {and} \bibinfo{person}{Phoebe Sengers}.} \bibinfo{year}{2005}\natexlab{}.
\newblock \showarticletitle{Making by making strange: Defamiliarization and the design of domestic technologies}.
\newblock \bibinfo{journal}{\emph{ACM Transactions on Computer-Human Interaction (TOCHI)}} \bibinfo{volume}{12}, \bibinfo{number}{2} (\bibinfo{year}{2005}), \bibinfo{pages}{149--173}.
\newblock


\bibitem[Bennett et~al\mbox{.}(2018)]%
        {bennett2018interdependence}
\bibfield{author}{\bibinfo{person}{Cynthia~L Bennett}, \bibinfo{person}{Erin Brady}, {and} \bibinfo{person}{Stacy~M Branham}.} \bibinfo{year}{2018}\natexlab{}.
\newblock \showarticletitle{Interdependence as a frame for assistive technology research and design}. In \bibinfo{booktitle}{\emph{Proceedings of the 20th international acm sigaccess conference on computers and accessibility}}. \bibinfo{pages}{161--173}.
\newblock


\bibitem[Berridge and Wetle(2020)]%
        {berridge2020older}
\bibfield{author}{\bibinfo{person}{Clara Berridge} {and} \bibinfo{person}{Terrie~Fox Wetle}.} \bibinfo{year}{2020}\natexlab{}.
\newblock \showarticletitle{Why older adults and their children disagree about in-home surveillance technology, sensors, and tracking}.
\newblock \bibinfo{journal}{\emph{The Gerontologist}} \bibinfo{volume}{60}, \bibinfo{number}{5} (\bibinfo{year}{2020}), \bibinfo{pages}{926--934}.
\newblock


\bibitem[Brown~Wilson(2007)]%
        {brown2007historical}
\bibfield{author}{\bibinfo{person}{Keren Brown~Wilson}.} \bibinfo{year}{2007}\natexlab{}.
\newblock \showarticletitle{Historical evolution of assisted living in the United States, 1979 to the present}.
\newblock \bibinfo{journal}{\emph{The Gerontologist}} \bibinfo{volume}{47}, \bibinfo{number}{suppl\_1} (\bibinfo{year}{2007}), \bibinfo{pages}{8--22}.
\newblock


\bibitem[Bulmer(2015)]%
        {bulmer2015social}
\bibfield{author}{\bibinfo{person}{Martin Bulmer}.} \bibinfo{year}{2015}\natexlab{}.
\newblock \bibinfo{booktitle}{\emph{The social basis of community care (routledge revivals)}}.
\newblock \bibinfo{publisher}{Routledge}.
\newblock


\bibitem[Bureau(2023)]%
        {virtualbed1}
\bibfield{author}{\bibinfo{person}{Shanghai Municipal Civil~Affairs Bureau}.} \bibinfo{year}{2023}\natexlab{}.
\newblock \bibinfo{booktitle}{\emph{Implementation Opinions on Actively Promoting the Standardized Development of Home Care Beds in the City}}.
\newblock
\urldef\tempurl%
\url{https://www.shanghai.gov.cn/gwk/search/content/639158860cf0402a92d2c3627e8a016b}
\showURL{%
\tempurl}


\bibitem[Butler(2006)]%
        {butler2006evaluating}
\bibfield{author}{\bibinfo{person}{Sandra~S Butler}.} \bibinfo{year}{2006}\natexlab{}.
\newblock \showarticletitle{Evaluating the Senior Companion Program: a mixed-method approach}.
\newblock \bibinfo{journal}{\emph{Journal of Gerontological Social Work}} \bibinfo{volume}{47}, \bibinfo{number}{1-2} (\bibinfo{year}{2006}), \bibinfo{pages}{45--70}.
\newblock


\bibitem[Caldeira et~al\mbox{.}(2017)]%
        {caldeira2017senior}
\bibfield{author}{\bibinfo{person}{Clara Caldeira}, \bibinfo{person}{Matthew Bietz}, \bibinfo{person}{Marisol Vidauri}, {and} \bibinfo{person}{Yunan Chen}.} \bibinfo{year}{2017}\natexlab{}.
\newblock \showarticletitle{Senior care for aging in place: balancing assistance and independence}. In \bibinfo{booktitle}{\emph{Proceedings of the 2017 ACM Conference on Computer Supported Cooperative Work and Social Computing}}. \bibinfo{pages}{1605--1617}.
\newblock


\bibitem[Caroux et~al\mbox{.}(2014)]%
        {caroux2014verification}
\bibfield{author}{\bibinfo{person}{Lo{\"\i}c Caroux}, \bibinfo{person}{Charles Consel}, \bibinfo{person}{Lucile Dupuy}, {and} \bibinfo{person}{H{\'e}l{\`e}ne Sauz{\'e}on}.} \bibinfo{year}{2014}\natexlab{}.
\newblock \showarticletitle{Verification of daily activities of older adults: a simple, non-intrusive, low-cost approach}. In \bibinfo{booktitle}{\emph{Proceedings of the 16th international ACM SIGACCESS conference on Computers \& accessibility}}. \bibinfo{pages}{43--50}.
\newblock


\bibitem[Carroll et~al\mbox{.}(2005)]%
        {carroll2005collective}
\bibfield{author}{\bibinfo{person}{John~M Carroll}, \bibinfo{person}{Mary~Beth Rosson}, {and} \bibinfo{person}{Jingying Zhou}.} \bibinfo{year}{2005}\natexlab{}.
\newblock \showarticletitle{Collective efficacy as a measure of community}. In \bibinfo{booktitle}{\emph{Proceedings of the SIGCHI conference on human factors in computing systems}}. \bibinfo{pages}{1--10}.
\newblock
\urldef\tempurl%
\url{https://doi.org/10.1145/1054972.1054974}
\showURL{%
\tempurl}


\bibitem[Chamie et~al\mbox{.}(2007)]%
        {chamie2007population}
\bibfield{author}{\bibinfo{person}{Joseph Chamie}, \bibinfo{person}{Lisa Berkman}, \bibinfo{person}{Adele~M Hayutin}, \bibinfo{person}{James~P Smith}, {and} \bibinfo{person}{Mary~Jo Hoeksema}.} \bibinfo{year}{2007}\natexlab{}.
\newblock \bibinfo{title}{Why Population Ageing Matters: A Global Perspectives}.
\newblock


\bibitem[Charmaz(2000)]%
        {charmaz2000grounded}
\bibfield{author}{\bibinfo{person}{Kathy Charmaz}.} \bibinfo{year}{2000}\natexlab{}.
\newblock \showarticletitle{Grounded theory: Objectivist and constructivist methods}.
\newblock \bibinfo{journal}{\emph{Handbook of qualitative research}} \bibinfo{volume}{2}, \bibinfo{number}{1} (\bibinfo{year}{2000}), \bibinfo{pages}{509--535}.
\newblock


\bibitem[Chen et~al\mbox{.}(2023)]%
        {chen2023maintainers}
\bibfield{author}{\bibinfo{person}{Yuchen Chen}, \bibinfo{person}{Yuling Sun}, {and} \bibinfo{person}{Silvia Lindtner}.} \bibinfo{year}{2023}\natexlab{}.
\newblock \showarticletitle{Maintainers of Stability: The Labor of China’s Data-Driven Governance and Dynamic Zero-COVID}. In \bibinfo{booktitle}{\emph{Proceedings of the 2023 CHI Conference on Human Factors in Computing Systems}}. \bibinfo{pages}{1--16}.
\newblock


\bibitem[Cheon(2023)]%
        {cheon2023powerful}
\bibfield{author}{\bibinfo{person}{EunJeong Cheon}.} \bibinfo{year}{2023}\natexlab{}.
\newblock \showarticletitle{Powerful Futures: How a Big Tech Company Envisions Humans and Technologies in the Workplace of the Future}.
\newblock \bibinfo{journal}{\emph{Proceedings of the ACM on Human-Computer Interaction}} \bibinfo{volume}{7}, \bibinfo{number}{CSCW2} (\bibinfo{year}{2023}), \bibinfo{pages}{1--35}.
\newblock


\bibitem[Cheung(2019)]%
        {cheung2019changing}
\bibfield{author}{\bibinfo{person}{Pui Ling~Ada Cheung}.} \bibinfo{year}{2019}\natexlab{}.
\newblock \showarticletitle{Changing perception of the rights and responsibilities in family care for older people in urban China}.
\newblock \bibinfo{journal}{\emph{Journal of Aging \& Social Policy}} \bibinfo{volume}{31}, \bibinfo{number}{4} (\bibinfo{year}{2019}), \bibinfo{pages}{298--320}.
\newblock
\urldef\tempurl%
\url{https://doi.org/10.1080/08959420.2019.1626324}
\showURL{%
\tempurl}


\bibitem[Ciolfi et~al\mbox{.}(2023)]%
        {ciolficscw}
\bibfield{author}{\bibinfo{person}{Luigina Ciolfi}, \bibinfo{person}{Myriam Lewkowicz}, {and} \bibinfo{person}{Kjeld Schmidt}.} \bibinfo{year}{2023}\natexlab{}.
\newblock \showarticletitle{CSCW: History, Core Issues, and Approaches in Computer-Supported Cooperative Work}.
\newblock  (\bibinfo{year}{2023}).
\newblock


\bibitem[Clark et~al\mbox{.}(2023)]%
        {clark2023place}
\bibfield{author}{\bibinfo{person}{William~AV Clark}, \bibinfo{person}{Rachel Ong~ViforJ}, {and} \bibinfo{person}{Christopher Phelps}.} \bibinfo{year}{2023}\natexlab{}.
\newblock \showarticletitle{Place Attachment and Aging in Place: Preferences and Disruptions}.
\newblock \bibinfo{journal}{\emph{Research on Aging}} (\bibinfo{year}{2023}), \bibinfo{pages}{01640275231209683}.
\newblock


\bibitem[Commission(2021)]%
        {Shanghaireport}
\bibfield{author}{\bibinfo{person}{Shanghai Municipal~Health Commission}.} \bibinfo{year}{2021}\natexlab{}.
\newblock \bibinfo{booktitle}{\emph{Statistics on Demographics and Programs of Senior Citizens in Shanghai}}.
\newblock
\urldef\tempurl%
\url{https://wsjkw.sh.gov.cn/cmsres/59/5983e677030440f18a034ce02a0c10e5/d55b3444b8c63b77d3b73231f4f66a28.pdf}
\showURL{%
\tempurl}


\bibitem[Commission(2022)]%
        {weijianwei}
\bibfield{author}{\bibinfo{person}{Shanghai Municipal~Health Commission}.} \bibinfo{year}{2022}\natexlab{}.
\newblock \bibinfo{booktitle}{\emph{Shanghai Healthy Aging Action Plan (2022-2025)}}.
\newblock
\urldef\tempurl%
\url{https://wsjkw.sh.gov.cn/gjhztgahz/20220930/68c6fc87be30409993b5311e93408254.html}
\showURL{%
\tempurl}


\bibitem[Consel et~al\mbox{.}(2015)]%
        {consel2015unifying}
\bibfield{author}{\bibinfo{person}{Charles Consel}, \bibinfo{person}{Lucile Dupuy}, {and} \bibinfo{person}{H{\'e}l{\`e}ne Sauz{\'e}on}.} \bibinfo{year}{2015}\natexlab{}.
\newblock \showarticletitle{A unifying notification system to scale up assistive services}. In \bibinfo{booktitle}{\emph{Proceedings of the 17th international ACM SIGACCESS conference on Computers \& accessibility}}. \bibinfo{pages}{77--87}.
\newblock


\bibitem[Czech et~al\mbox{.}(2023)]%
        {czech2023independence}
\bibfield{author}{\bibinfo{person}{Elaine Czech}, \bibinfo{person}{Ewan Soubutts}, \bibinfo{person}{Rachel Eardley}, {and} \bibinfo{person}{Aisling~Ann O'Kane}.} \bibinfo{year}{2023}\natexlab{}.
\newblock \showarticletitle{Independence for Whom? A Critical Discourse Analysis of Onboarding a Home Health Monitoring System for Older Adult Care}. In \bibinfo{booktitle}{\emph{Proceedings of the 2023 CHI Conference on Human Factors in Computing Systems}}. \bibinfo{pages}{1--15}.
\newblock


\bibitem[D'haeseleer et~al\mbox{.}(2019)]%
        {d2019ageing}
\bibfield{author}{\bibinfo{person}{Ine D'haeseleer}, \bibinfo{person}{Kathrin Gerling}, \bibinfo{person}{Dominique Schreurs}, \bibinfo{person}{Bart Vanrumste}, {and} \bibinfo{person}{Vero Vanden~Abeele}.} \bibinfo{year}{2019}\natexlab{}.
\newblock \showarticletitle{Ageing is not a disease: Pitfalls for the acceptance of self-management health systems supporting healthy ageing}. In \bibinfo{booktitle}{\emph{Proceedings of the 21st International ACM SIGACCESS Conference on Computers and Accessibility}}. \bibinfo{pages}{286--298}.
\newblock


\bibitem[Doyle et~al\mbox{.}(2012)]%
        {doyle2012yourwellness}
\bibfield{author}{\bibinfo{person}{Julie Doyle}, \bibinfo{person}{Brian O'Mullane}, \bibinfo{person}{Shauna McGee}, {and} \bibinfo{person}{R~Benjamin Knapp}.} \bibinfo{year}{2012}\natexlab{}.
\newblock \showarticletitle{YourWellness: designing an application to support positive emotional wellbeing in older adults}. In \bibinfo{booktitle}{\emph{International British Computer Society Human Computer Interaction Conference}}. Electronic Workshops in Computing.
\newblock


\bibitem[Durick et~al\mbox{.}(2013)]%
        {durick2013dispelling}
\bibfield{author}{\bibinfo{person}{Jeannette Durick}, \bibinfo{person}{Toni Robertson}, \bibinfo{person}{Margot Brereton}, \bibinfo{person}{Frank Vetere}, {and} \bibinfo{person}{Bjorn Nansen}.} \bibinfo{year}{2013}\natexlab{}.
\newblock \showarticletitle{Dispelling ageing myths in technology design}. In \bibinfo{booktitle}{\emph{Proceedings of the 25th Australian Computer-Human Interaction Conference: Augmentation, Application, Innovation, Collaboration}}. \bibinfo{pages}{467--476}.
\newblock
\urldef\tempurl%
\url{https://dl.acm.org/doi/10.1145/2541016.2541040}
\showURL{%
\tempurl}


\bibitem[Economic and Commission(2022)]%
        {jingxinwei}
\bibfield{author}{\bibinfo{person}{Shanghai~Municipal Economic} {and} \bibinfo{person}{Informatization Commission}.} \bibinfo{year}{2022}\natexlab{}.
\newblock \bibinfo{booktitle}{\emph{Notice on the Collection of Shanghai Smart Health and Elderly Care Product and Service Promotion Catalog}}.
\newblock
\urldef\tempurl%
\url{https://sheitc.sh.gov.cn/xxfw/20220209/85e0773171d64372b7f8fc86ef270b21.html}
\showURL{%
\tempurl}


\bibitem[Edwards et~al\mbox{.}(2007)]%
        {edwards2007understanding}
\bibfield{author}{\bibinfo{person}{Paul~N Edwards}, \bibinfo{person}{Steven~J Jackson}, \bibinfo{person}{Geoffrey~C Bowker}, {and} \bibinfo{person}{Cory~Philip Knobel}.} \bibinfo{year}{2007}\natexlab{}.
\newblock \showarticletitle{Understanding infrastructure: Dynamics, tensions, and design}.
\newblock  (\bibinfo{year}{2007}).
\newblock


\bibitem[Emerson et~al\mbox{.}(1995)]%
        {emerson1995processing}
\bibfield{author}{\bibinfo{person}{Robert~M Emerson}, \bibinfo{person}{Rachel~I Fretz}, {and} \bibinfo{person}{Linda~L Shaw}.} \bibinfo{year}{1995}\natexlab{}.
\newblock \showarticletitle{Processing fieldnotes: Coding and memoing}.
\newblock \bibinfo{journal}{\emph{Writing ethnographic fieldnotes}} (\bibinfo{year}{1995}), \bibinfo{pages}{142--168}.
\newblock


\bibitem[Fausset et~al\mbox{.}(2011)]%
        {fausset2011challenges}
\bibfield{author}{\bibinfo{person}{Cara~Bailey Fausset}, \bibinfo{person}{Andrew~J Kelly}, \bibinfo{person}{Wendy~A Rogers}, {and} \bibinfo{person}{Arthur~D Fisk}.} \bibinfo{year}{2011}\natexlab{}.
\newblock \showarticletitle{Challenges to aging in place: Understanding home maintenance difficulties}.
\newblock \bibinfo{journal}{\emph{Journal of Housing for the Elderly}} \bibinfo{volume}{25}, \bibinfo{number}{2} (\bibinfo{year}{2011}), \bibinfo{pages}{125--141}.
\newblock


\bibitem[Fedosov et~al\mbox{.}(2021)]%
        {fedosov2021dozen}
\bibfield{author}{\bibinfo{person}{Anton Fedosov}, \bibinfo{person}{Airi Lampinen}, \bibinfo{person}{William Odom}, {and} \bibinfo{person}{Elaine~M Huang}.} \bibinfo{year}{2021}\natexlab{}.
\newblock \showarticletitle{A dozen stickers on a mailbox: Physical encounters and digital interactions in a local sharing community}.
\newblock \bibinfo{journal}{\emph{Proceedings of the ACM on Human-Computer Interaction}} \bibinfo{volume}{4}, \bibinfo{number}{CSCW3} (\bibinfo{year}{2021}), \bibinfo{pages}{1--23}.
\newblock
\urldef\tempurl%
\url{https://doi.org/10.1145/3432939}
\showURL{%
\tempurl}


\bibitem[Felt et~al\mbox{.}(2016)]%
        {felt2016handbook}
\bibfield{author}{\bibinfo{person}{Ulrike Felt}, \bibinfo{person}{Rayvon Fouch{\'e}}, \bibinfo{person}{Clark~A Miller}, {and} \bibinfo{person}{Laurel Smith-Doerr}.} \bibinfo{year}{2016}\natexlab{}.
\newblock \bibinfo{booktitle}{\emph{The handbook of science and technology studies}}.
\newblock \bibinfo{publisher}{Mit Press}.
\newblock


\bibitem[Friedman et~al\mbox{.}(2019)]%
        {friedman2019aging}
\bibfield{author}{\bibinfo{person}{Carli Friedman}, \bibinfo{person}{Joe Caldwell}, \bibinfo{person}{Angela Rapp~Kennedy}, {and} \bibinfo{person}{Mary~C Rizzolo}.} \bibinfo{year}{2019}\natexlab{}.
\newblock \showarticletitle{Aging in place: A national analysis of home-and community-based Medicaid services for older adults}.
\newblock \bibinfo{journal}{\emph{Journal of Disability Policy Studies}} \bibinfo{volume}{29}, \bibinfo{number}{4} (\bibinfo{year}{2019}), \bibinfo{pages}{245--256}.
\newblock


\bibitem[Gardner and Zodikoff(2003)]%
        {gardner2003meeting}
\bibfield{author}{\bibinfo{person}{Daniel~S Gardner} {and} \bibinfo{person}{Bradley~D Zodikoff}.} \bibinfo{year}{2003}\natexlab{}.
\newblock \showarticletitle{Meeting the challenges of social work practice in health care and aging in the 21st century}.
\newblock \bibinfo{journal}{\emph{Social work and health care in an aging society}} (\bibinfo{year}{2003}), \bibinfo{pages}{377--392}.
\newblock


\bibitem[Golant(2015)]%
        {golant2015aging}
\bibfield{author}{\bibinfo{person}{Stephen~M Golant}.} \bibinfo{year}{2015}\natexlab{}.
\newblock \bibinfo{booktitle}{\emph{Aging in the right place}}.
\newblock \bibinfo{publisher}{HPP, Health Professions Press}.
\newblock


\bibitem[Goldwater and Harris(2011)]%
        {AIPdigitalcaretech1}
\bibfield{author}{\bibinfo{person}{Jason Goldwater} {and} \bibinfo{person}{Yael Harris}.} \bibinfo{year}{2011}\natexlab{}.
\newblock \bibinfo{booktitle}{\emph{Using technology to enhance the aging experience: a market analysis of existing technologies}}.
\newblock
\urldef\tempurl%
\url{https://link.springer.com/article/10.1007/s12126-010-9071-2}
\showURL{%
\tempurl}


\bibitem[Government(2020)]%
        {shanghaitech4older}
\bibfield{author}{\bibinfo{person}{Shanghai Government}.} \bibinfo{year}{2020}\natexlab{}.
\newblock \bibinfo{booktitle}{\emph{List of Demands for Smart Elderly Care Application Scenarios in Shanghai}}.
\newblock
\urldef\tempurl%
\url{https://www.shanghai.gov.cn/nw31406/20200820/0001-31406\_1441030.html}
\showURL{%
\tempurl}


\bibitem[Government(2021a)]%
        {shanghaidigitaltrans}
\bibfield{author}{\bibinfo{person}{Shanghai Government}.} \bibinfo{year}{2021}\natexlab{a}.
\newblock \bibinfo{booktitle}{\emph{Comprehensively Promoting the Digital Transformation of Shanghai's Urban Areas}}.
\newblock
\urldef\tempurl%
\url{https://stcsm.sh.gov.cn/xwzx/zt/shcsszx/}
\showURL{%
\tempurl}


\bibitem[Government(2022a)]%
        {olderpartner2}
\bibfield{author}{\bibinfo{person}{Shanghai Government}.} \bibinfo{year}{2022}\natexlab{a}.
\newblock \bibinfo{booktitle}{\emph{`Old Partners' Program Ensures Peace of Mind for Elderly Individuals Living Alone}}.
\newblock
\urldef\tempurl%
\url{https://www.shanghai.gov.cn/nw15343/20220708/5d26125580294b1eb9db113926589535.html}
\showURL{%
\tempurl}


\bibitem[Government(2022b)]%
        {shanghai145}
\bibfield{author}{\bibinfo{person}{Shanghai Government}.} \bibinfo{year}{2022}\natexlab{b}.
\newblock \bibinfo{booktitle}{\emph{Outline of the Fourteenth Five-Year Plan and Long-Range Objectives Through the Year 2035 for Economic and Social Development of Shanghai Municipality}}.
\newblock
\urldef\tempurl%
\url{https://www.shanghai.gov.cn/2035nyjmbgy/index.html}
\showURL{%
\tempurl}


\bibitem[Government(2023a)]%
        {human1}
\bibfield{author}{\bibinfo{person}{Shanghai Government}.} \bibinfo{year}{2023}\natexlab{a}.
\newblock \bibinfo{booktitle}{\emph{Home Care Services}}.
\newblock
\urldef\tempurl%
\url{https://www.shanghai.gov.cn/jjylfw/index.html}
\showURL{%
\tempurl}


\bibitem[Government(2023b)]%
        {human2}
\bibfield{author}{\bibinfo{person}{Shanghai Government}.} \bibinfo{year}{2023}\natexlab{b}.
\newblock \bibinfo{booktitle}{\emph{Implementation Plan for Advancing the Construction of the Basic Elderly Care Service System}}.
\newblock
\urldef\tempurl%
\url{https://www.shanghai.gov.cn/nw12344/20230712/7bad958f61fb435c8417bd5e8361df92.html}
\showURL{%
\tempurl}


\bibitem[Government(2024)]%
        {olderpartner4}
\bibfield{author}{\bibinfo{person}{Shanghai~Baoshan Government}.} \bibinfo{year}{2024}\natexlab{}.
\newblock \bibinfo{booktitle}{\emph{The most beautiful volunteer of Shanghai Old Partner Initiative}}.
\newblock
\urldef\tempurl%
\url{http://www.shwmsj.gov.cn/bsq/2024/01/05/4a45dc28-54d4-40f4-bcf1-e10938fc5054.shtml}
\showURL{%
\tempurl}


\bibitem[Government(2021b)]%
        {laohuobanpolicy}
\bibfield{author}{\bibinfo{person}{Shanghai~Minhang Government}.} \bibinfo{year}{2021}\natexlab{b}.
\newblock \bibinfo{booktitle}{\emph{Policy interpretation materials on the Old Partner Program document}}.
\newblock
\urldef\tempurl%
\url{https://zwgk.shmh.gov.cn/mh-xxgk-cms/website/mh\_xxgk/xxgk\_mzj\_zhzw\_zcwj\_zcjd/content/3F802C5D-3697-11ED-A56E-B0FBDE767DFC.htm}
\showURL{%
\tempurl}


\bibitem[Government(2023c)]%
        {olderpartner1}
\bibfield{author}{\bibinfo{person}{Shanghai~Putuo Government}.} \bibinfo{year}{2023}\natexlab{c}.
\newblock \bibinfo{booktitle}{\emph{Older Partner Initiative}}.
\newblock
\urldef\tempurl%
\url{https://www.shpt.gov.cn/shpt/eslshsh2022-llgz/20230210/885482.html}
\showURL{%
\tempurl}


\bibitem[Government(2022c)]%
        {olderpartner3}
\bibfield{author}{\bibinfo{person}{Shanghai~Songjiang Government}.} \bibinfo{year}{2022}\natexlab{c}.
\newblock \bibinfo{booktitle}{\emph{'Notice on Implementing the 'Old Partners' Program Project for the Year 2023}}.
\newblock
\urldef\tempurl%
\url{https://www.songjiang.gov.cn/govxxgk/SHSJ11/2022-11-15/0290b7ed-1063-4c44-9e85-045f32136e20.html}
\showURL{%
\tempurl}


\bibitem[Greig et~al\mbox{.}(2019)]%
        {greig2019transforming}
\bibfield{author}{\bibinfo{person}{Jenni Greig}, \bibinfo{person}{Sabih-Ur Rehman}, \bibinfo{person}{Anwaar Ul-Haq}, \bibinfo{person}{Greg Dresser}, {and} \bibinfo{person}{Oliver~K Burmeister}.} \bibinfo{year}{2019}\natexlab{}.
\newblock \showarticletitle{Transforming Ageing in Community: addressing global ageing vulnerabilities through smart communities}. In \bibinfo{booktitle}{\emph{Proceedings of the 9th International Conference on Communities \& Technologies-Transforming Communities}}. \bibinfo{pages}{228--238}.
\newblock


\bibitem[Gui and Chen(2019)]%
        {gui2019making}
\bibfield{author}{\bibinfo{person}{Xinning Gui} {and} \bibinfo{person}{Yunan Chen}.} \bibinfo{year}{2019}\natexlab{}.
\newblock \showarticletitle{Making healthcare infrastructure work: Unpacking the infrastructuring work of individuals}. In \bibinfo{booktitle}{\emph{Proceedings of the 2019 CHI Conference on Human Factors in Computing Systems}}. \bibinfo{pages}{1--14}.
\newblock


\bibitem[Guo et~al\mbox{.}(2022)]%
        {guo2022caremap}
\bibfield{author}{\bibinfo{person}{Jiancong Guo}, \bibinfo{person}{Bin Wang}, \bibinfo{person}{Jin Zhao}, {and} \bibinfo{person}{Yuling Sun}.} \bibinfo{year}{2022}\natexlab{}.
\newblock \showarticletitle{CareMap: Human-Space-Service Based Healthcare Modeling and Quantifying for the Elderly Aging in Place}. In \bibinfo{booktitle}{\emph{2022 IEEE 25th International Conference on Computer Supported Cooperative Work in Design (CSCWD)}}. IEEE, \bibinfo{pages}{1317--1322}.
\newblock


\bibitem[Harvey and Thurnwald(2009)]%
        {harvey2009ageing}
\bibfield{author}{\bibinfo{person}{Peter~W Harvey} {and} \bibinfo{person}{Ian Thurnwald}.} \bibinfo{year}{2009}\natexlab{}.
\newblock \showarticletitle{Ageing well, ageing productively: The essential contribution of Australia's ageing population to the social and economic prosperity of the nation}.
\newblock \bibinfo{journal}{\emph{Health Sociology Review: The Journal of the Health Section of the Australian Sociological Association}} \bibinfo{volume}{18}, \bibinfo{number}{4} (\bibinfo{year}{2009}), \bibinfo{pages}{379--386}.
\newblock


\bibitem[Holb{\o} et~al\mbox{.}(2013)]%
        {holbo2013safe}
\bibfield{author}{\bibinfo{person}{Kristine Holb{\o}}, \bibinfo{person}{Silje B{\o}thun}, {and} \bibinfo{person}{Yngve Dahl}.} \bibinfo{year}{2013}\natexlab{}.
\newblock \showarticletitle{Safe walking technology for people with dementia: what do they want?}. In \bibinfo{booktitle}{\emph{Proceedings of the 15th international acm sigaccess conference on computers and accessibility}}. \bibinfo{pages}{1--8}.
\newblock


\bibitem[Jackson(2014)]%
        {jackson2014rethinking}
\bibfield{author}{\bibinfo{person}{Steven~J Jackson}.} \bibinfo{year}{2014}\natexlab{}.
\newblock \showarticletitle{Rethinking repair}.
\newblock  (\bibinfo{year}{2014}).
\newblock


\bibitem[Kaziunas et~al\mbox{.}(2019)]%
        {kaziunas2019precarious}
\bibfield{author}{\bibinfo{person}{Elizabeth Kaziunas}, \bibinfo{person}{Michael~S Klinkman}, {and} \bibinfo{person}{Mark~S Ackerman}.} \bibinfo{year}{2019}\natexlab{}.
\newblock \showarticletitle{Precarious interventions: Designing for ecologies of care}.
\newblock \bibinfo{journal}{\emph{Proceedings of the ACM on Human-Computer Interaction}} \bibinfo{volume}{3}, \bibinfo{number}{CSCW} (\bibinfo{year}{2019}), \bibinfo{pages}{1--27}.
\newblock


\bibitem[Kim et~al\mbox{.}(2017)]%
        {kim2017digital}
\bibfield{author}{\bibinfo{person}{Kwang-il Kim}, \bibinfo{person}{Shreya~S Gollamudi}, {and} \bibinfo{person}{Steven Steinhubl}.} \bibinfo{year}{2017}\natexlab{}.
\newblock \showarticletitle{Digital technology to enable aging in place}.
\newblock \bibinfo{journal}{\emph{Experimental gerontology}}  \bibinfo{volume}{88} (\bibinfo{year}{2017}), \bibinfo{pages}{25--31}.
\newblock


\bibitem[King et~al\mbox{.}(2021)]%
        {DemographicCareBurden}
\bibfield{author}{\bibinfo{person}{Elizabeth~M King}, \bibinfo{person}{Hannah~L Randolph}, \bibinfo{person}{Maria~S Floro}, {and} \bibinfo{person}{Jooyeoun Suh}.} \bibinfo{year}{2021}\natexlab{}.
\newblock \showarticletitle{Demographic, health, and economic transitions and the future care burden}.
\newblock \bibinfo{journal}{\emph{World Development}}  \bibinfo{volume}{140} (\bibinfo{year}{2021}), \bibinfo{pages}{105371}.
\newblock


\bibitem[Kropczynski et~al\mbox{.}(2021)]%
        {kropczynski2021towards}
\bibfield{author}{\bibinfo{person}{Jess Kropczynski}, \bibinfo{person}{Zaina Aljallad}, \bibinfo{person}{Nathan~Jeffrey Elrod}, \bibinfo{person}{Heather Lipford}, {and} \bibinfo{person}{Pamela~J Wisniewski}.} \bibinfo{year}{2021}\natexlab{}.
\newblock \showarticletitle{Towards building community collective efficacy for managing digital privacy and security within older adult communities}.
\newblock \bibinfo{journal}{\emph{Proceedings of the ACM on Human-Computer Interaction}} \bibinfo{volume}{4}, \bibinfo{number}{CSCW3} (\bibinfo{year}{2021}), \bibinfo{pages}{1--27}.
\newblock
\urldef\tempurl%
\url{https://doi.org/10.1145/3432954}
\showURL{%
\tempurl}


\bibitem[Kumar et~al\mbox{.}(2017)]%
        {kumar2017hci}
\bibfield{author}{\bibinfo{person}{Neha Kumar}, \bibinfo{person}{Susan~M Dray}, \bibinfo{person}{Christian Sturm}, \bibinfo{person}{Nithya Sambasivan}, \bibinfo{person}{Laura~S Gayt{\'a}n-Lugo}, \bibinfo{person}{Leonel~V Morales~Diaz}, \bibinfo{person}{Negin Dahya}, {and} \bibinfo{person}{Nova Ahmed}.} \bibinfo{year}{2017}\natexlab{}.
\newblock \showarticletitle{HCI Across Borders}. In \bibinfo{booktitle}{\emph{Proceedings of the 2017 CHI Conference Extended Abstracts on Human Factors in Computing Systems}}. \bibinfo{pages}{3065--3072}.
\newblock


\bibitem[Lancet(2022)]%
        {lancet2022population}
\bibfield{author}{\bibinfo{person}{The Lancet}.} \bibinfo{year}{2022}\natexlab{}.
\newblock \bibinfo{title}{Population ageing in China: crisis or opportunity?}
\newblock \bibinfo{numpages}{1821}~pages.
\newblock


\bibitem[Larkin(2013)]%
        {larkin2013politics}
\bibfield{author}{\bibinfo{person}{Brian Larkin}.} \bibinfo{year}{2013}\natexlab{}.
\newblock \showarticletitle{The politics and poetics of infrastructure}.
\newblock \bibinfo{journal}{\emph{Annual review of anthropology}} \bibinfo{volume}{42}, \bibinfo{number}{1} (\bibinfo{year}{2013}), \bibinfo{pages}{327--343}.
\newblock


\bibitem[Lau et~al\mbox{.}(2007)]%
        {lau2007health}
\bibfield{author}{\bibinfo{person}{Denys~T Lau}, \bibinfo{person}{Karen~Glasser Scandrett}, \bibinfo{person}{Mary Jarzebowski}, \bibinfo{person}{Kami Holman}, {and} \bibinfo{person}{Linda Emanuel}.} \bibinfo{year}{2007}\natexlab{}.
\newblock \showarticletitle{Health-related safety: a framework to address barriers to aging in place}.
\newblock \bibinfo{journal}{\emph{The Gerontologist}} \bibinfo{volume}{47}, \bibinfo{number}{6} (\bibinfo{year}{2007}), \bibinfo{pages}{830--837}.
\newblock


\bibitem[Lazar et~al\mbox{.}(2018)]%
        {Lazar2018}
\bibfield{author}{\bibinfo{person}{Amanda Lazar}, \bibinfo{person}{Hilaire~J Thompson}, \bibinfo{person}{Shih-Yin Lin}, {and} \bibinfo{person}{George Demiris}.} \bibinfo{year}{2018}\natexlab{}.
\newblock \showarticletitle{Negotiating relation work with telehealth home care companionship technologies that support aging in place}.
\newblock \bibinfo{journal}{\emph{Proceedings of the ACM on Human-Computer Interaction}} \bibinfo{volume}{2}, \bibinfo{number}{CSCW} (\bibinfo{year}{2018}), \bibinfo{pages}{1--19}.
\newblock
\urldef\tempurl%
\url{https://dl.acm.org/doi/10.1145/3274372}
\showURL{%
\tempurl}


\bibitem[Legislature(2022)]%
        {Services2022}
\bibfield{author}{\bibinfo{person}{Michigan Legislature}.} \bibinfo{year}{2022}\natexlab{}.
\newblock \showarticletitle{Services for Seniors: Laws \& Programs for Senior Adults}.
\newblock  (\bibinfo{year}{2022}).
\newblock


\bibitem[Lehning(2012)]%
        {lehning2012city}
\bibfield{author}{\bibinfo{person}{Amanda~J Lehning}.} \bibinfo{year}{2012}\natexlab{}.
\newblock \showarticletitle{City governments and aging in place: Community design, transportation and housing innovation adoption}.
\newblock \bibinfo{journal}{\emph{The Gerontologist}} \bibinfo{volume}{52}, \bibinfo{number}{3} (\bibinfo{year}{2012}), \bibinfo{pages}{345--356}.
\newblock


\bibitem[Li et~al\mbox{.}(2023b)]%
        {li2023dynamic}
\bibfield{author}{\bibinfo{person}{Bangyan Li}, \bibinfo{person}{Junyan Mao}, \bibinfo{person}{Xingjiao Wu}, \bibinfo{person}{Yuling Sun}, {and} \bibinfo{person}{Liang He}.} \bibinfo{year}{2023}\natexlab{b}.
\newblock \showarticletitle{A Dynamic Composite Ensemble Learning Framework for Multi-Stage Dementia Prediction}. In \bibinfo{booktitle}{\emph{2023 IEEE International Conference on Bioinformatics and Biomedicine (BIBM)}}. IEEE, \bibinfo{pages}{867--872}.
\newblock


\bibitem[Li(2023)]%
        {shanghai9073}
\bibfield{author}{\bibinfo{person}{Jing Li}.} \bibinfo{year}{2023}\natexlab{}.
\newblock \bibinfo{booktitle}{\emph{Home-based elderly care remains the most crucial elderly care model in China}}.
\newblock
\urldef\tempurl%
\url{http://www.crca.cn/index.php/16-research/984-2023-11-28-03-25-03.html}
\showURL{%
\tempurl}


\bibitem[Li et~al\mbox{.}(2023a)]%
        {li2023any}
\bibfield{author}{\bibinfo{person}{Lin Li}, \bibinfo{person}{Vitica Arnold}, {and} \bibinfo{person}{Anne~Marie Piper}.} \bibinfo{year}{2023}\natexlab{a}.
\newblock \showarticletitle{“Any bit of help, helps”: Understanding how older caregivers use carework platforms for caregiving support}. In \bibinfo{booktitle}{\emph{Proceedings of the 2023 CHI Conference on Human Factors in Computing Systems}}. \bibinfo{pages}{1--17}.
\newblock


\bibitem[Lindley et~al\mbox{.}(2008)]%
        {lindley2008designing}
\bibfield{author}{\bibinfo{person}{Si{\^a}n~E Lindley}, \bibinfo{person}{Richard Harper}, {and} \bibinfo{person}{Abigail Sellen}.} \bibinfo{year}{2008}\natexlab{}.
\newblock \showarticletitle{Designing for elders: exploring the complexity of relationships in later life}.
\newblock \bibinfo{journal}{\emph{People and Computers XXII Culture, Creativity, Interaction 22}} (\bibinfo{year}{2008}), \bibinfo{pages}{77--86}.
\newblock


\bibitem[Lu et~al\mbox{.}(2023)]%
        {lu2023shifting}
\bibfield{author}{\bibinfo{person}{Alex~Jiahong Lu}, \bibinfo{person}{Shruti Sannon}, \bibinfo{person}{Cameron Moy}, \bibinfo{person}{Savana Brewer}, \bibinfo{person}{Jaye Green}, \bibinfo{person}{Kisha~N Jackson}, \bibinfo{person}{Daivon Reeder}, \bibinfo{person}{Camaria Wafer}, \bibinfo{person}{Mark~S Ackerman}, {and} \bibinfo{person}{Tawanna~R Dillahunt}.} \bibinfo{year}{2023}\natexlab{}.
\newblock \showarticletitle{Shifting from Surveillance-as-Safety to Safety-through-Noticing: A Photovoice Study with Eastside Detroit Residents}. In \bibinfo{booktitle}{\emph{Proceedings of the 2023 CHI Conference on Human Factors in Computing Systems}}. \bibinfo{pages}{1--19}.
\newblock


\bibitem[Maciuszek et~al\mbox{.}(2005)]%
        {maciuszek2005help}
\bibfield{author}{\bibinfo{person}{Dennis Maciuszek}, \bibinfo{person}{Johan Aberg}, {and} \bibinfo{person}{Nahid Shahmehri}.} \bibinfo{year}{2005}\natexlab{}.
\newblock \showarticletitle{What help do older people need? Constructing a functional design space of electronic assistive technology applications}. In \bibinfo{booktitle}{\emph{Proceedings of the 7th international ACM SIGACCESS conference on Computers and accessibility}}. \bibinfo{pages}{4--11}.
\newblock


\bibitem[Manda and Backhouse(2018)]%
        {manda2018inclusive}
\bibfield{author}{\bibinfo{person}{More~Ickson Manda} {and} \bibinfo{person}{Judy Backhouse}.} \bibinfo{year}{2018}\natexlab{}.
\newblock \showarticletitle{Inclusive digital transformation in South Africa: An institutional perspective}. In \bibinfo{booktitle}{\emph{Proceedings of the 11th International Conference on Theory and Practice of Electronic Governance}}. \bibinfo{pages}{464--470}.
\newblock


\bibitem[Mattern(2018)]%
        {mattern2018maintenance}
\bibfield{author}{\bibinfo{person}{Shannon Mattern}.} \bibinfo{year}{2018}\natexlab{}.
\newblock \showarticletitle{Maintenance and care}.
\newblock \bibinfo{journal}{\emph{Places Journal}} (\bibinfo{year}{2018}).
\newblock


\bibitem[Mezuk et~al\mbox{.}(2015)]%
        {mezuk2015suicide}
\bibfield{author}{\bibinfo{person}{Briana Mezuk}, \bibinfo{person}{Matthew Lohman}, \bibinfo{person}{Marc Leslie}, {and} \bibinfo{person}{Virginia Powell}.} \bibinfo{year}{2015}\natexlab{}.
\newblock \showarticletitle{Suicide risk in nursing homes and assisted living facilities: 2003--2011}.
\newblock \bibinfo{journal}{\emph{American journal of public health}} \bibinfo{volume}{105}, \bibinfo{number}{7} (\bibinfo{year}{2015}), \bibinfo{pages}{1495--1502}.
\newblock


\bibitem[Mosconi et~al\mbox{.}(2017)]%
        {mosconi2017facebook}
\bibfield{author}{\bibinfo{person}{Gaia Mosconi}, \bibinfo{person}{Matthias Korn}, \bibinfo{person}{Christian Reuter}, \bibinfo{person}{Peter Tolmie}, \bibinfo{person}{Maurizio Teli}, {and} \bibinfo{person}{Volkmar Pipek}.} \bibinfo{year}{2017}\natexlab{}.
\newblock \showarticletitle{From facebook to the neighbourhood: Infrastructuring of hybrid community engagement}.
\newblock \bibinfo{journal}{\emph{Computer Supported Cooperative Work (CSCW)}}  \bibinfo{volume}{26} (\bibinfo{year}{2017}), \bibinfo{pages}{959--1003}.
\newblock


\bibitem[Murthy et~al\mbox{.}(2021)]%
        {murthy2021individually}
\bibfield{author}{\bibinfo{person}{Savanthi Murthy}, \bibinfo{person}{Karthik~S Bhat}, \bibinfo{person}{Sauvik Das}, {and} \bibinfo{person}{Neha Kumar}.} \bibinfo{year}{2021}\natexlab{}.
\newblock \showarticletitle{Individually vulnerable, collectively safe: The security and privacy practices of households with older adults}.
\newblock \bibinfo{journal}{\emph{Proceedings of the ACM on Human-Computer Interaction}} \bibinfo{volume}{5}, \bibinfo{number}{CSCW1} (\bibinfo{year}{2021}), \bibinfo{pages}{1--24}.
\newblock


\bibitem[Nations(2019)]%
        {agingdefination}
\bibfield{author}{\bibinfo{person}{United Nations}.} \bibinfo{year}{2019}\natexlab{}.
\newblock \showarticletitle{World Population Ageing 2019}.
\newblock
\urldef\tempurl%
\url{https://www.un.org/en/development/desa/population/publications/pdf/ageing/WorldPopulationAgeing2019-Highlights.pdf}
\showURL{%
\tempurl}


\bibitem[News(2023a)]%
        {4tech2}
\bibfield{author}{\bibinfo{person}{Jiefang~Daily News}.} \bibinfo{year}{2023}\natexlab{a}.
\newblock \bibinfo{booktitle}{\emph{Virtual Nursing Home: Intelligent Devices Promptly 'Alarm' in Case of Unexpected Incidents}}.
\newblock
\urldef\tempurl%
\url{http://news.cjn.cn/zjjjdpd/yw\_20048/202203/t3975322.htm}
\showURL{%
\tempurl}


\bibitem[News(2022)]%
        {contribution}
\bibfield{author}{\bibinfo{person}{People News}.} \bibinfo{year}{2022}\natexlab{}.
\newblock \bibinfo{booktitle}{\emph{Promoting the spirit of dedication, friendship, mutual assistance, and progress}}.
\newblock
\urldef\tempurl%
\url{http://politics.people.com.cn/n1/2022/1205/c1001-32581063.html}
\showURL{%
\tempurl}


\bibitem[News(2023b)]%
        {4tech1}
\bibfield{author}{\bibinfo{person}{Sohu News}.} \bibinfo{year}{2023}\natexlab{b}.
\newblock \bibinfo{booktitle}{\emph{Smart Water Meters Enhance Safety for Solitary Elderly in Shanghai: Introduction of the 'Smart Four-Piece Set}}.
\newblock
\urldef\tempurl%
\url{https://www.sohu.com/a/697162771_100140408}
\showURL{%
\tempurl}


\bibitem[News(2020)]%
        {shanghaiaging}
\bibfield{author}{\bibinfo{person}{Xinhua News}.} \bibinfo{year}{2020}\natexlab{}.
\newblock \bibinfo{booktitle}{\emph{The proportion of elderly residents with Shanghai exceeds 35\%, further deepening the aging population}}.
\newblock
\urldef\tempurl%
\url{http://www.xinhuanet.com/politics/2020-05/24/c\_1126025478.htm}
\showURL{%
\tempurl}


\bibitem[Ni et~al\mbox{.}(2019)]%
        {ni2019human}
\bibfield{author}{\bibinfo{person}{Liuqian Ni}, \bibinfo{person}{Yuling Sun}, \bibinfo{person}{Yanqin Yang}, {and} \bibinfo{person}{Liang He}.} \bibinfo{year}{2019}\natexlab{}.
\newblock \showarticletitle{Human-Engaged Health Care Services Recommendation for Aging and Long-term Care}. In \bibinfo{booktitle}{\emph{2019 IEEE 23rd International Conference on Computer Supported Cooperative Work in Design (CSCWD)}}. IEEE, \bibinfo{pages}{339--344}.
\newblock


\bibitem[Nurain et~al\mbox{.}(2021)]%
        {nurain2021hugging}
\bibfield{author}{\bibinfo{person}{Novia Nurain}, \bibinfo{person}{Chia-Fang Chung}, \bibinfo{person}{Clara Caldeira}, {and} \bibinfo{person}{Kay Connelly}.} \bibinfo{year}{2021}\natexlab{}.
\newblock \showarticletitle{Hugging with a Shower Curtain: Older Adults' Social Support Realities During the COVID-19 Pandemic}.
\newblock \bibinfo{journal}{\emph{Proceedings of the ACM on Human-Computer Interaction}} \bibinfo{volume}{5}, \bibinfo{number}{CSCW2} (\bibinfo{year}{2021}), \bibinfo{pages}{1--31}.
\newblock


\bibitem[on~Aging(2023)]%
        {aging}
\bibfield{author}{\bibinfo{person}{National~Institute on Aging}.} \bibinfo{year}{2023}\natexlab{}.
\newblock \bibinfo{booktitle}{\emph{Aging in Place: Growing Older at Home}}.
\newblock
\urldef\tempurl%
\url{https://www.nia.nih.gov/health/aging-place/aging-place-growing-older-home}
\showURL{%
\tempurl}


\bibitem[on~Healthy~Aging(2024)]%
        {national2024most}
\bibfield{author}{\bibinfo{person}{National~Poll on Healthy~Aging}.} \bibinfo{year}{2024}\natexlab{}.
\newblock \bibinfo{booktitle}{\emph{Most older adults don’t know about resources that can help them navigate aging \& caregiving}}.
\newblock
\urldef\tempurl%
\url{https://ihpi.umich.edu/news/most-older-adults-dont-know-about-resources-can-help-them-navigate-aging-caregiving}
\showURL{%
\tempurl}


\bibitem[Oogjes et~al\mbox{.}(2018)]%
        {oogjes2018designing}
\bibfield{author}{\bibinfo{person}{Doenja Oogjes}, \bibinfo{person}{William Odom}, {and} \bibinfo{person}{Pete Fung}.} \bibinfo{year}{2018}\natexlab{}.
\newblock \showarticletitle{Designing for an other home: Expanding and speculating on different forms of domestic life}. In \bibinfo{booktitle}{\emph{Proceedings of the 2018 Designing Interactive Systems Conference}}. \bibinfo{pages}{313--326}.
\newblock


\bibitem[Organization({[n.\,d.]})]%
        {ChinaAging}
\bibfield{author}{\bibinfo{person}{World~Health Organization}.} \bibinfo{year}{[n.\,d.]}\natexlab{}.
\newblock \bibinfo{booktitle}{\emph{Ageing and health in China}}.
\newblock
\urldef\tempurl%
\url{https://www.who.int/china/health-topics/ageing}
\showURL{%
\tempurl}


\bibitem[Organization(2017)]%
        {world2017}
\bibfield{author}{\bibinfo{person}{World~Health Organization}.} \bibinfo{year}{2017}\natexlab{}.
\newblock \bibinfo{booktitle}{\emph{Global Strategy and Action Plan on Ageing and Health}}.
\newblock \bibinfo{publisher}{World Health Organization}.
\newblock
\urldef\tempurl%
\url{https://iris.who.int/handle/10665/329960}
\showURL{%
\tempurl}


\bibitem[Peek et~al\mbox{.}(2016)]%
        {peek2016older}
\bibfield{author}{\bibinfo{person}{Sebastiaan~TM Peek}, \bibinfo{person}{Katrien~G Luijkx}, \bibinfo{person}{Maurice~D Rijnaard}, \bibinfo{person}{Marianne~E Nieboer}, \bibinfo{person}{Claire~S Van Der~Voort}, \bibinfo{person}{Sil Aarts}, \bibinfo{person}{Joost Van~Hoof}, \bibinfo{person}{Hubertus~JM Vrijhoef}, {and} \bibinfo{person}{Eveline~JM Wouters}.} \bibinfo{year}{2016}\natexlab{}.
\newblock \showarticletitle{Older adults' reasons for using technology while aging in place}.
\newblock \bibinfo{journal}{\emph{Gerontology}} \bibinfo{volume}{62}, \bibinfo{number}{2} (\bibinfo{year}{2016}), \bibinfo{pages}{226--237}.
\newblock


\bibitem[Petrakaki et~al\mbox{.}(2018)]%
        {petrakaki2018between}
\bibfield{author}{\bibinfo{person}{Dimitra Petrakaki}, \bibinfo{person}{Eva Hilberg}, {and} \bibinfo{person}{Justin Waring}.} \bibinfo{year}{2018}\natexlab{}.
\newblock \showarticletitle{Between empowerment and self-discipline: governing patients' conduct through technological self-care}.
\newblock \bibinfo{journal}{\emph{Social Science \& Medicine}}  \bibinfo{volume}{213} (\bibinfo{year}{2018}), \bibinfo{pages}{146--153}.
\newblock


\bibitem[Pine(2012)]%
        {pine2012fragmentation}
\bibfield{author}{\bibinfo{person}{Katie Pine}.} \bibinfo{year}{2012}\natexlab{}.
\newblock \showarticletitle{Fragmentation and choreography: caring for a patient and a chart during childbirth}. In \bibinfo{booktitle}{\emph{Proceedings of the ACM 2012 conference on computer supported cooperative work}}. \bibinfo{pages}{887--896}.
\newblock


\bibitem[Piper et~al\mbox{.}(2016)]%
        {piper2016technological}
\bibfield{author}{\bibinfo{person}{Anne~Marie Piper}, \bibinfo{person}{Raymundo Cornejo}, \bibinfo{person}{Lisa Hurwitz}, {and} \bibinfo{person}{Caitlin Unumb}.} \bibinfo{year}{2016}\natexlab{}.
\newblock \showarticletitle{Technological caregiving: Supporting online activity for adults with cognitive impairments}. In \bibinfo{booktitle}{\emph{Proceedings of the 2016 chi conference on human factors in computing systems}}. \bibinfo{pages}{5311--5323}.
\newblock


\bibitem[Powell(2006)]%
        {powell2006social}
\bibfield{author}{\bibinfo{person}{Jason~L Powell}.} \bibinfo{year}{2006}\natexlab{}.
\newblock \bibinfo{booktitle}{\emph{Social theory and aging}}.
\newblock \bibinfo{publisher}{Rowman \& Littlefield}.
\newblock


\bibitem[Ribes and Lee(2010)]%
        {ribes2010sociotechnical}
\bibfield{author}{\bibinfo{person}{David Ribes} {and} \bibinfo{person}{Charlotte~P Lee}.} \bibinfo{year}{2010}\natexlab{}.
\newblock \showarticletitle{Sociotechnical studies of cyberinfrastructure and e-research: Current themes and future trajectories}.
\newblock \bibinfo{journal}{\emph{Computer Supported Cooperative Work (CSCW)}}  \bibinfo{volume}{19} (\bibinfo{year}{2010}), \bibinfo{pages}{231--244}.
\newblock


\bibitem[Riger and Sigurvinsdottir(2016)]%
        {riger2016thematic}
\bibfield{author}{\bibinfo{person}{Stephanie Riger} {and} \bibinfo{person}{Rannveig Sigurvinsdottir}.} \bibinfo{year}{2016}\natexlab{}.
\newblock \showarticletitle{Thematic analysis}.
\newblock \bibinfo{journal}{\emph{Handbook of methodological approaches to community-based research: Qualitative, quantitative, and mixed methods}} (\bibinfo{year}{2016}), \bibinfo{pages}{33--41}.
\newblock


\bibitem[Robinson-Lane et~al\mbox{.}(2022)]%
        {robinson2022older}
\bibfield{author}{\bibinfo{person}{Sheria Robinson-Lane}, \bibinfo{person}{Erica Solway}, \bibinfo{person}{Dianne Singer}, \bibinfo{person}{Matthias Kirch}, \bibinfo{person}{Jeffrey Kullgren}, {and} \bibinfo{person}{Preeti Malani}.} \bibinfo{year}{2022}\natexlab{}.
\newblock \showarticletitle{OLDER ADULTS'PREPAREDNESS TO AGE IN PLACE: FINDINGS FROM THE NATIONAL POLL ON HEALTHY AGING}.
\newblock \bibinfo{journal}{\emph{Innovation in Aging}} \bibinfo{volume}{6}, \bibinfo{number}{Supplement\_1} (\bibinfo{year}{2022}), \bibinfo{pages}{227--227}.
\newblock


\bibitem[Rosen(2014)]%
        {Rosen2014}
\bibfield{author}{\bibinfo{person}{Stanley Rosen}.} \bibinfo{year}{2014}\natexlab{}.
\newblock \showarticletitle{Michel Bonnin. The Lost Generation: The Rustication of China's Educated Youth (1968–1980).}
\newblock  \bibinfo{volume}{119}, \bibinfo{number}{5} (\bibinfo{year}{2014}), \bibinfo{pages}{1676}.
\newblock
\showISSN{0002-8762}
\href{https://doi.org/10.1093/ahr/119.5.1676}{doi:\nolinkurl{10.1093/ahr/119.5.1676}}


\bibitem[Schorr and Khalaila(2018)]%
        {vitmanschorrAgingPlaceQuality}
\bibfield{author}{\bibinfo{person}{Adi~Vitman Schorr} {and} \bibinfo{person}{Rabia Khalaila}.} \bibinfo{year}{2018}\natexlab{}.
\newblock \showarticletitle{Aging in place and quality of life among the elderly in Europe: A moderated mediation model}.
\newblock \bibinfo{journal}{\emph{Archives of gerontology and geriatrics}}  \bibinfo{volume}{77} (\bibinfo{year}{2018}), \bibinfo{pages}{196--204}.
\newblock


\bibitem[Semke(2003)]%
        {semke2003older}
\bibfield{author}{\bibinfo{person}{Jeanette Semke}.} \bibinfo{year}{2003}\natexlab{}.
\newblock \showarticletitle{Older adults with dementia: Community based long term care alternatives}.
\newblock \bibinfo{journal}{\emph{Social work and health care in an aging society}} (\bibinfo{year}{2003}), \bibinfo{pages}{49--72}.
\newblock


\bibitem[Shen and Sun(2023)]%
        {shen2023privacy}
\bibfield{author}{\bibinfo{person}{Jifan Shen} {and} \bibinfo{person}{Yuling Sun}.} \bibinfo{year}{2023}\natexlab{}.
\newblock \showarticletitle{Privacy-preserved video monitoring method with 3d human pose estimation}. In \bibinfo{booktitle}{\emph{2023 26th International Conference on Computer Supported Cooperative Work in Design (CSCWD)}}. IEEE, \bibinfo{pages}{1502--1507}.
\newblock


\bibitem[Shenggao(2022)]%
        {shanghaidigital}
\bibfield{author}{\bibinfo{person}{Yuan Shenggao}.} \bibinfo{year}{2022}\natexlab{}.
\newblock \bibinfo{booktitle}{\emph{Shanghai takes aim to become global capital for digitization}}.
\newblock
\urldef\tempurl%
\url{https://epaper.chinadaily.com.cn/a/202211/10/WS636c31b4a310407c0b52cb1d.html}
\showURL{%
\tempurl}


\bibitem[Simone(2004)]%
        {simone2004people}
\bibfield{author}{\bibinfo{person}{AbdouMaliq Simone}.} \bibinfo{year}{2004}\natexlab{}.
\newblock \showarticletitle{People as infrastructure: Intersecting fragments in Johannesburg}.
\newblock \bibinfo{journal}{\emph{Public culture}} \bibinfo{volume}{16}, \bibinfo{number}{3} (\bibinfo{year}{2004}), \bibinfo{pages}{407--429}.
\newblock


\bibitem[Simone(2021)]%
        {simone2021ritornello}
\bibfield{author}{\bibinfo{person}{AbdouMaliq Simone}.} \bibinfo{year}{2021}\natexlab{}.
\newblock \showarticletitle{Ritornello:“People as infrastructure”}.
\newblock \bibinfo{journal}{\emph{Urban Geography}} \bibinfo{volume}{42}, \bibinfo{number}{9} (\bibinfo{year}{2021}), \bibinfo{pages}{1341--1348}.
\newblock


\bibitem[Soubutts et~al\mbox{.}(2021)]%
        {soubutts2021aging}
\bibfield{author}{\bibinfo{person}{Ewan Soubutts}, \bibinfo{person}{Amid Ayobi}, \bibinfo{person}{Rachel Eardley}, \bibinfo{person}{Kirsten Cater}, {and} \bibinfo{person}{Aisling~Ann O'Kane}.} \bibinfo{year}{2021}\natexlab{}.
\newblock \showarticletitle{Aging in place together: the journey towards adoption and acceptance of stairlifts in multi-resident homes}.
\newblock \bibinfo{journal}{\emph{Proceedings of the ACM on Human-Computer Interaction}} \bibinfo{volume}{5}, \bibinfo{number}{CSCW2} (\bibinfo{year}{2021}), \bibinfo{pages}{1--26}.
\newblock


\bibitem[Star and Griesemer(1989)]%
        {star1989institutional}
\bibfield{author}{\bibinfo{person}{Susan~Leigh Star} {and} \bibinfo{person}{James~R Griesemer}.} \bibinfo{year}{1989}\natexlab{}.
\newblock \showarticletitle{Institutional ecology,translations' and boundary objects: Amateurs and professionals in Berkeley's Museum of Vertebrate Zoology, 1907-39}.
\newblock \bibinfo{journal}{\emph{Social studies of science}} \bibinfo{volume}{19}, \bibinfo{number}{3} (\bibinfo{year}{1989}), \bibinfo{pages}{387--420}.
\newblock


\bibitem[Star and Ruhleder(1994)]%
        {star1994steps}
\bibfield{author}{\bibinfo{person}{Susan~Leigh Star} {and} \bibinfo{person}{Karen Ruhleder}.} \bibinfo{year}{1994}\natexlab{}.
\newblock \showarticletitle{Steps towards an ecology of infrastructure: complex problems in design and access for large-scale collaborative systems}. In \bibinfo{booktitle}{\emph{Proceedings of the 1994 ACM conference on Computer supported cooperative work}}. \bibinfo{pages}{253--264}.
\newblock


\bibitem[Stones and Gullifer(2016)]%
        {stones2016home}
\bibfield{author}{\bibinfo{person}{Damien Stones} {and} \bibinfo{person}{Judith Gullifer}.} \bibinfo{year}{2016}\natexlab{}.
\newblock \showarticletitle{‘At home it's just so much easier to be yourself’: older adults' perceptions of ageing in place}.
\newblock \bibinfo{journal}{\emph{Ageing \& Society}} \bibinfo{volume}{36}, \bibinfo{number}{3} (\bibinfo{year}{2016}), \bibinfo{pages}{449--481}.
\newblock


\bibitem[Strauss and Corbin(1990)]%
        {strauss1990basics}
\bibfield{author}{\bibinfo{person}{Anselm Strauss} {and} \bibinfo{person}{Juliet Corbin}.} \bibinfo{year}{1990}\natexlab{}.
\newblock \bibinfo{booktitle}{\emph{Basics of qualitative research}}.
\newblock \bibinfo{publisher}{Sage publications}.
\newblock


\bibitem[Suchman(1987)]%
        {suchman1987plans}
\bibfield{author}{\bibinfo{person}{Lucille~Alice Suchman}.} \bibinfo{year}{1987}\natexlab{}.
\newblock \bibinfo{booktitle}{\emph{Plans and situated actions: The problem of human-machine communication}}.
\newblock \bibinfo{publisher}{Cambridge university press}.
\newblock


\bibitem[Sun et~al\mbox{.}(2014)]%
        {sun2014being}
\bibfield{author}{\bibinfo{person}{Yuling Sun}, \bibinfo{person}{Xianghua Ding}, \bibinfo{person}{Silvia Lindtner}, \bibinfo{person}{Tun Lu}, {and} \bibinfo{person}{Ning Gu}.} \bibinfo{year}{2014}\natexlab{}.
\newblock \showarticletitle{Being senior and ICT: A study of seniors using ICT in China}. In \bibinfo{booktitle}{\emph{Proceedings of the SIGCHI Conference on Human factors in Computing Systems}}. \bibinfo{pages}{3933--3942}.
\newblock
\urldef\tempurl%
\url{https://doi.org/10.1145/2556288.2557248}
\showURL{%
\tempurl}


\bibitem[Sun et~al\mbox{.}(2015)]%
        {sun2015reliving}
\bibfield{author}{\bibinfo{person}{Yuling Sun}, \bibinfo{person}{Silvia Lindtner}, \bibinfo{person}{Xianghua Ding}, \bibinfo{person}{Tun Lu}, {and} \bibinfo{person}{Ning Gu}.} \bibinfo{year}{2015}\natexlab{}.
\newblock \showarticletitle{Reliving the past \& making a harmonious society today: A study of elderly electronic hackers in China}. In \bibinfo{booktitle}{\emph{Proceedings of the 18th ACM Conference on Computer Supported Cooperative Work \& Social Computing}}. \bibinfo{pages}{44--55}.
\newblock
\urldef\tempurl%
\url{https://doi-org.proxy.lib.umich.edu/10.1145/2675133.2675195}
\showURL{%
\tempurl}


\bibitem[Sun et~al\mbox{.}(2017)]%
        {sun2017method}
\bibfield{author}{\bibinfo{person}{Yuling Sun}, \bibinfo{person}{Tun Lu}, {and} \bibinfo{person}{Ning Gu}.} \bibinfo{year}{2017}\natexlab{}.
\newblock \showarticletitle{A method of electronic health data quality assessment: Enabling data provenance}. In \bibinfo{booktitle}{\emph{2017 IEEE 21st International Conference on Computer Supported Cooperative Work in Design (CSCWD)}}. IEEE, \bibinfo{pages}{233--238}.
\newblock


\bibitem[Sun et~al\mbox{.}(2023a)]%
        {sun2023care}
\bibfield{author}{\bibinfo{person}{Yuling Sun}, \bibinfo{person}{Xiaojuan Ma}, \bibinfo{person}{Silvia Lindtner}, {and} \bibinfo{person}{Liang He}.} \bibinfo{year}{2023}\natexlab{a}.
\newblock \showarticletitle{Care Workers' Wellbeing in Data-Driven Healthcare Workplace: Identity, Agency, and Social Justice}.
\newblock \bibinfo{journal}{\emph{Proceedings of the ACM on Human-Computer Interaction}} \bibinfo{volume}{7}, \bibinfo{number}{CSCW2} (\bibinfo{year}{2023}), \bibinfo{pages}{1--29}.
\newblock


\bibitem[Sun et~al\mbox{.}(2023b)]%
        {sun2023data}
\bibfield{author}{\bibinfo{person}{Yuling Sun}, \bibinfo{person}{Xiaojuan Ma}, \bibinfo{person}{Silvia Lindtner}, {and} \bibinfo{person}{Liang He}.} \bibinfo{year}{2023}\natexlab{b}.
\newblock \showarticletitle{Data Work of Frontline Care Workers: Practices, Problems, and Opportunities in the Context of Data-Driven Long-Term Care}.
\newblock \bibinfo{journal}{\emph{Proceedings of the ACM on Human-Computer Interaction}} \bibinfo{volume}{7}, \bibinfo{number}{CSCW1} (\bibinfo{year}{2023}), \bibinfo{pages}{1--28}.
\newblock


\bibitem[Taylor et~al\mbox{.}(2013)]%
        {taylor2013ethnography}
\bibfield{author}{\bibinfo{person}{TL Taylor}, \bibinfo{person}{Tom Boellstorff}, \bibinfo{person}{Bonnie Nardi}, {and} \bibinfo{person}{Celia Pearce}.} \bibinfo{year}{2013}\natexlab{}.
\newblock \bibinfo{booktitle}{\emph{Ethnography and virtual worlds: A handbook of method}}.
\newblock \bibinfo{publisher}{Princeton university press}.
\newblock


\bibitem[Thach et~al\mbox{.}(2023)]%
        {thach2023key}
\bibfield{author}{\bibinfo{person}{Kong~Saoane Thach}, \bibinfo{person}{Reeva Lederman}, {and} \bibinfo{person}{Jenny Waycott}.} \bibinfo{year}{2023}\natexlab{}.
\newblock \showarticletitle{Key Considerations for The Design of Technology for Enrichment in Residential Aged Care: An Ethnographic Study}. In \bibinfo{booktitle}{\emph{Proceedings of the 2023 CHI Conference on Human Factors in Computing Systems}}. \bibinfo{pages}{1--16}.
\newblock


\bibitem[Thomas et~al\mbox{.}(2020)]%
        {thomas2020s}
\bibfield{author}{\bibinfo{person}{Kali~S Thomas}, \bibinfo{person}{Emily~A Gadbois}, \bibinfo{person}{Renee~R Shield}, \bibinfo{person}{Ucheoma Akobundu}, \bibinfo{person}{Andrea~M Morris}, {and} \bibinfo{person}{David~M Dosa}.} \bibinfo{year}{2020}\natexlab{}.
\newblock \showarticletitle{“It’s not just a simple meal. It’s so much more”: Interactions between meals on wheels clients and drivers}.
\newblock \bibinfo{journal}{\emph{Journal of Applied Gerontology}} \bibinfo{volume}{39}, \bibinfo{number}{2} (\bibinfo{year}{2020}), \bibinfo{pages}{151--158}.
\newblock


\bibitem[Thomas and Blanchard(2009)]%
        {thomas2009moving}
\bibfield{author}{\bibinfo{person}{William Thomas} {and} \bibinfo{person}{Janice Blanchard}.} \bibinfo{year}{2009}\natexlab{}.
\newblock \showarticletitle{Moving beyond place: Aging in community}.
\newblock \bibinfo{journal}{\emph{Generations}} \bibinfo{volume}{33}, \bibinfo{number}{2} (\bibinfo{year}{2009}), \bibinfo{pages}{12--17}.
\newblock


\bibitem[Toombs et~al\mbox{.}(2018)]%
        {toombs2018sociotechnical}
\bibfield{author}{\bibinfo{person}{Austin Toombs}, \bibinfo{person}{Laura Devendorf}, \bibinfo{person}{Patrick Shih}, \bibinfo{person}{Elizabeth Kaziunas}, \bibinfo{person}{David Nemer}, \bibinfo{person}{Helena Mentis}, {and} \bibinfo{person}{Laura Forlano}.} \bibinfo{year}{2018}\natexlab{}.
\newblock \showarticletitle{Sociotechnical systems of care}. In \bibinfo{booktitle}{\emph{Companion of the 2018 ACM conference on computer supported cooperative work and social computing}}. \bibinfo{pages}{479--485}.
\newblock


\bibitem[Trothen(2022)]%
        {trothen2022intelligent}
\bibfield{author}{\bibinfo{person}{Tracy~J Trothen}.} \bibinfo{year}{2022}\natexlab{}.
\newblock \showarticletitle{Intelligent Assistive Technology Ethics for Aging Adults: Spiritual Impacts as a Necessary Consideration}.
\newblock \bibinfo{journal}{\emph{Religions}} \bibinfo{volume}{13}, \bibinfo{number}{5} (\bibinfo{year}{2022}), \bibinfo{pages}{452}.
\newblock


\bibitem[Vacher et~al\mbox{.}(2015)]%
        {vacher2015evaluation}
\bibfield{author}{\bibinfo{person}{Michel Vacher}, \bibinfo{person}{Sybille Caffiau}, \bibinfo{person}{Fran{\c{c}}ois Portet}, \bibinfo{person}{Brigitte Meillon}, \bibinfo{person}{Camille Roux}, \bibinfo{person}{Elena Elias}, \bibinfo{person}{Benjamin Lecouteux}, {and} \bibinfo{person}{Pedro Chahuara}.} \bibinfo{year}{2015}\natexlab{}.
\newblock \showarticletitle{Evaluation of a context-aware voice interface for ambient assisted living: qualitative user study vs. quantitative system evaluation}.
\newblock \bibinfo{journal}{\emph{ACM Transactions on Accessible Computing (TACCESS)}} \bibinfo{volume}{7}, \bibinfo{number}{2} (\bibinfo{year}{2015}), \bibinfo{pages}{1--36}.
\newblock


\bibitem[Vines et~al\mbox{.}(2015)]%
        {vines2015age}
\bibfield{author}{\bibinfo{person}{John Vines}, \bibinfo{person}{Gary Pritchard}, \bibinfo{person}{Peter Wright}, \bibinfo{person}{Patrick Olivier}, {and} \bibinfo{person}{Katie Brittain}.} \bibinfo{year}{2015}\natexlab{}.
\newblock \showarticletitle{An age-old problem: Examining the discourses of ageing in HCI and strategies for future research}.
\newblock \bibinfo{journal}{\emph{ACM Transactions on Computer-Human Interaction (TOCHI)}} \bibinfo{volume}{22}, \bibinfo{number}{1} (\bibinfo{year}{2015}), \bibinfo{pages}{1--27}.
\newblock


\bibitem[Vitak et~al\mbox{.}(2021)]%
        {vitak2021designing}
\bibfield{author}{\bibinfo{person}{Jessica Vitak}, \bibinfo{person}{Michael Zimmer}, \bibinfo{person}{Anna Lenhart}, \bibinfo{person}{Sunyup Park}, \bibinfo{person}{Richmond Y.~Wong}, {and} \bibinfo{person}{Yaxing Yao}.} \bibinfo{year}{2021}\natexlab{}.
\newblock \showarticletitle{Designing for data awareness: addressing privacy and security concerns about “smart” technologies}. In \bibinfo{booktitle}{\emph{Companion Publication of the 2021 Conference on Computer Supported Cooperative Work and Social Computing}}. \bibinfo{pages}{364--367}.
\newblock


\bibitem[Vivacqua and Garcia(2018)]%
        {vivacqua2018personal}
\bibfield{author}{\bibinfo{person}{Adriana~S Vivacqua} {and} \bibinfo{person}{Ana Cristina~Bicharra Garcia}.} \bibinfo{year}{2018}\natexlab{}.
\newblock \showarticletitle{Personal neighborhood networks for senior citizen support}. In \bibinfo{booktitle}{\emph{Companion of the 2018 ACM Conference on Computer Supported Cooperative Work and Social Computing}}. \bibinfo{pages}{293--296}.
\newblock


\bibitem[Wallace et~al\mbox{.}(2017)]%
        {wallace2017technologies}
\bibfield{author}{\bibinfo{person}{James~R Wallace}, \bibinfo{person}{Saba Oji}, {and} \bibinfo{person}{Craig Anslow}.} \bibinfo{year}{2017}\natexlab{}.
\newblock \showarticletitle{Technologies, methods, and values: changes in empirical research at CSCW 1990-2015}.
\newblock \bibinfo{journal}{\emph{Proceedings of the ACM on Human-Computer Interaction}} \bibinfo{volume}{1}, \bibinfo{number}{CSCW} (\bibinfo{year}{2017}), \bibinfo{pages}{1--18}.
\newblock


\bibitem[Wang et~al\mbox{.}(2022)]%
        {wang2022lightweight}
\bibfield{author}{\bibinfo{person}{Bin Wang}, \bibinfo{person}{Xingjiao Wu}, \bibinfo{person}{Miaomiao Gong}, \bibinfo{person}{Jin Zhao}, {and} \bibinfo{person}{Yuling Sun}.} \bibinfo{year}{2022}\natexlab{}.
\newblock \showarticletitle{Lightweight Network Based Real-time Anomaly Detection Method for Caregiving at Home}. In \bibinfo{booktitle}{\emph{2022 IEEE 25th International Conference on Computer Supported Cooperative Work in Design (CSCWD)}}. IEEE, \bibinfo{pages}{1323--1328}.
\newblock


\bibitem[Wang et~al\mbox{.}(2019)]%
        {wang2019technology}
\bibfield{author}{\bibinfo{person}{Shengzhi Wang}, \bibinfo{person}{Khalisa Bolling}, \bibinfo{person}{Wenlin Mao}, \bibinfo{person}{Jennifer Reichstadt}, \bibinfo{person}{Dilip Jeste}, \bibinfo{person}{Ho-Cheol Kim}, {and} \bibinfo{person}{Camille Nebeker}.} \bibinfo{year}{2019}\natexlab{}.
\newblock \showarticletitle{Technology to support aging in place: Older adults’ perspectives}. In \bibinfo{booktitle}{\emph{Healthcare}}, Vol.~\bibinfo{volume}{7}. MDPI, \bibinfo{pages}{60}.
\newblock


\bibitem[Wang(2023)]%
        {xiaohui}
\bibfield{author}{\bibinfo{person}{Xiaohui Wang}.} \bibinfo{year}{2023}\natexlab{}.
\newblock \showarticletitle{The development track, trend and approach of intelligent elderly care}.
\newblock \bibinfo{journal}{\emph{Decision \& Information}}  \bibinfo{volume}{2} (\bibinfo{year}{2023}), \bibinfo{pages}{62--73}.
\newblock


\bibitem[Wiles et~al\mbox{.}(2012)]%
        {wilesMeaningAgingPlace}
\bibfield{author}{\bibinfo{person}{Janine~L Wiles}, \bibinfo{person}{Annette Leibing}, \bibinfo{person}{Nancy Guberman}, \bibinfo{person}{Jeanne Reeve}, {and} \bibinfo{person}{Ruth~ES Allen}.} \bibinfo{year}{2012}\natexlab{}.
\newblock \showarticletitle{The meaning of “aging in place” to older people}.
\newblock \bibinfo{journal}{\emph{The gerontologist}} \bibinfo{volume}{52}, \bibinfo{number}{3} (\bibinfo{year}{2012}), \bibinfo{pages}{357--366}.
\newblock


\bibitem[Xiao et~al\mbox{.}(2024)]%
        {xiao2024chinese}
\bibfield{author}{\bibinfo{person}{He Xiao}, \bibinfo{person}{Xingjiao Wu}, \bibinfo{person}{Jialiang Tong}, \bibinfo{person}{Bangyan Li}, {and} \bibinfo{person}{Yuling Sun}.} \bibinfo{year}{2024}\natexlab{}.
\newblock \showarticletitle{Chinese Elderly Healthcare-Oriented Conversation: CareQA Dataset and Its Knowledge Distillation Based Generation Framework}. In \bibinfo{booktitle}{\emph{2024 IEEE International Conference on Bioinformatics and Biomedicine (BIBM)}}. IEEE, \bibinfo{pages}{3866--3871}.
\newblock


\bibitem[Zhang(2006)]%
        {zhang2006family}
\bibfield{author}{\bibinfo{person}{Hong Zhang}.} \bibinfo{year}{2006}\natexlab{}.
\newblock \showarticletitle{Family care or residential care? The moral and practical dilemmas facing the elderly in urban China}.
\newblock \bibinfo{journal}{\emph{Asian Anthropology}} \bibinfo{volume}{5}, \bibinfo{number}{1} (\bibinfo{year}{2006}), \bibinfo{pages}{57--83}.
\newblock


\bibitem[Zhou et~al\mbox{.}(2018)]%
        {zhou2018data}
\bibfield{author}{\bibinfo{person}{Jie Zhou}, \bibinfo{person}{Yuling Sun}, {and} \bibinfo{person}{Liang He}.} \bibinfo{year}{2018}\natexlab{}.
\newblock \showarticletitle{A data driven collaborative caring framework for aging in place in China}. In \bibinfo{booktitle}{\emph{2018 IEEE 22nd International Conference on Computer Supported Cooperative Work in Design (CSCWD)}}. IEEE, \bibinfo{pages}{359--364}.
\newblock


\bibitem[Zimmerman(2001)]%
        {zimmermanAssistedLivingNeeds2001}
\bibfield{author}{\bibinfo{person}{Sheryl Zimmerman}.} \bibinfo{year}{2001}\natexlab{}.
\newblock \bibinfo{booktitle}{\emph{Assisted Living: Needs, Practices, and Policies in Residential Care for the Elderly}}.
\newblock \bibinfo{publisher}{JHU Press}.
\newblock


\end{thebibliography}



\end{document}